\documentclass[12pt, a4paper]{article}

\usepackage{subfigure}
\usepackage{graphicx}
\usepackage{amsmath}
\usepackage{amssymb}
\usepackage{ccaption}
\numberwithin{equation}{section}

\captionnamefont{\bfseries}
  \captiontitlefont{\small\sffamily}
  \captiondelim{: }
  \hangcaption

\else
\fi \topmargin -30pt

\begin{document}
\title{Quantum complex sine-Gordon \\ dressed boundaries}
\author{P. Bowcock\footnote{peter.bowcock@durham.ac.uk}\;\; and J. M. Umpleby\footnote{j.m.umpleby@durham.ac.uk} \\   Centre for Particle Theory\\ Department of Mathematical Sciences \\ University of Durham \\ Durham, DH1 3LE, U.K.  }
\maketitle

\begin{abstract}
In this paper we investigate the quantum reflection factor for the CSG dressed boundary, previously constructed by dressing the Dirichlet boundary with the integrable CSG defect \cite{Bowcock:2008jf}. We analyse classical bound states and use semi-classical methods to investigate the quantum 
boundary spectrum.
We conjecture  a fully quantum reflection matrix for  a particle reflecting from an unexcited boundary. By
using the reflection and boundary bootstrap equations,  the reflection matrix for a charge $Q=+n$ soliton reflecting from the $m^{th}$ excited boundary is constructed. Evidence supporting our conjecture is given by checking that the bootstrap closes and that the reflection matrices agrees with known results in the classical limit. A partial analysis of the poles in the reflection matrices which arise
from Coleman-Thun diagrams is given.
\end{abstract}

\newpage

\section{Introduction}

 The CSG model was first introduced independently by  Lund and Regge as a model of relativistic vortices in a superfluid \cite{Lund:1976ze,Lund:1977dt} and by Pohlmeyer in a dimensional reduction of a $O(4)$ non-linear $\sigma$-model\cite{Pohlmeyer:1979ch}. It belongs to the class of homogeneous sine-Gordon theories, which are $G/U(1)$ gauged Wess-Zumino-Witten Models perturbed by a potential. For CSG the group $G=SU(2)$. More recently, the quantum case has been studied by Dorey and Hollowood\cite{Dorey:1994mg} and Maillet and de Vega\cite{deVega:1982sh}. The CSG model has been studied with a boundary\cite{Bowcock:2002vz,Bowcock:2006hj}. Boundary conditions were found which preserved integrability and soliton solutions were analysed. 

In \cite{Bowcock:2008jf} we constructed a integrable CSG defect which we then analysed classically. It was found that charge can be transferred to and from the defect allowing soliton absorption and emission processes. The classical time-delay for soliton-defect scattering was calculated and also the reflectionless particle transmission factor was given. Using the CSG defect, we constructed a more general CSG boundary theory by dressing the Dirichlet boundary. This led to boundaries described by two parameters. As for the defect, the dressed boundaries can allow soliton absorption and emission, and in these cases the two boundary parameters are related to the charge and rapidity of the emitted/absorbed soliton. In addition, some boundaries allow for a one-parameter family of classical bound states.  The soliton-boundary reflection time-delay was also calculated for these excited boundaries.

Most recently the CSG model has attracted some attention in the context of magnons in string theory.
The CSG equation is equivalent to the equations of motion of a string moving on an $\mathbb{R}$ x $S^{3}$ subspace of $AdS_{5}$ x $S^{5}$ \cite{Pohlmeyer:1975nb}. This equivalence is used in current work verifying the prediction of the $A$d$S$/CFT correspondence that the spectrum of operator dimensions in planar $\mathcal{N}=4$ SUSY Yang-Mills and the spectrum of a free strings on $A$d$S_{5}$ x $S^{5}$\cite{Chen:2006gea} are the same. Integrable boundaries have already been used in this context \cite{Hofman:2007xp, Ahn:2007bq}.

The complex sine-Gordon theory is a 1+1 dimensional field theory described by the following Lagrangian

\begin{eqnarray}\label{eq:bulklagrangian}
 \mathcal{L}_{CSG}& =&  \frac{ \partial_{t} u \partial_{t}u^{*} -\partial_{x} u \partial_{x} u^{*} }{1-  \lambda^{2} u u^{*}} - 4\beta uu^{*} \, .
\end{eqnarray}
Here $u$ is a complex field, $\lambda$ is the coupling constant  and $\beta$ the mass parameter. The constant $\lambda$ can be absorbed into $u$ and $u^{*}$ by scaling the field, in which case it appears as an overall factor multiplying the Lagrangian.  Note that the Lagrangian has a global $U(1)$ symmetry, this leads to a conserved charge by Noether's theorem. 

\begin{figure}[htb]
\fbox{
  \centering
  \begin{minipage}[c]{0.45\textwidth}
    \centering
 \resizebox{\textwidth}{!}{
   \begin{tabular}{cc}
		
		$0\ <\ a\ <\ \frac{\pi}{2}$& $ Q_{sol} =\ \frac{1}{\lambda^{2}}(-4a + 2\pi)$\\
		&\\
		$\frac{\pi}{2}\ <\ a\ <\ \pi$ &$ Q_{sol} =\ \frac{1}{\lambda^{2}} (-2\pi+4a)$\\ 
		&\\
		$-\frac{\pi}{2}\ <\ a <\ 0 $& $ Q_{sol} =\ \frac{1}{\lambda^{2}}(-4a - 2\pi) $\\
		&\\
		$-\pi\ <\ a\ <\ -\frac{\pi}{2}$&$ Q_{sol} =\ \frac{1}{\lambda^{2}}(4a + 2\pi)$\\ 
		
	\end{tabular}}
  \end{minipage}
  \begin{minipage}[c]{0.45\textwidth}
    \includegraphics[width=0.8\textwidth,height=0.8\textwidth]{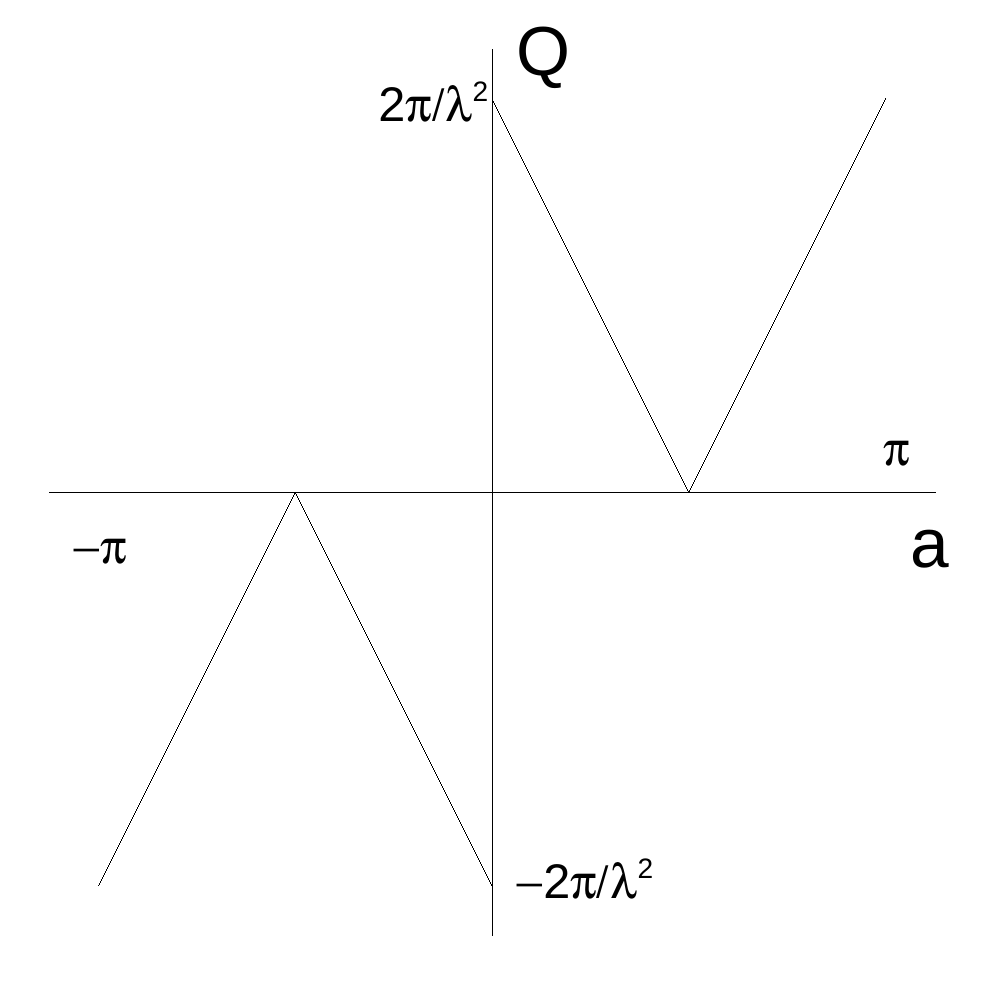}
  \end{minipage}
}
  \caption{Charge $Q(a)$ of the complex sine-Gordon soliton.}\label{fig:solitoncharge}
\end{figure}

The complex sine-Gordon equation of motion (and its complex conjugate)
\begin{eqnarray}\label{eq:bulkeom}
 \partial_{tt} u - \partial_{xx} u  + \frac{u^{*} ((\partial_{t} u)^{2}-(\partial_{x}u)^{2})}{1-uu^{*}} + 4\beta u(1-uu^{*}) &=&0
\end{eqnarray}
are derived by varying the action $S = \int \ dt L$ in the usual way. One can straightforwardly derive the forms of the conserved energy, momentum and charge to respectively be

\begin{eqnarray}\label{eq:energy}
 E &=& \frac{1}{\lambda^{2}}\int dx \ \ \frac{ \partial_{t}u\partial_{t}u^{*}+\partial_{x}u\partial_{x}u^{*}}{1-uu^{*}} + 4\beta uu^{*}\, ,\nonumber\\
 P&=&-\frac{1}{\lambda^{2}}\int dx \ \ \ \frac{\partial_{x}u\partial_{t}u^{*} +\partial_{x}u^{*}\partial_{t}u}{1-uu^{*}}\, , \nonumber\\
Q &=& \frac{i}{\lambda^{2}} \ \int dx \ \frac{u\partial_{t}u^{*}-u^{*}\partial_{t}u}{1-uu^{*}}\, .
\end{eqnarray}
In \cite{Bowcock:2008jf} we showed that it is possible to write the B\"{a}cklund transformation (BT) in the form

\begin{eqnarray}\label{eq:DefectCond}
0&=& \frac{u_{t}-u_{x}}{\sqrt{1-uu^{*}}} - \frac{w_{t}-w_{x}}{\sqrt{1-ww^{*}}}e^{i\alpha} +2\sqrt{\beta}\delta\left(w \sqrt{1-uu^{*}} + u\sqrt{1-ww^{*}} e^{i\alpha}\right)\, ,\nonumber\\ 
0&=& \frac{u_{t}+u_{x}}{\sqrt{1-uu^{*}}}e^{i\alpha} + \frac{w_{t}+w_{x}}{\sqrt{1-ww^{*}}} -\frac{2\sqrt{\beta}}{\delta}\left(u \sqrt{1-ww^{*}} - w\sqrt{1-uu^{*}} e^{i\alpha}\right)\, ,
\end{eqnarray}
where $\alpha$ is given by
\begin{equation}\label{eq:alpha}
\alpha =\mathrm{arcsin}\left[\frac{i}{2}\left(\frac{uw^{*}-wu^{*} +2i\mathrm{sin}A}{\sqrt{1-ww^{*}}\sqrt{1-uu^{*}}}      \right) \right] \, .
\end{equation}
Like the SG theory the CSG theory admits soliton solutions. The form of a one-soliton solution can be found by substituting $u=0,\ u^{*}=0$ into the BT and solving for $w$, which yields

\begin{eqnarray}\label{eq:1sol}
w=u_{1-sol}&=&\frac{ \mathrm{cos}(a) e^{2i\sqrt{\beta}\mathrm{sin}(a)(t\, \mathrm{cosh}(\theta)-x\,\mathrm{sinh}(\theta))}}{\mathrm{cosh}(2\sqrt{\beta}\mathrm{cos}(a)(x\ ,\mathrm{cosh}(\theta)-t\ ,\mathrm{sinh}(\theta)))} \, .
\end{eqnarray}
The parameters in the soliton solution are related to the parameters in the BT by $a=A$ and  $ e^{\theta} = \delta$. It is noted that unlike the SG soliton the CSG soliton is non-topological. The CSG soliton has a conserved charge due to the $U(1)$ invariance of the theory. Classically this charge lies can take any value in a (finite) continuous range, in contrast to the discrete topological charge that the SG soliton holds. The energy of the soliton is always positive, while the momentum can be positive or negative. They both depend on the charge parameter $a$ and the rapidity of the soliton $\theta$  

\begin{eqnarray}\label{eq:energysol}
E_{sol} = \frac{8\sqrt{\beta}}{\lambda^{2}}\left|\mathrm{cos}(a)\right|\mathrm{cosh}(\theta) \, , \hspace{0.5in}
P_{sol} = \frac{8\sqrt{\beta}}{\lambda^{2}}\left|\mathrm{cos}(a)\right| \mathrm{sinh}(\theta)\, . 
\end{eqnarray}
The charge of the soliton $Q_{sol}$, graphically illustrated in figure \ref{fig:solitoncharge}, is $2\pi$ periodic in $a$.

The next section summarises the CSG dressed boundary theory introduced in \cite{Bowcock:2008jf}. . We briefly recall the use of soliton solutions in the boundary theory which describe both scattering and also classical bound states. The classical particle reflection factor from this boundary bound state is calculated. In the third section we review some aspects of the quantum $S$-matrix in the bulk CSG theory including some analysis of the pole structure.

The fourth section contains the majority of the new work in this paper. We use semi-classical methods to get a handle on the quantum spectrum of bound states. We use this information to  conjecture a quantum reflection matrix for the $Q=+1$ CSG soliton (or particle) from the unexcited boundary consistent with the classical particle reflection factor. By implementing the reflection and boundary bootstrap, we generate the general quantum reflection matrix for any charged CSG soliton from any excited boundary. In the following section we offer some analysis and explanation for the pole structure in the quantum reflection matrices.

In the final section, we summarise the work covered and provide some discussion and thoughts about future directions.

\section{Complex sine-Gordon dressed boundary}\label{sec:CSGdb}

In \cite{Bowcock:2008jf}  we constructed a new CSG boundary by placing an integrable CSG defect in front of a Dirichlet boundary and moving the defect up to the boundary to create the dressed boundary.
\begin{figure}[!h]
\begin{center}
\includegraphics[width=0.75\textwidth]{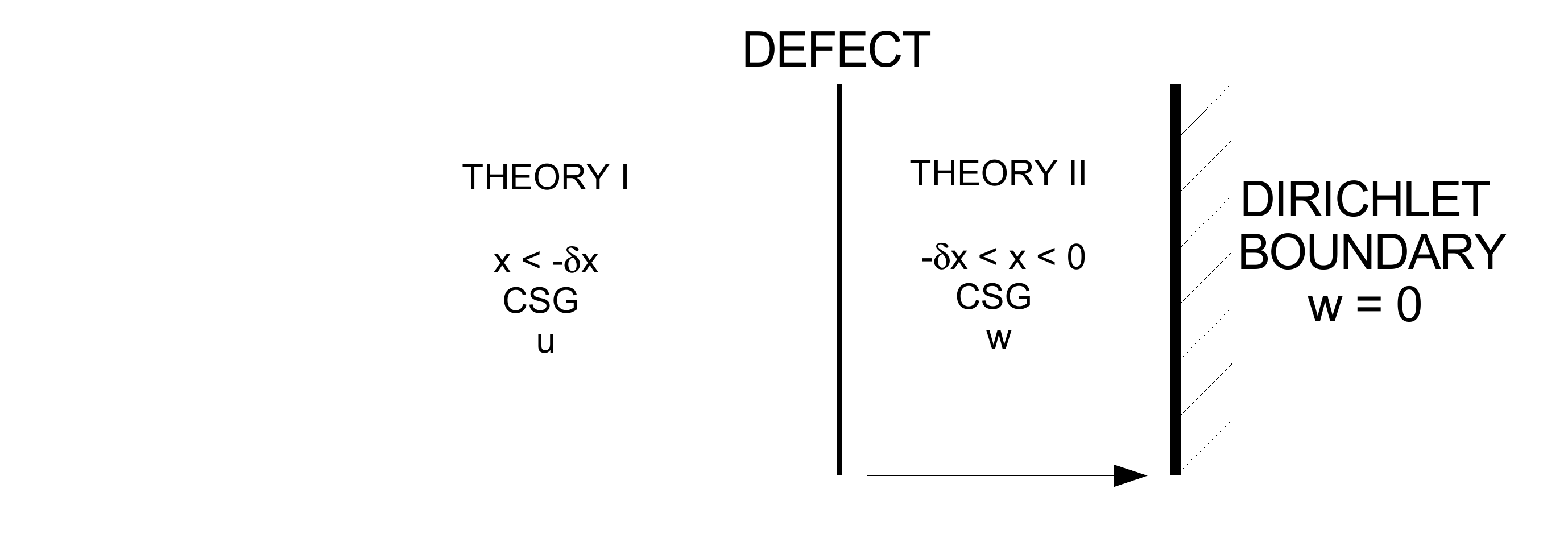}
\end{center}
\caption{Dressed boundary model set up.}
\label{Fig:dressedboundary}
\end{figure}

The dressed boundary theory is described by the Lagrangian
\begin{eqnarray}\label{eq:dressedboundarylagrangianA}
 \mathcal{L}& =& \frac{1}{\lambda^{2}} \int^{0}_{-\infty} dx\ \  \frac{ \partial_{t} u \partial_{t}u^{*} -\partial_{x}u \partial_{x} u^{*} }{1-   u u^{*}} - 4\beta uu^{*}+ \frac{1}{\lambda^{2}} \left.\bigl[A_{1} \partial_{t}u + A_{2} \partial_{t}u^{*}  -\mathcal{L}_{db}\bigr]\right|_{x=0} \, ,
\end{eqnarray}
with the standard bulk Lagrangian and a boundary piece made up of a boundary potential term and terms in linear in the time derivative of the field. The boundary terms are 
\begin{equation}
\mathcal{L}_{db} =  2\sqrt{\beta}\left(\delta + \frac{1}{\delta}\right) \mathrm{cos}(\alpha')\sqrt{ 1 -uu^{*}}\, , \hspace{0.5in}
A_{1} = -\frac{i}{u}\alpha' \, , \hspace{0.5in} A_{2} = \frac{i}{u^{*}}\alpha'\, , 
\end{equation}
where
\begin{equation}
\alpha' = \mathrm{arcsin}\left(\frac{-\mathrm{sin}A}{\sqrt{1-uu^{*}}}      \right) \, .
\end{equation}
Varying the action gives the dressed boundary conditions
\begin{eqnarray}\label{eq:dressedBC1}
\partial_{x} u &=& -\partial_{t}u \ i\  \mathrm{tan}(\alpha') + \frac{\sqrt{\beta}}{\mathrm{cos}(\alpha')}\left(\delta+\frac{1}{\delta}\right)u\sqrt{1-uu^{*}} \, ,\nonumber\\
\partial_{x} u^{*} &=& \partial_{t}u^{*} \ i\  \mathrm{tan}(\alpha') + \frac{\sqrt{\beta}}{\mathrm{cos}(\alpha')}\left(\delta+\frac{1}{\delta}\right)u^{*}\sqrt{1-uu^{*}} \, ,
\end{eqnarray}
which depend on the two parameters $\delta\, , A$ that appear in the CSG BT. The conservation of the energy 
\begin{eqnarray}\label{eq:boundaryenergy}
E_{db} &=& \frac{1}{\lambda^{2}}\int^{0}_{-\infty} dx \ \ \frac{ \partial_{t}u \partial_{t} u^{*} + \partial_{x}u\partial_{x} u^{*}}{1-uu^{*}} + 4\beta uu^{*} + \frac{1}{\lambda^{2}}\left.\left[ 2\sqrt{\beta}\left(\delta + \frac{1}{\delta}\right) \mathrm{cos}(\alpha')\sqrt{ 1 -uu^{*}} \right]\right|_{x=0}\nonumber \\
\end{eqnarray}
and charge 
\begin{eqnarray}\label{eq:boundarycharge}
Q_{db} &=& \frac{i}{\lambda^{2}}\int^{0}_{-\infty} dx \ \ \frac{u\partial_{t}u^{*} -u^{*}\partial_{t}u}{1-uu^{*}}  +\frac{1}{\lambda^{2}}\left.\bigl[2\alpha'  \right]\bigr|_{x=0}\, ,
\end{eqnarray}
can be easily checked. As is usual for boundary theories there is no conserved momentum. In \cite{Bowcock:2008jf} we showed that the lack of conserved momentum does not stop the theory from being classically integrable, by explicitly constructing the next higher-spin energy-like conserved charge. 

We continue by introducing the different soliton solutions in the dressed boundary theory. The vacuum of the dressed theory is $u=0$, with energy and charge
\begin{eqnarray}\label{eq:unexenergy}
E_{vac}& =&\frac{2\sqrt{\beta}}{\lambda^{2}}\left(\delta + \frac{1}{\delta}\right)  \mathrm{cos}(\alpha_0)\, ,\nonumber\\
Q_{vac} &=& \frac{2\alpha_0}{\lambda^{2}} \, ,
\end{eqnarray}
where $\alpha_0 = \alpha'(u=0)$. The energy and charge of the vacuum depend on the two parameters that appear in the BT and explicitly on $\alpha_0$. This $\alpha_0$ dependence means that the dressed boundary can have different values of energy and charge for the same values of $\delta$ and $A$. Whether the energy of the dressed boundary is positive or negative determines its properties. Namely it is found that if the boundary has positive energy then it can emit a soliton and negative energy boundaries can absorb a soliton \cite{Bowcock:2008jf}. We found that solitons which do not have their parameters matched with the boundary and are therefore not absorbed, reflect from the boundary with the non-zero time-delay 
\begin{eqnarray}\label{eq:dbtimedelay}
\Delta t &=& \frac{1}{ 2\sqrt{\beta}\mathrm{cos}(a)\mathrm{sinh}(\theta)}\mathrm{ln} \left| \frac{\mathrm{sinh}\left(\frac{\theta-\chi}{2} + i\frac{a-A}{2}\right)\mathrm{sinh}\left(\frac{\theta+\chi}{2} + i\frac{a-A}{2}\right)}{\mathrm{cosh}\left(\frac{\theta-\chi}{2} + i\frac{a+A}{2}\right)\mathrm{cosh}\left(\frac{\theta+\chi}{2} + i\frac{a+A}{2}\right)}\right| \, ,\nonumber\\
&&
\end{eqnarray}
where $\delta = e^{\chi}$ and phase shift
\begin{eqnarray}\label{eq:dbphase}
e^{i\phi} &=& -\left(\frac{\delta +e^{iA}e^{\theta}e^{ia}}{e^{\theta}e^{ia}-\delta e^{iA}}\right)\left(\frac{1 +\delta e^{iA} e^{ia} e^{\theta}}{e^{iA} - \delta e^{\theta}e^{ia}}\right)e^{2\sqrt{\beta}(\mathrm{sinh}(\theta+ia)\Delta t} \, .\nonumber \\
&&
\end{eqnarray}   
Both the emission and absorption processes appear as limits of the classical time-delay. The CSG particle also reflects from the dressed boundary with the reflection factor
\begin{eqnarray}\label{eq:particlereflection}
R_{particle} &=& \frac{2i\ \mathrm{sinh}(\theta+iA) + (\delta+\frac{1}{\delta})}{2i\ \mathrm{sinh}(\theta-iA) - (\delta+\frac{1}{\delta})}\, .
\end{eqnarray}
This particle reflection factor can also be obtained as the $a = \frac{\pi}{2}$ limit of the soliton reflection phase factor.

We found that there exists classical dressed boundary bound states. These can be constructed by solving the boundary conditions with a stationary soliton solution, (\ref{eq:1sol}) with $\theta=0$, in the bulk with its position shifted $x \rightarrow x-c$. We find that the boundary conditions are satisfied when 
\begin{eqnarray}\label{eq:boundconstraint}
\left( \delta+\frac{1}{\delta}\right) &=& 2\frac{ \mathrm{cos}(a) \mathrm{sinh}(C) \sqrt{ \mathrm{cos}^{2}(A) \mathrm{cosh}(C)^{2} - \mathrm{cos}^{2}(a)} + \mathrm{sin}(a)\mathrm{sin}(A)\mathrm{cosh}(C)^{2}}{\mathrm{cosh}(C)^{2} -\mathrm{cos}^{2}(a)}\, ,\nonumber \\
\end{eqnarray} 
where $C = 2\sqrt{\beta}\mathrm{cos}(a) c$. This constraint is only valid when the argument in the square root is greater than zero and
\begin{equation}\label{eq:bsconstraint}
\frac{1}{2}\left(\delta + \frac{1}{\delta}\right) \in \Bigl[ \mathrm{min}\left\{ -\cos(A+a), \cos(A-a)\right\},\ \mathrm{max}\left\{ -\cos(A+a), \cos(A-a) \right\}  \Bigr] \, , 
\end{equation}
which implies that $\delta$ is a pure phase within the range specified. We can solve the constraint (\ref{eq:boundconstraint}) to find
\begin{equation}\label{eq:tanhC}
\tanh(2\sqrt{\beta}\cos(a)c) = \pm \frac{(\cos(b)\, \sin(a) - \sin(A))}{\cos(a)\, \sin(b)} \, , \ \ \ \pm i \tan(a)\, ,
\end{equation}
with the last two solutions discounted as they infer $c$ is complex. Here we have defined $b$ through the relation $ \frac{1}{2}(\delta+\frac{1}{\delta}) = \mathrm{cos}(b)$. There are therefore two positions where the bound soliton can be placed for the boundary conditions to be satisfied, and these are related by the parity transformation 
$x\rightarrow -x$. To understand these solutions, it is useful to rewrite the boundary conditions (\ref{eq:dressedBC1}) in the form 
\begin{eqnarray}\label{eq:dressedBC10}
\sqrt{\cos(A)-uu^{*}}\,\partial_{x} u &=& \partial_{t}u \ i\  \sin(A)+ \sqrt{\beta}\left(\delta+\frac{1}{\delta}\right)u(1-uu^{*})\, .
\end{eqnarray}
Using the expression (\ref{eq:tanhC}) for $\tanh(2\sqrt{\beta}\cos(a)c)$, it is easy to show that 
\begin{eqnarray}\label{eq:cosalpha}
\sqrt{\cos(A)-uu^{*}}=\pm \frac{(\cos(b)\, \sin(a)\, -\sin(A))}{\sin(b)}
\end{eqnarray}
and for the  boundary condition (\ref{eq:dressedBC10}) to be satisfied, the signs in (\ref{eq:tanhC}) and 
(\ref{eq:cosalpha}) must be correlated.
Thus the two solutions for the position of the soliton correspond to different choices of the sign of the square root in the boundary conditions (\ref{eq:dressedBC10}). The same square root appears in the boundary Lagrangian, and thus the two different solutions correspond to bound states of different 
boundaries. We calculate the energy and charge of the bound state by substituting the stationary soliton solution into the total energy (\ref{eq:boundaryenergy}) and charge (\ref{eq:boundarycharge}) respectively and simplify using the two valid solutions for $\tanh(2\sqrt{\beta}\cos(a)c)$ to give
\begin{eqnarray}\label{eq:bsenergy}
E^{\pm}_{bs} &=&  4\sqrt{\beta}\left(|\mathrm{cos}(a)| \pm \mathrm{sin}(A)\, \mathrm{sin}(b)\right)\, , \nonumber\\
Q^{\pm}_{bs} &=& Q_{bulk} \ \pm \ 2\left(b-\frac{\pi}{2}\right)\, ,
\end{eqnarray}
where
\begin{equation}
Q_{bulk} = \left\{ \begin{array}{lcrcl}2a - \pi &;&\frac{\pi}{2}&<a<&\pi \\ \pi -2a &;&0&<a<&\frac{\pi}{2}  \\ -2a -\pi &;&-\frac{\pi}{2}&<a<&0 \\ 2a +\pi &;&-\pi&<a<&-\frac{\pi}{2}\end{array} \right\}\, .
\end{equation}

\subsection{Particle reflection from bound state}

In this section we calculate the classical particle reflection factor from the dressed boundary bound state. Later we shall demand that  our conjectured quantum reflection factors tend to these in the classical limit. To find the classical reflection factor, we begin by finding the solution for a small perturbation around a stationary soliton solution, for a right-moving plane wave
\begin{equation}
e_{R}(x,t) = f(x) e^{-2i\sqrt{\beta}(\cosh(\theta)t-\sinh(\theta)x)}e^{4i\sqrt{\beta}\sin(a)t} +g_{R}(x) e^{2i\sqrt{\beta}(\cosh(\theta)t-\sinh(\theta)x)}\, ,
\end{equation}
where
\begin{eqnarray}
f(x) &=& \frac{1}{\cosh^{2}(2\sqrt{\beta}\cos(a)x)}\, , \nonumber \\
g_{R}(x) &=& c_{1} + c_{2} \tanh(2\sqrt{\beta}\cos(a)x) + \frac{1}{\cosh^{2}(2\sqrt{\beta}\cos(a)x)}
\end{eqnarray}
and
\begin{eqnarray}
c_{1} &=& -\frac{2i (e^{\theta}e^{ia}-e^{\theta}-ie^{ia}-i)(e^{\theta}e^{ia}+e^{\theta}-i+ie^{ia})e^{ia}}{e^{\theta}(e^{ia}-i)^{2}(e^{ia}+i)^{2}}\, ,\nonumber \\
c_{2} &=& -\frac{2i(e^{\theta}+1)(e^{\theta}-1)e^{ia}}{e^{\theta}(e^{ia}-i)(e^{ia}+i)} \, .
\end{eqnarray}
Similarly for a left-moving plane wave
\begin{equation}
e_{L}(x,t) = f(x) e^{-2i\sqrt{\beta}(\cosh(\theta)t+\sinh(\theta)x)}e^{4i\sqrt{\beta}\sin(a)t} +g_{L}(x) e^{2i\sqrt{\beta}(\cosh(\theta)t+\sinh(\theta)x)}\, ,
\end{equation}
where
\begin{eqnarray}
g_{L}(x) &=& c_{1} - c_{2} \tanh(2\sqrt{\beta}\cos(a)x) + \frac{1}{\cosh^{2}(2\sqrt{\beta}\cos(a)x)} \, .
\end{eqnarray}
By substituting a small perturbation around the stationary soliton into the boundary conditions we find the linearised boundary conditions around the bound state. This is a differential equation in the linearised bulk solution involving the one-soliton solution bound to the boundary. The linearised solution
which represents a particle reflecting off the boundary bound state is of the form
\begin{equation}
 E(x,t) = e_{R}(x-c,t) + {\rho}\  e_{L}(x-c,t)\, ,
\end{equation}
where $c$ is the position of the bound soliton determined by (\ref{eq:tanhC}). ${\rho}$ is a constant which can be found by demanding that the perturbation $E(x,t)$ satisfies the linearised boundary 
conditions.  By inspecting the $x\rightarrow -\infty$ limit of the resulting plane wave solution, the classical reflection factor is found to be
\begin{equation}\label{eq:ParticleBoundStateRef}
R_{bs} = \frac{(e^{\theta}+ie^{ib}e^{iA})(e^{iA}e^{\theta}-ie^{ib})(e^{\theta}e^{ia}-i)^{2}}{(e^{ib}e^{\theta}e^{iA}-i)(ie^{iA}+e^{ib}e^{\theta})(ie^{ia}+e^{\theta})^{2}}\, .
\end{equation}

\subsection{Descriptions of charged boundaries}\label{ref:chargebound}
We begin our examination of the spectrum of boundaries by comparing the two descriptions for charged boundaries. For the CSG dressed boundary, both unexcited boundary with the properties (\ref{eq:unexenergy}) and the boundary bound states with the properties (\ref{eq:bsenergy}) can carry charge. One idea is that these might provide two alternative descriptions for a single tower of charged
boundary states. However this turns out not to be the case. Despite having the freedom to set the charge and energies of unexcited and boundary bound states to coincide, we find that the particle reflection factors do not then coincide. We conclude that the unexcited boundary and bound state are not the same object, expect in particular cases. 

As an example of such a coincidence, consider the unexcited boundary with charge $Q$ which is described by the charge parameter $A = -\frac{Q}{2}$ .If we consider a bound state (using $E_{bs}^{+}$ and $Q_{bs}^{+}$ (\ref{eq:bsenergy})) described by the same $A$ with $0 < a <\frac{\pi}{2}$ then the bound state has charge $Q$ when $a = b +A$ and the energy
\begin{equation}
E_{bs}^{+} =   4\sqrt{\beta}\left(\cos(b +A) +  \sin(A)\sin(b)\right) = 4\sqrt{\beta}\cos(A)\cos(b)\, ,
\end{equation}
equals the energy of the unexcited boundary. We find that the particle reflection factors also agree in this limit. Therefore the bound state and unexcited boundary are the same object when the charge parameter of the bound soliton is the specific value $a=A+b$. There is a similar limit when using $E_{bs}^{-}$ and $Q_{bs}^{-}$ (\ref{eq:bsenergy}), in this case to bound state reduces to the unexcited boundary when $a= A-b$.

\begin{figure}[!h]
\begin{center}
\subfigure[]{\label{fig:bsrange1}\includegraphics[width=0.3\textwidth]{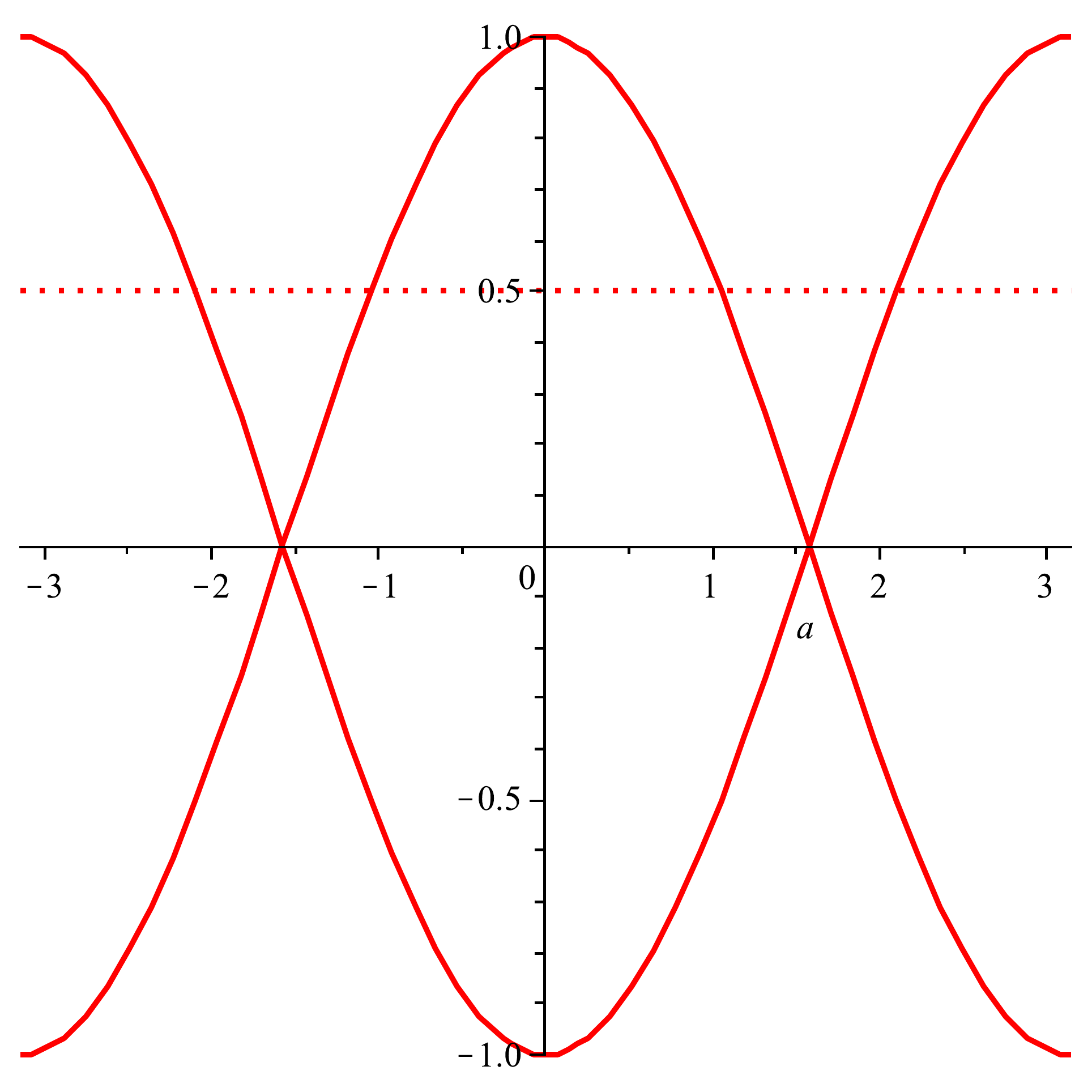}}
\hspace{0.5in}
\subfigure[]{\label{fig:bsrange2}\includegraphics[width=0.3\textwidth]{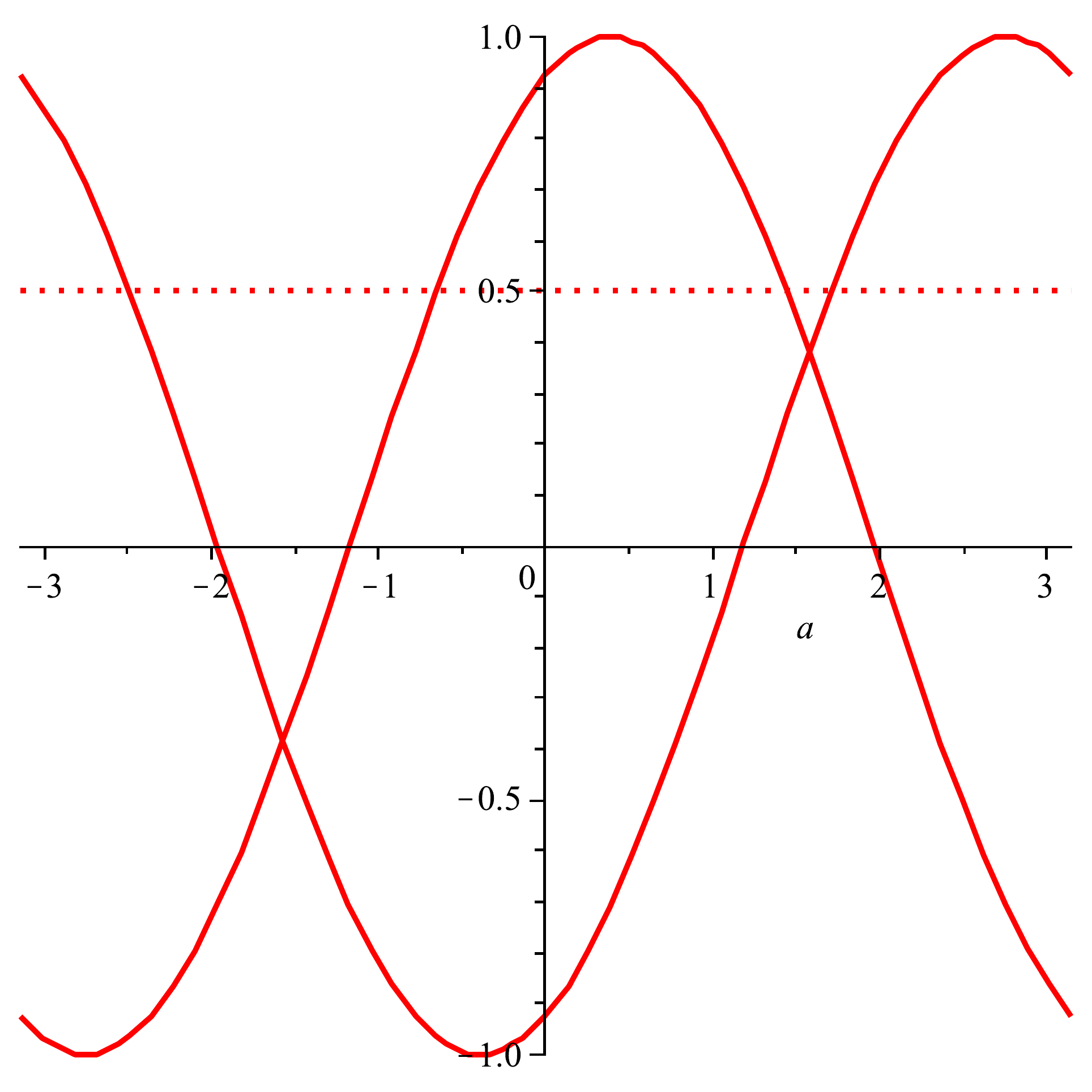}}
\end{center}
\caption[Values of $a$ where bound states exist.]{Plots of $\cos(b), \ \cos(A-a), \ -\cos(A+a),$ for (a) $A=0, \ b= \frac{\pi}{3}$, (b) $A=\frac{\pi}{8}, \ b= \frac{\pi}{3}$.}
\label{fig:bsrange}
\end{figure}

To understand these limits we analyse the bound state solutions. We reinterpret the allowed range of values for $\cos(b)$ (\ref{eq:bsconstraint}) as a constraint on the charge parameter $a$ of the bound soliton. Figure \ref{fig:bsrange} illustrates the values of $a$ for two boundaries, with the dotted line on the figures $\cos(b)$  and the two solid curves $\cos(A-a)$ and $-\cos(A+a)$.
Classical bound states exist for the values of $a$ when the two curves lie either side the dotted line. For both boundaries there are two separate regions, we concentrate on the region that includes $a=0$. Figure \ref{fig:bsrange1} shows that there exists bound states if $ -\frac{\pi}{3} < a < \frac{\pi}{3}$ for the boundary described by $A=0, \ b=\frac{\pi}{3}$. Similarly, figure \ref{fig:bsrange2} shows that there exists bound states if $ -\frac{\pi}{3} + \frac{\pi}{8} < a < \frac{\pi}{3} + \frac{\pi}{8}$ for the boundary described by $A=\frac{\pi}{8}, \ b=\frac{\pi}{3}$. We note that both ranges are between $A+b$ and $A-b$, which are precisely the values where the bound state reduces to the unexcited boundary.

\begin{figure}[!h]
\begin{center}
\subfigure[]{\label{fig:tanhC1}\includegraphics[width=0.3\textwidth]{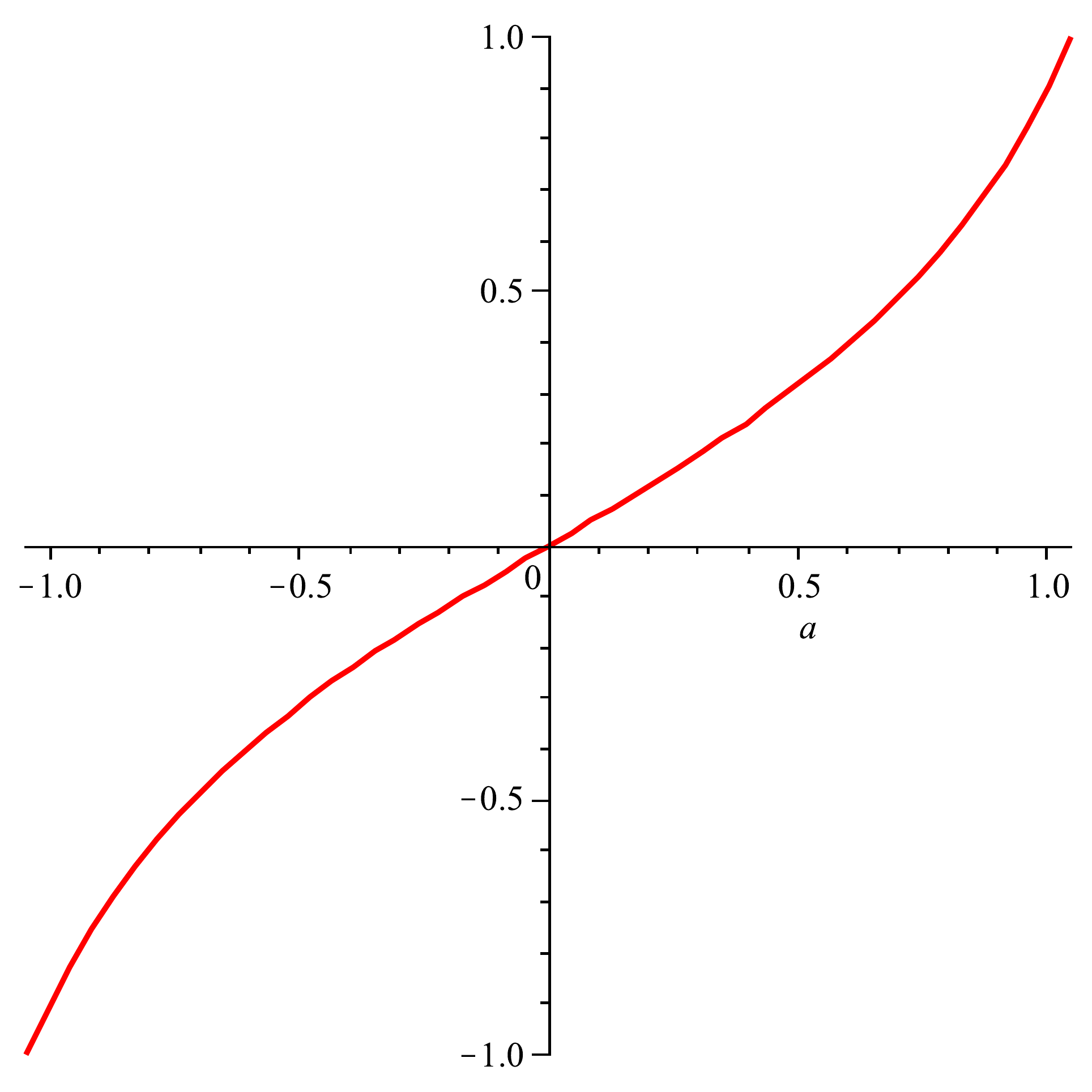}}
\hspace{0.5in}
\subfigure[]{\label{fig:tanhC2}\includegraphics[width=0.3\textwidth]{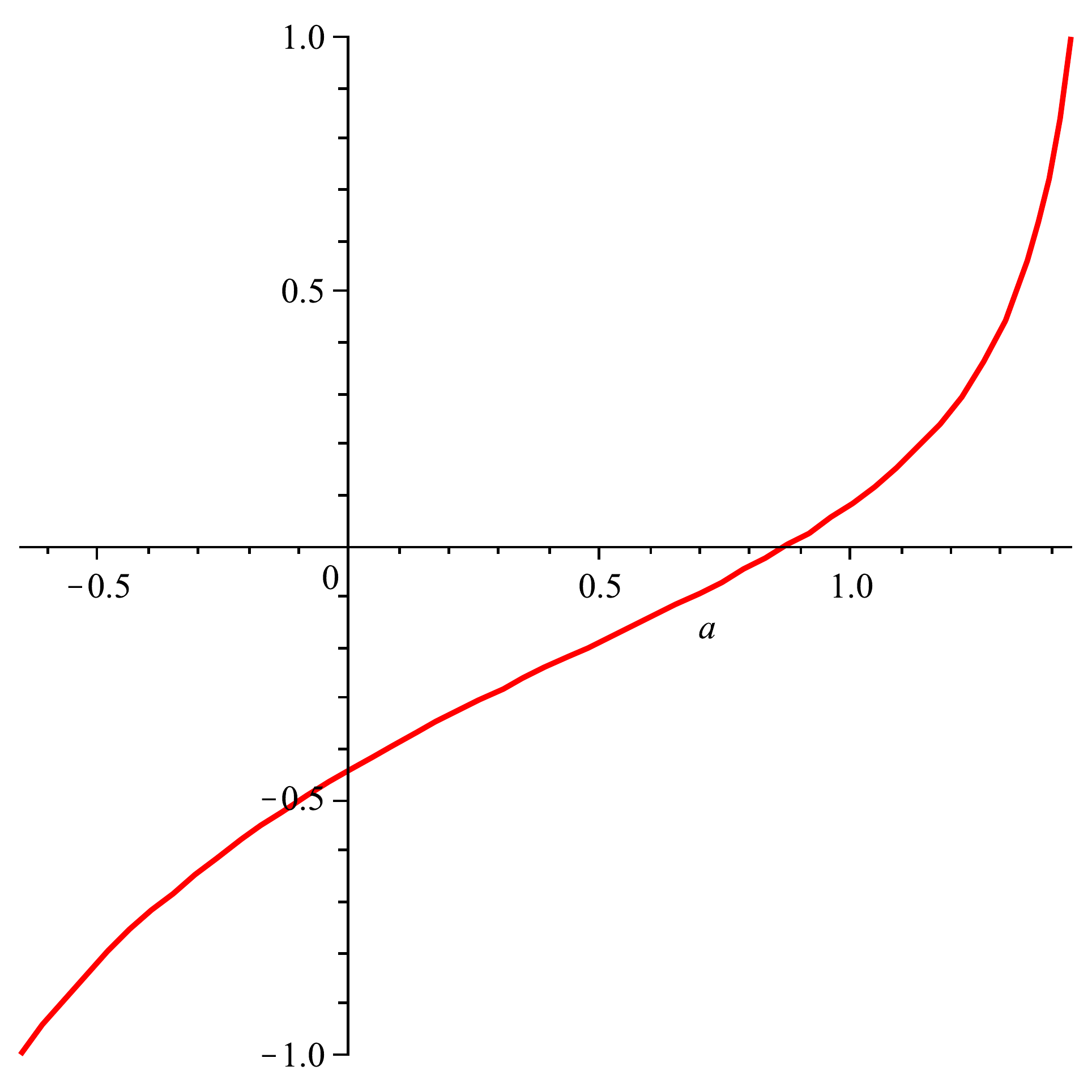}}
\end{center}
\caption[Position of bound soliton.]{Plots of $\tanh(2\sqrt{\beta}\cos(a)c)$ using plus sign with $\beta=1$ for (a) $A=0, \ b= \frac{\pi}{3}$, (b) $A=\frac{\pi}{8}, \ b= \frac{\pi}{3}$.}
\label{fig:tanhC}
\end{figure}

\begin{figure}[!h]
\begin{center}
\subfigure[]{\includegraphics[width=0.3\textwidth]{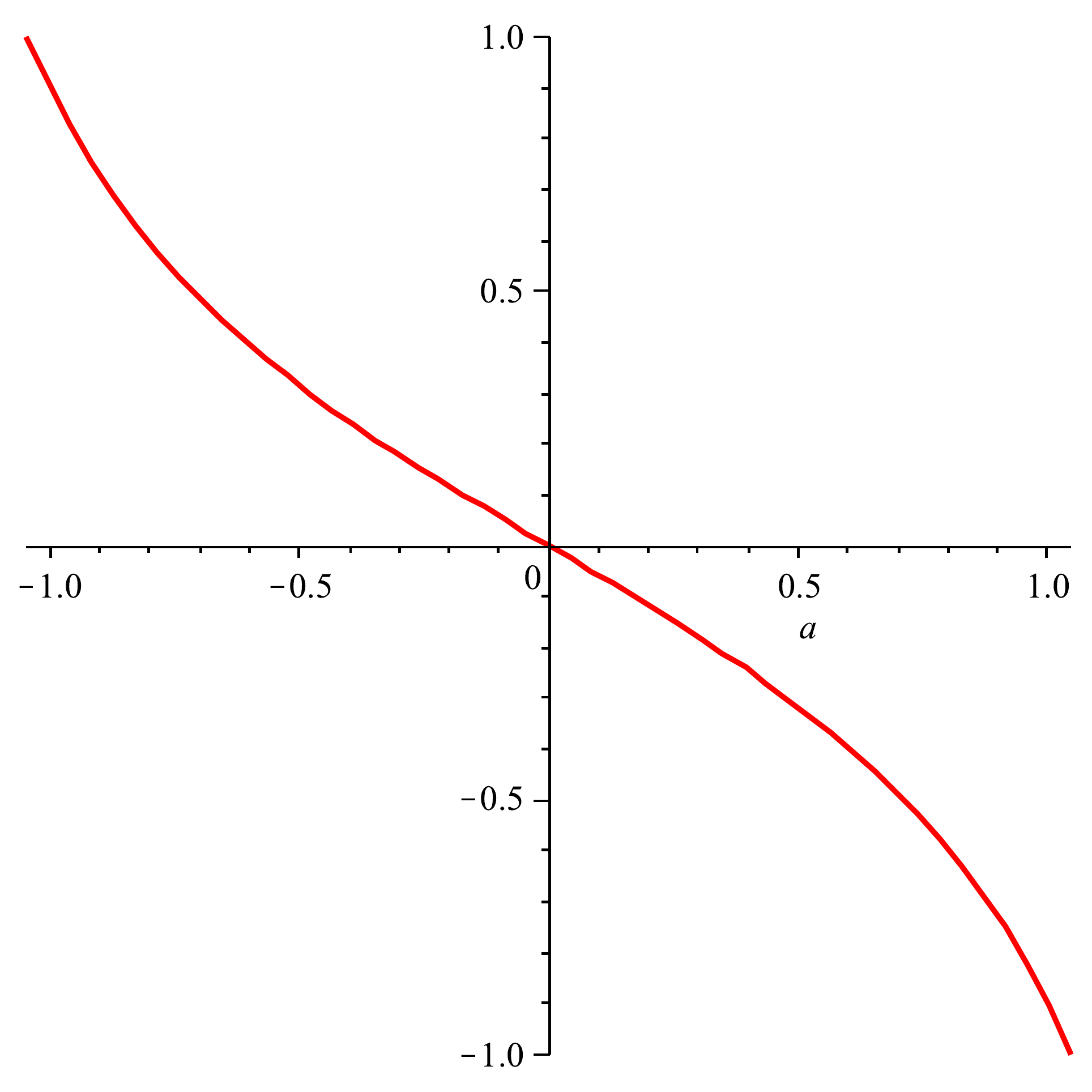}}
\hspace{0.5in}
\subfigure[]{\includegraphics[width=0.3\textwidth]{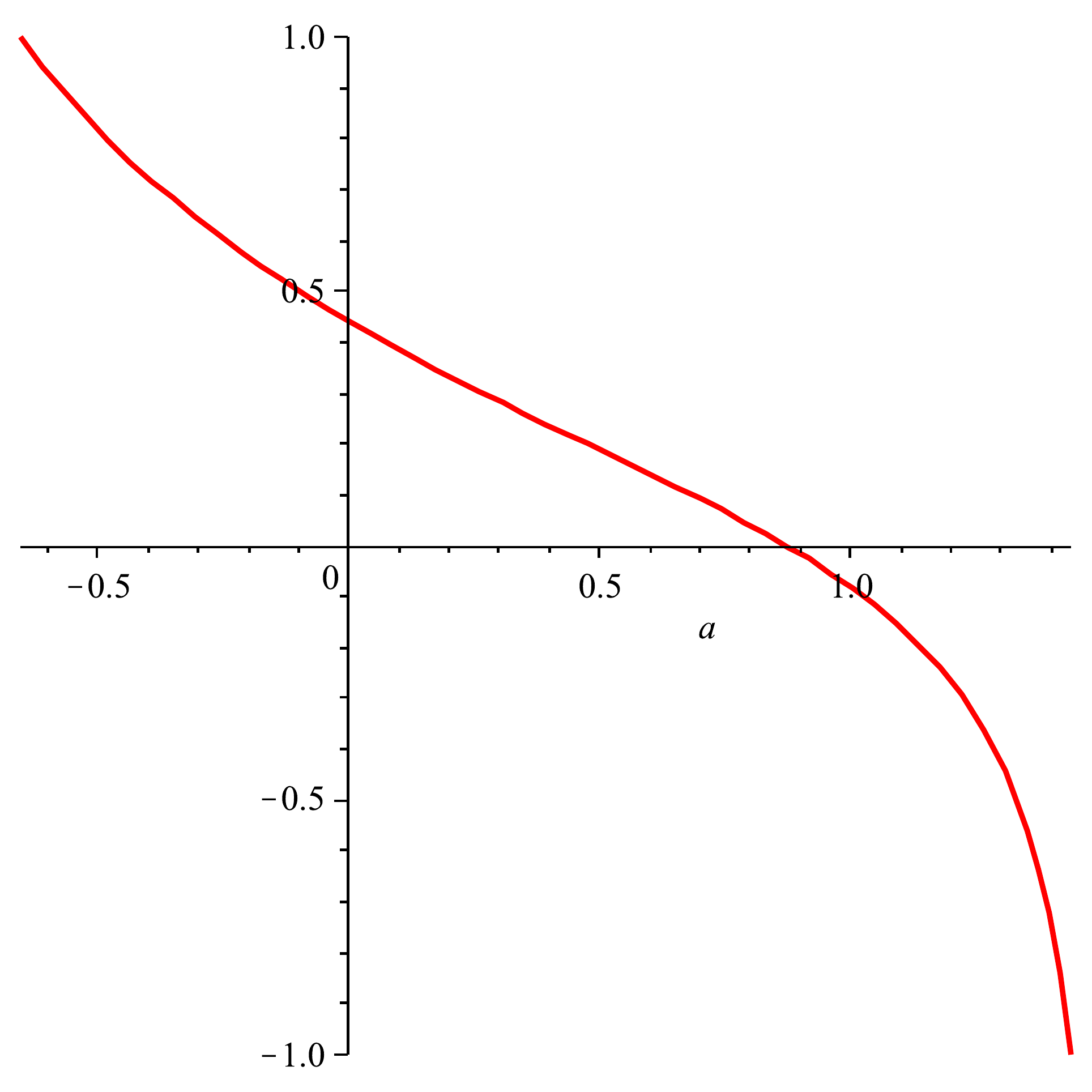}}
\end{center}
\caption[Position of bound soliton.]{Plots of $\tanh(2\sqrt{\beta}\cos(a)c)$ using minus sign with $\beta=1$ for (a) $A=0, \ b= \frac{\pi}{3}$, (b) $A=\frac{\pi}{8}, \ b= \frac{\pi}{3}$.}
\label{fig:tanhCA}
\end{figure}

\begin{figure}[!h]
\begin{center}
\subfigure[]{\label{fig:bsEQ1}\includegraphics[width=0.3\textwidth]{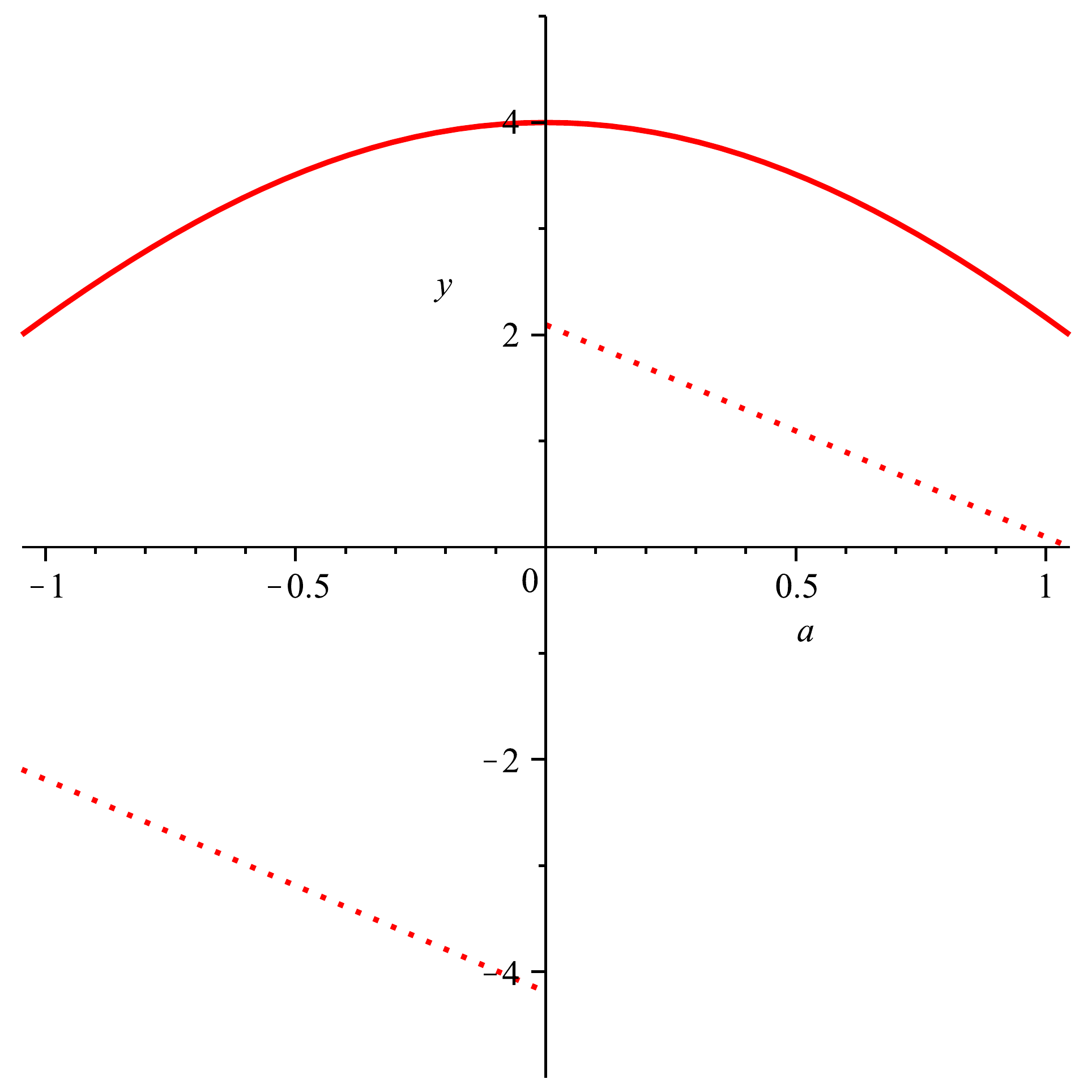}}
\hspace{0.5in}
\subfigure[]{\label{fig:bsEQ1A}\includegraphics[width=0.3\textwidth]{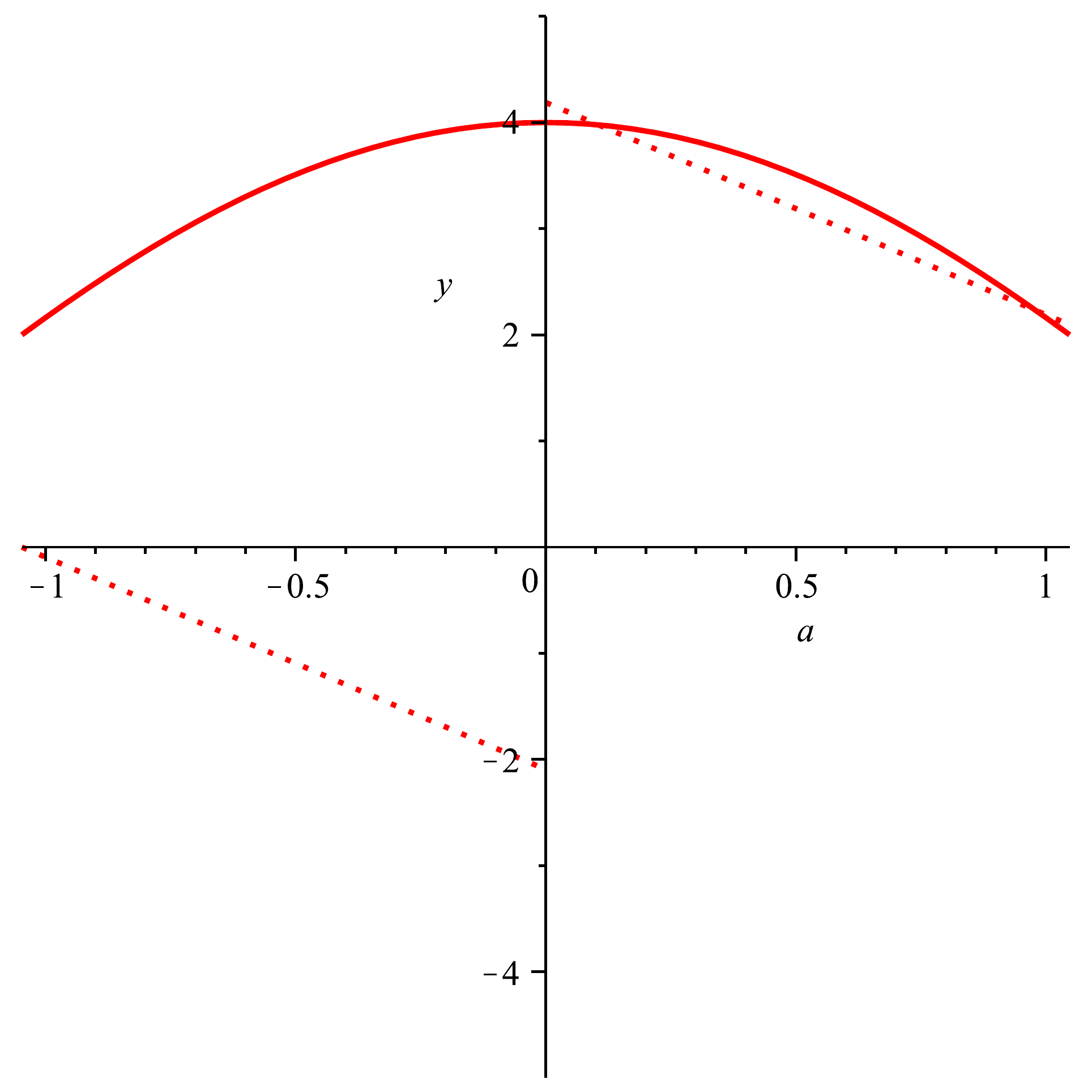}}
\end{center}
\caption[Charge and Energy of bound states.]{Charge (dotted) and energy (solid) of bound states with $\beta=1$ for $A=0, \ b= \frac{\pi}{3}$ using (a) $E^{+}\, , Q^{+}$ (b) $E^{-}\, , Q^{-}$.}
\label{fig:bsEQ}
\end{figure}

\begin{figure}[!h]
\begin{center}
\subfigure[]{\label{fig:bsEQ2}\includegraphics[width=0.3\textwidth]{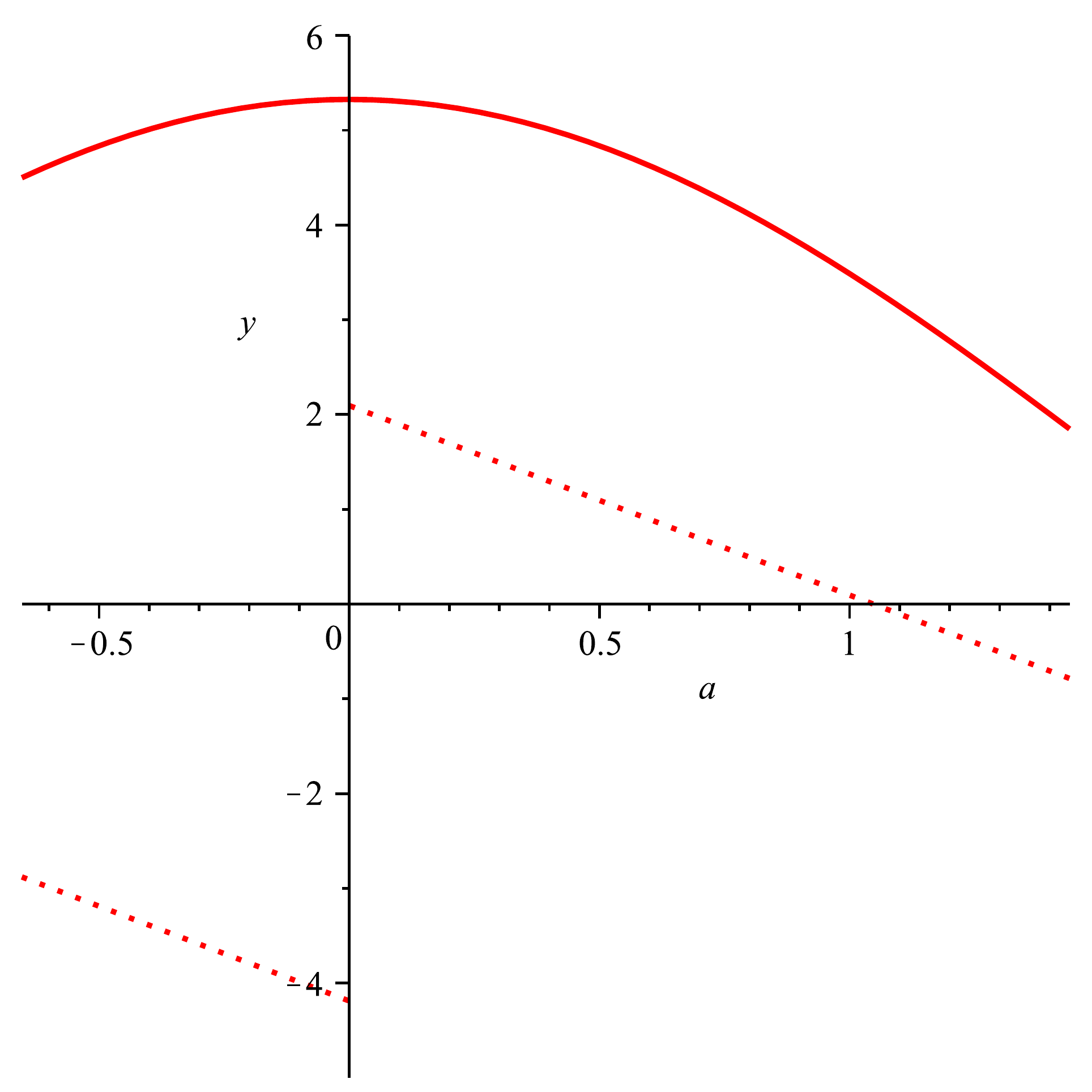}}
\hspace{0.5in}
\subfigure[]{\label{fig:bsEQ2A}\includegraphics[width=0.3\textwidth]{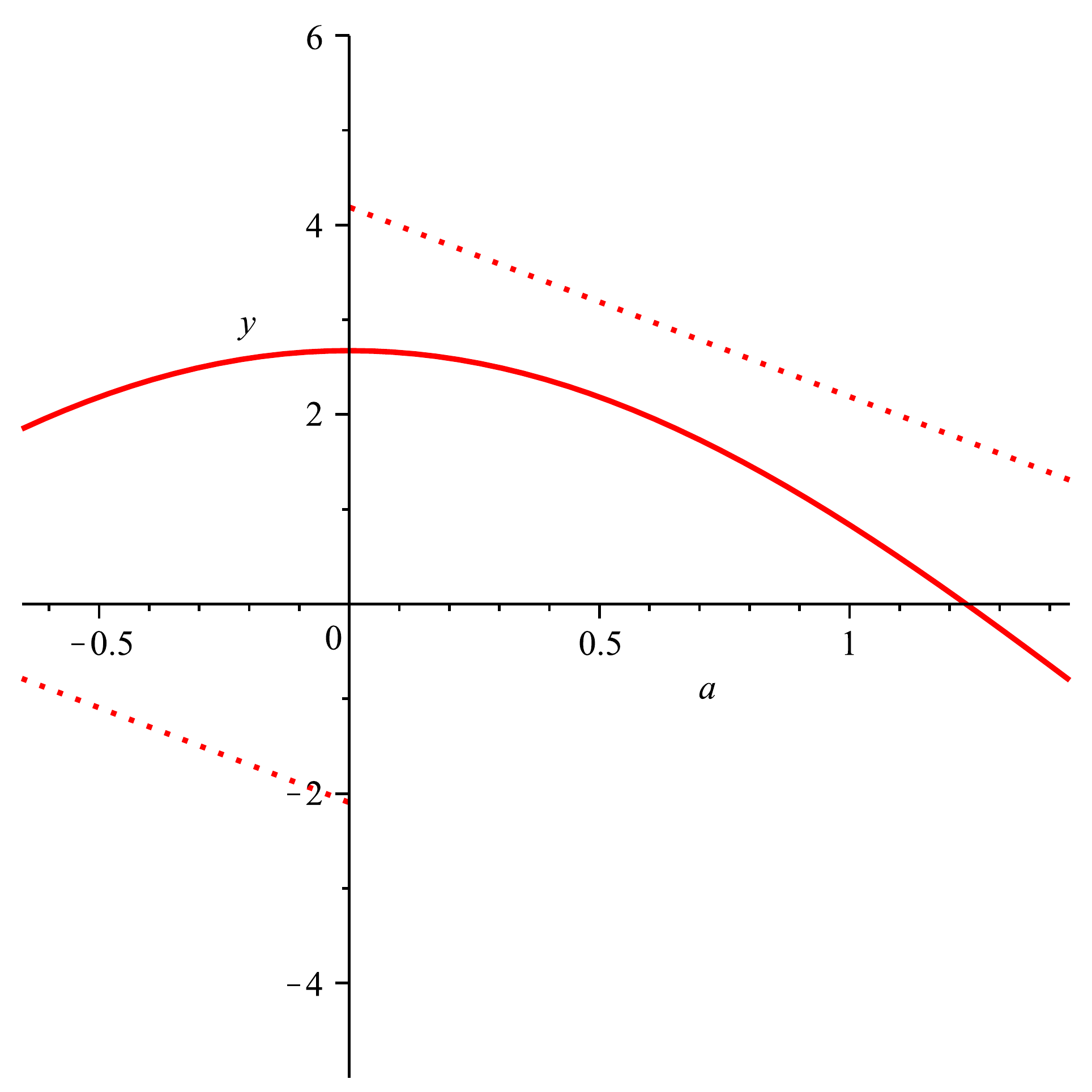}}
\end{center}
\caption[Charge and Energy of bound states.]{Charge (dotted) and energy (solid) of bound states with $\beta=1$ for $A=\frac{\pi}{8}, \ b= \frac{\pi}{3}$ using (a) $E^{+}\, , Q^{+}$ (b) $E^{-}\, , Q^{-}$.}
\label{fig:bsEQtwo}
\end{figure}

Figure \ref{fig:tanhC} plots the function $\tanh(2\sqrt{\beta}\cos(a)c)$ when the plus solution in (\ref{eq:tanhC}) is used for two particular choices of boundary parameters. It shows that in both cases, as $a\rightarrow A+b$, the bound soliton is positioned away at positive infinity $c \rightarrow + \infty$ whilst as  $a\rightarrow A-b$ the bound soliton approaches negative infinity.
On the other hand figure \ref{fig:tanhCA} illustrates the behaviour of $\tanh(2\sqrt{\beta}\cos(a)c)$ when the minus solution in (\ref{eq:tanhC}) is used. The figure shows that in both examples the bound soliton is positioned away at negative infinity $c \rightarrow - \infty$  when $a=A+b$  and positioned at positive infinity  when $a=A-b$  These figures illustrate that in both charge limits where the bound state reduces to the unexcited boundary, the bound soliton is positioned at right infinity behind the boundary. As the charge parameter moves away from the unexcited boundary limit, either decreasing from $a=A+b$ or increasing from $a=A-b$, the bound soliton moves from right infinity to left infinity when it reaches the other end of the range. The soliton being positioned at right infinity and hidden behind the boundary, fits with the fact that the bound state reduces to the unexcited boundary when the soliton is in this position.
In figures \ref{fig:bsEQ} and \ref{fig:bsEQtwo} we graph the energy and charge for the bound states in the range of $a$ where the constraint is satisfied, using both forms of the energy and charge. They show that the energy and charge are simply shifted by a constant between the different energy and charge formulae. In both examples the energy increases from both ends of the classical region with the maximum energy of the bound state at $a=0$. The charge of the bound state increases as $a$ decreases from $A+b$, but the charge decreases when $a$ increases from $A-b$. This conflicting behaviour suggests the two ends of the region, despite both limiting to an unexcited boundary when one of the energy formulae is used, have some different properties. We come back to this point later in the paper.

\begin{figure}[!h]
\begin{center}
\subfigure[]{\label{fig:bsEQ3a}\includegraphics[width=0.3\textwidth]{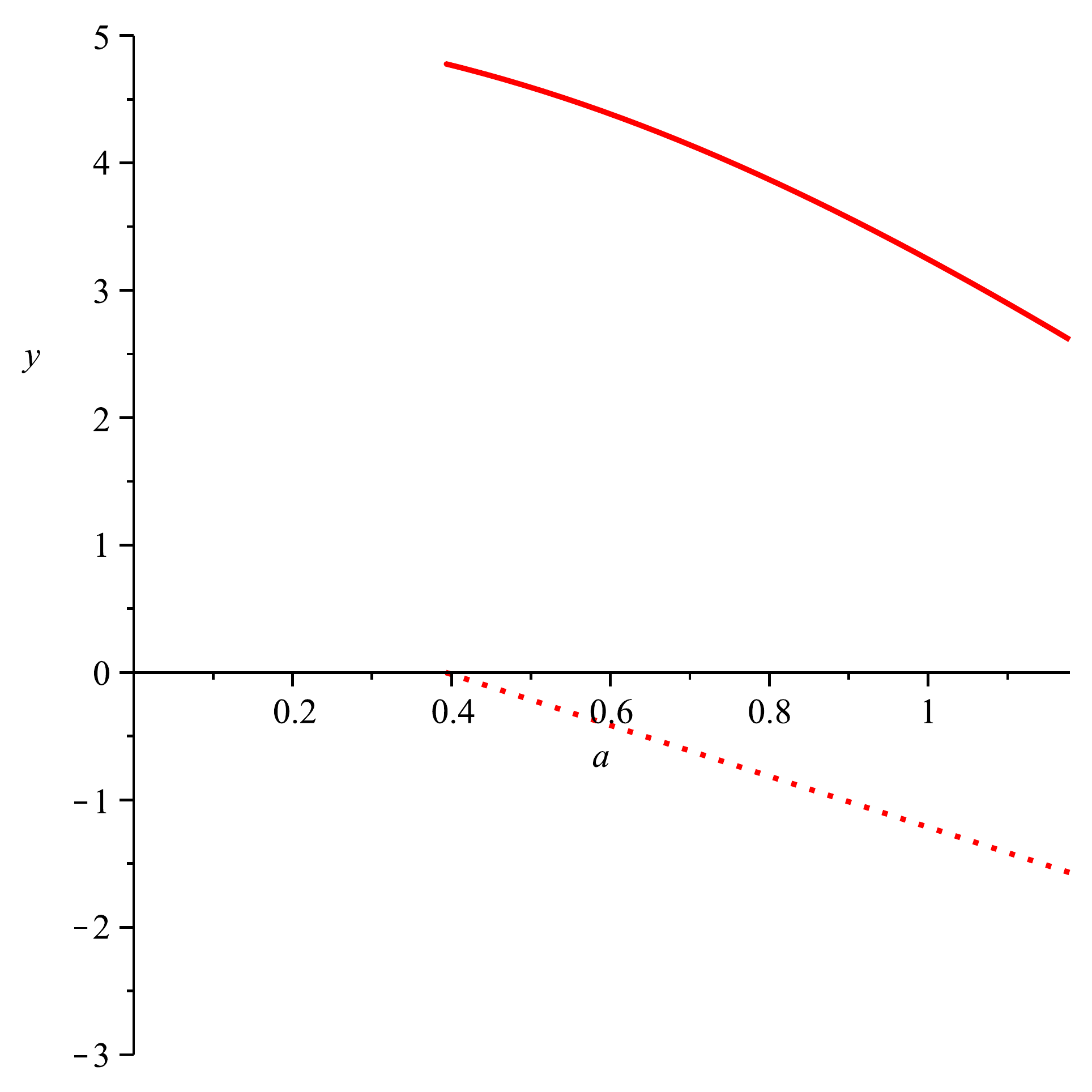}}
\hspace{0.5in}
\subfigure[]{\label{fig:bsEQ3A}\includegraphics[width=0.3\textwidth]{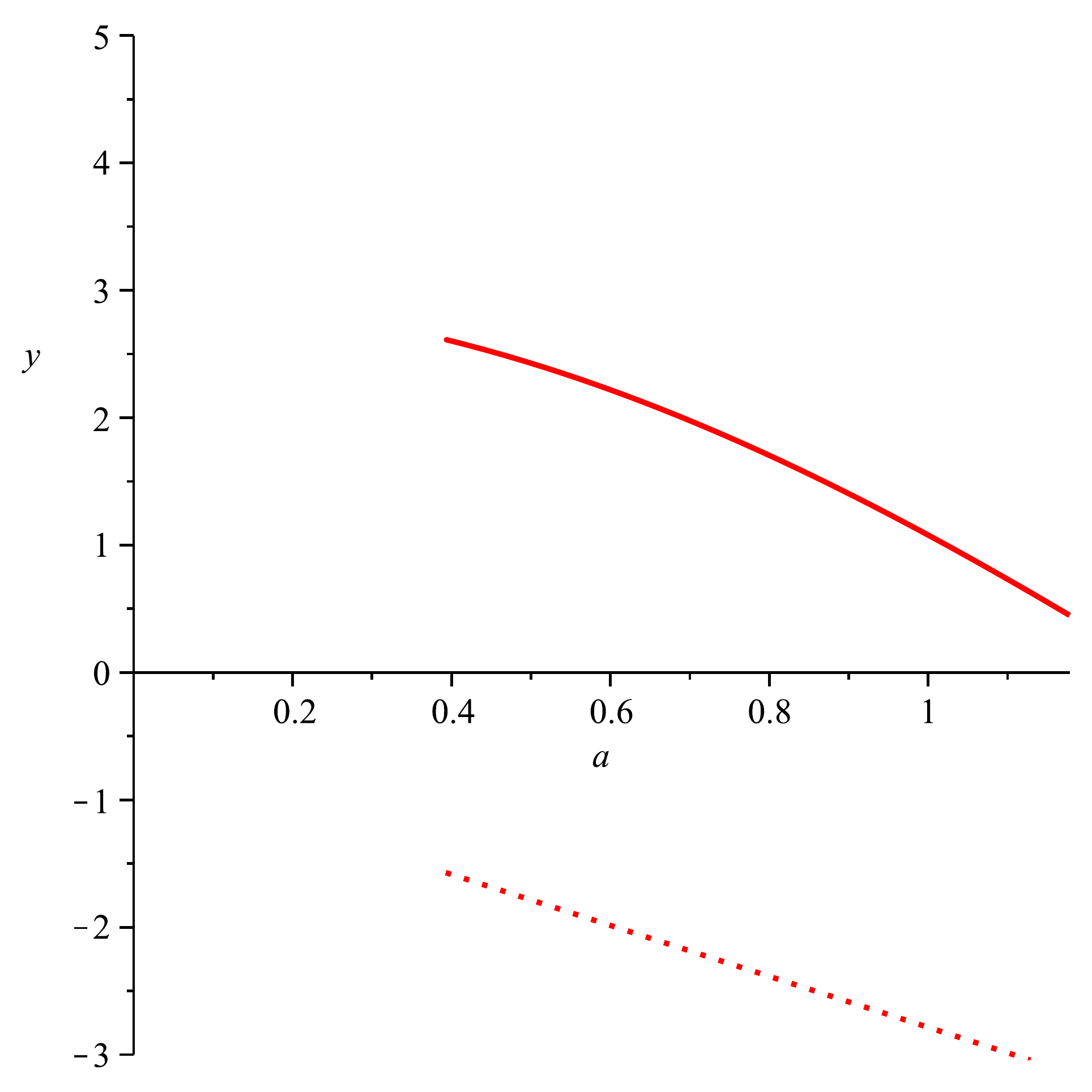}}
\end{center}
\caption{Charge (dotted) and energy (solid) of bound states with $\beta=1$ for $A=\frac{\pi}{4}, \ b= \frac{\pi}{8}$ using (a) $E^{+}\, , Q^{+}$ (b) $E^{-}\, , Q^{-}$}
\label{fig:bsEQ3}
\end{figure}

To complete the analysis of the classical bound states we examine one further example with $A=\frac{\pi}{4}, \ b= \frac{\pi}{8}$. In figure \ref{fig:bsEQ3} the energy and charge are plotted for two different energy and charge formulae. We note that in figure \ref{fig:bsEQ3A} that as the bound soliton moves out from right infinity, this corresponds to the charge parameter increasing from $A-b$, the energy of the bound state decreases. This is the case because $A-b >0$ and as with the previous examples the energy would reach its maximum at $a=0$ and therefore the energy is still increasing, from right to left, at $a=A-b$. 

Analysis of the particle reflection factor (\ref{eq:particlereflection}) shows that it has two poles at
\begin{equation}
\theta = A+b -\frac{\pi}{2}\, , \ \ \  A-b -\frac{\pi}{2}\, ,
\end{equation}
which correspond respectively to the field taking the values
\begin{eqnarray}
u = \epsilon\ e^{-2i\sqrt{\beta}\sin(A-b)t}\, e^{2\sqrt{\beta}\cos(A-b)x}\, , \ \ \  \epsilon\ e^{-2i\sqrt{\beta}\sin(A+b)t}\, e^{2\sqrt{\beta}\cos(A+b)x}\, .
\end{eqnarray}
for infinitesimal $\epsilon$.
These suggest the existence of bound states when the bound soliton has either charge parameter $a=A+b$ or $a=A-b$. This is in agreement with our analysis of bound solitons; the above exponential solutions correspond to the tails of bound solitons which are hidden far behind the boundary at right infinity. This concludes the analysis of classical solutions to the CSG dressed boundary theory.

\section{Quantum CSG bulk theory}
We begin by reviewing quantum aspects of the bulk theory. The quantised bulk theory was first considered by Maillet and de Vega \cite{deVega:1982sh}, with the results reviewed and expanded on by Dorey and Hollowood \cite{Dorey:1994mg} to the point where they conjecture a $S$-matrix to describe the quantum scattering of charged solitons in CSG theory. The next section deals with the semi-classical results, with the $S$-matrix introduced in the following section.

\subsection{Semi-classical quantisation}
\begin{figure}[!h]
\begin{center}
\subfigure[$k$ even]{\label{fig:SolChargeEven}\includegraphics[width=0.4\textwidth]{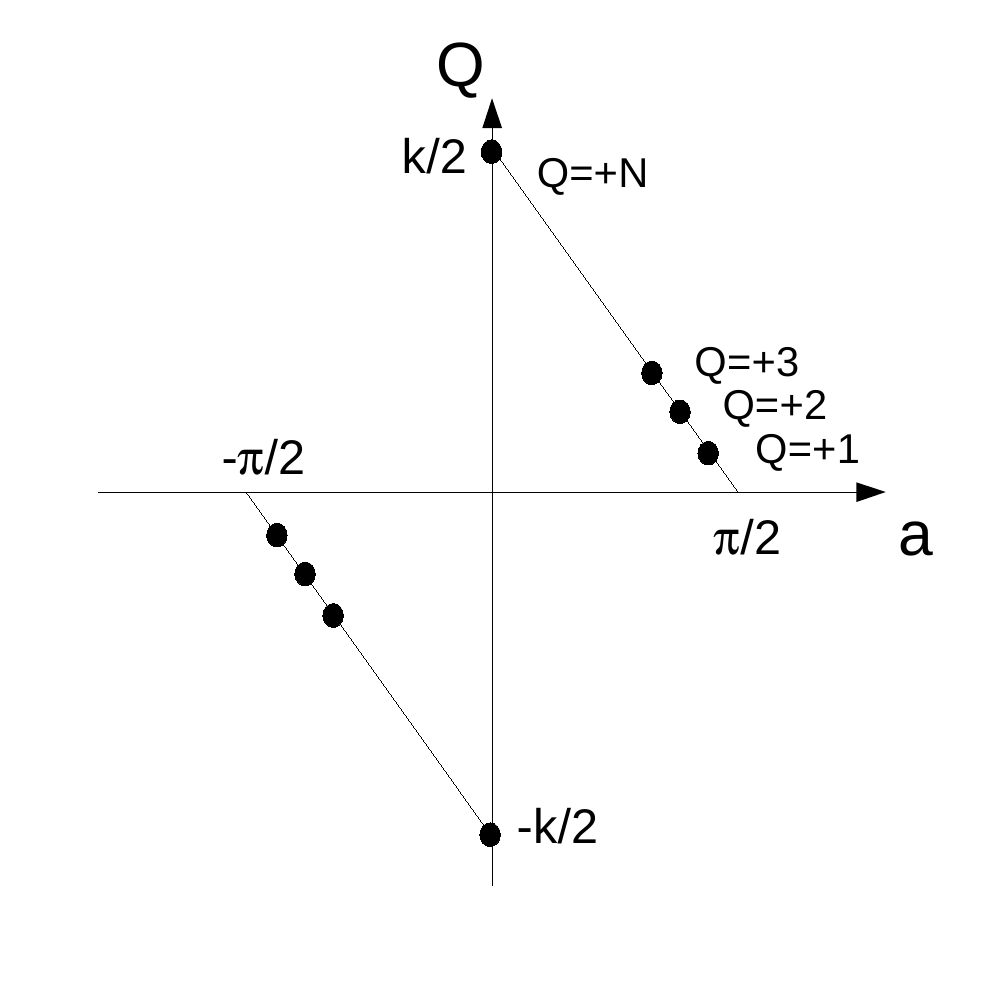}}
\hspace{0.3in}
\subfigure[$k$ odd]{\label{fig:SolChargeOdd}\includegraphics[width=0.4\textwidth]{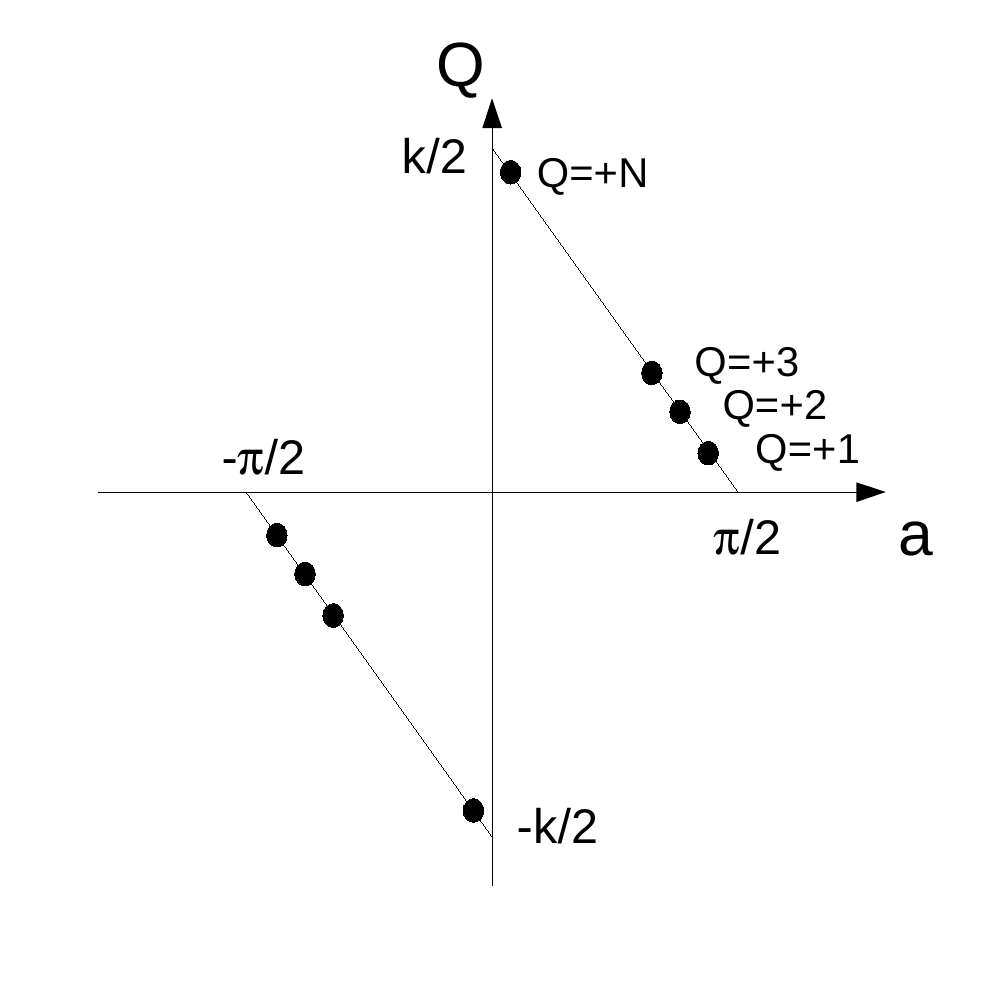}}
\end{center}
\caption{The quantisation of the charge of a CSG soliton.}
\label{fig:SolCharge}
\end{figure}

To begin the discussion a few properties of the CSG soliton (\ref{eq:1sol}) are noted. As already commented on the CSG soliton rotates in the internal $U(1)$ space with the constant angular velocity $w=2\sqrt{\beta}\mathrm{sin}(a)$. We re-express the energy 
\begin{equation}
E(w)=\frac{8\sqrt{\beta}}{\lambda^{2}}\sqrt{1-\frac{w^{2}}{4\beta}}\mathrm{cosh}(\theta)= E(0)\sqrt{1-\frac{w^{2}}{4\beta}}\, , 
\end{equation}
and charge
\begin{equation}
Q(w) = \frac{4}{\lambda^{2}}\mathrm{arccos}\left(\frac{w}{2\sqrt{\beta}}\right)\, , 
\end{equation}
of the soliton in terms of its angular velocity. These expressions highlight the property that the energy and charge of the soliton decreases as the angular velocity in the internal space increases. When the angular velocity reaches its maximum $w=2\sqrt{\beta}$, the energy and charge vanish but also in this limit the soliton is damped to zero by the $\cos(a)$ factor. Due to the periodic time-dependent nature of the stationary soliton solution
\begin{equation}\label{eq:1solstat}
u^{stat}_{1-sol}=\frac{ \mathrm{cos}(a) e^{2i\sqrt{\beta}\mathrm{sin}(a)t}}{\mathrm{cosh}(2\sqrt{\beta}\mathrm{cos}(a)x)} \, ,
\end{equation} 
the Bohr-Sommerfeld Quantisation (B-S) condition 
\begin{equation}\label{eq:BScond}
S[u] + E[u]\tau =2\pi n \, ,
\end{equation}
can be applied \cite{Ventura:1976vi, Montonen:1976yk}. Where $n \in \mathbb{Z}$, $S$ is the action functional, $E$ the energy and $\tau=\frac{2\pi}{w}$ the period of the solution $u$. Using the explicit form of the stationary soliton (\ref{eq:1solstat}) the left hand side of the B-S condition becomes
\begin{equation}\label{eq:SolBS}
 8\beta \mathrm{sin}^{2}(a)\ \frac{\pi}{\sqrt{\beta}\mathrm{sin}(a)} \int^{0}_{-\infty} dx \  \frac{uu^{*}}{1-uu^{*}}\, , 
\end{equation}
which is proportional to the charge of the stationary soliton (\ref{fig:solitoncharge}) and the B-S condition reduces to
\begin{equation}
2\pi Q\ = 2\pi n \, .
\end{equation}
Hence the charge is restricted to integer values $ Q = \pm 1,\ \pm 2, \dots, \pm N =\pm\lfloor \frac{2\pi}{\lambda^{2}}\rfloor $. The classical charge formula, illustrated in figure \ref{fig:solitoncharge}, shows the multi-valued nature of the charge when $a=0$. Dorey and Hollowood \cite{Dorey:1994mg} resolve this issue by stating that the charge should be identified mod $2N$. More generally only specific values of the coupling constant should be considered $\lambda^{2} = \frac{4\pi}{k}$, where $k \in \mathbb{Z} > 1$ and the charge is now identified mod $k$. The spectrum of the charge becomes
\begin{eqnarray}\label{eq:ChargeSpec}
Q &=& 0,\  \pm 1, \ \pm 2, \dots, \ \pm \frac{k}{2} \ \ \ \ \ \ \ \ \ k \ \ \mathrm{even} \, ,\nonumber\\
Q &=& 0,\  \pm 1, \ \pm 2, \dots, \ \pm \frac{k-1}{2} \ \ \ \ k \ \ \mathrm{odd} \, .
\end{eqnarray}
If $k$ is even then the solitons with charge $Q = \pm \frac{k}{2}$ are identified, but if $k$ is odd then no solitons are identified. However when incrementing up from $Q=+1$ in single units of charge the step from $Q = \frac{k-1}{2}$ leads to $Q = -\frac{k-1}{2}$, from where it continues up to the $Q=-1$. Figure \ref{fig:SolChargeEven} shows the case when $k$ is even and figure \ref{fig:SolChargeOdd} when $k$ is odd.  These two cases illustrate why the coupling constant $\lambda$ is restricted in the way it is.

The quantisation of the charge can be equivalently described as the quantisation of the soliton charge parameter 
\begin{equation}
a = \frac{\pi}{2} - \frac{n\pi}{k}\, ,
\end{equation}
where $n$ is the charge of the soliton. This gives the semi-classical energy spectrum of the stationary soliton with the charge $n$ soliton having the energy
\begin{equation}
 E_{n} = \frac{2\sqrt{\beta}k}{\pi}\ \mathrm{sin}\left(\frac{n\pi}{k}\right)\, . 
\end{equation}
Maillet and de Vega \cite{deVega:1982sh} computed the one-loop corrections to this energy spectrum. They found these corrections were obtained by a renormalisation of the coupling constant
\begin{equation}
\lambda^{2} \rightarrow \lambda_{R}^{2}= \frac{4\pi\lambda^{2}}{4\pi-\lambda^{2}}\, , \ \ \ \ \ \ \ \ k \rightarrow k_{R}= k -1\, .
\end{equation}

\subsection{Quantum CSG $S$-matrix}

In this section we review the conjectured exact form for the CSG $S$-matrix by Dorey and Hollowood \cite{Dorey:1994mg}. From this point the coupling constant $k$ will be taken to be the renormalised $k_{R}$. The $S$-matrix which describes the scattering of two charged solitons is given as
\begin{equation}\label{eq:SMatrix}
\mathcal{S}_{Q_{1},Q_{2}}(\theta) = F_{Q_{1}-Q_{2}}(\theta)\left[\prod_{n=1}^{Q_{2}-1} F_{Q_{1}+Q_{2}-2n}(\theta) \right]^{2}F_{Q_{1}+Q_{2}}(\theta)\, , 
\end{equation}
where 
\begin{equation}\label{eqFFactor}
F_{x}(\theta) = \frac{\sinh\left(\frac{\theta}{2}+i\frac{\pi x}{2k}\right)}{\sinh\left(\frac{\theta}{2}-i\frac{\pi x}{2k}\right)} \, .
\end{equation}
It is constructed from products of $F$ factors so it automatically satisfies the analyticity and unitarity $F_{x}(\theta)\ F_{x}(-\theta)=1$  constraints. Each of the F  factors has a pole at $\theta = i\frac{\pi x}{k}$. This $S$-matrix is the minimal choice which has the correct pole structure, explicitly poles are expected at the rapidities where the scattering solitons form bound states. Charge conservation suggests that two solitons with charge $Q_1$ and $Q_2$ bind to form solitons with charge $Q_{1} \pm Q_{2}$ in the forward and crossed channels respectively. Other poles are expected to coincide with processes introduced by Coleman and Thun \cite{coleman:1978}. 

\begin{figure}[!h]
\begin{center}
\subfigure[]{\label{fig:SolBound11}\includegraphics[width=0.4\textwidth]{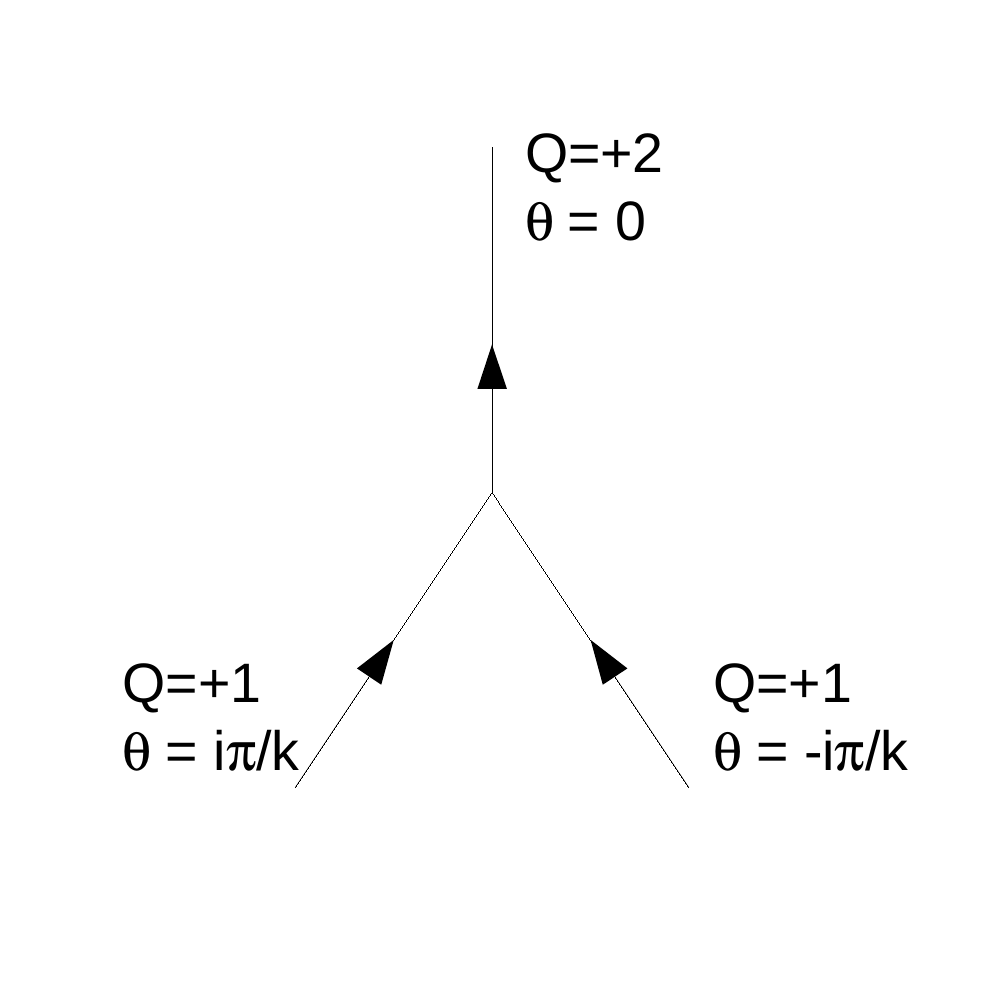}}
\hspace{0.3in}
\subfigure[]{\label{fig:SolBoundnm}\includegraphics[width=0.4\textwidth]{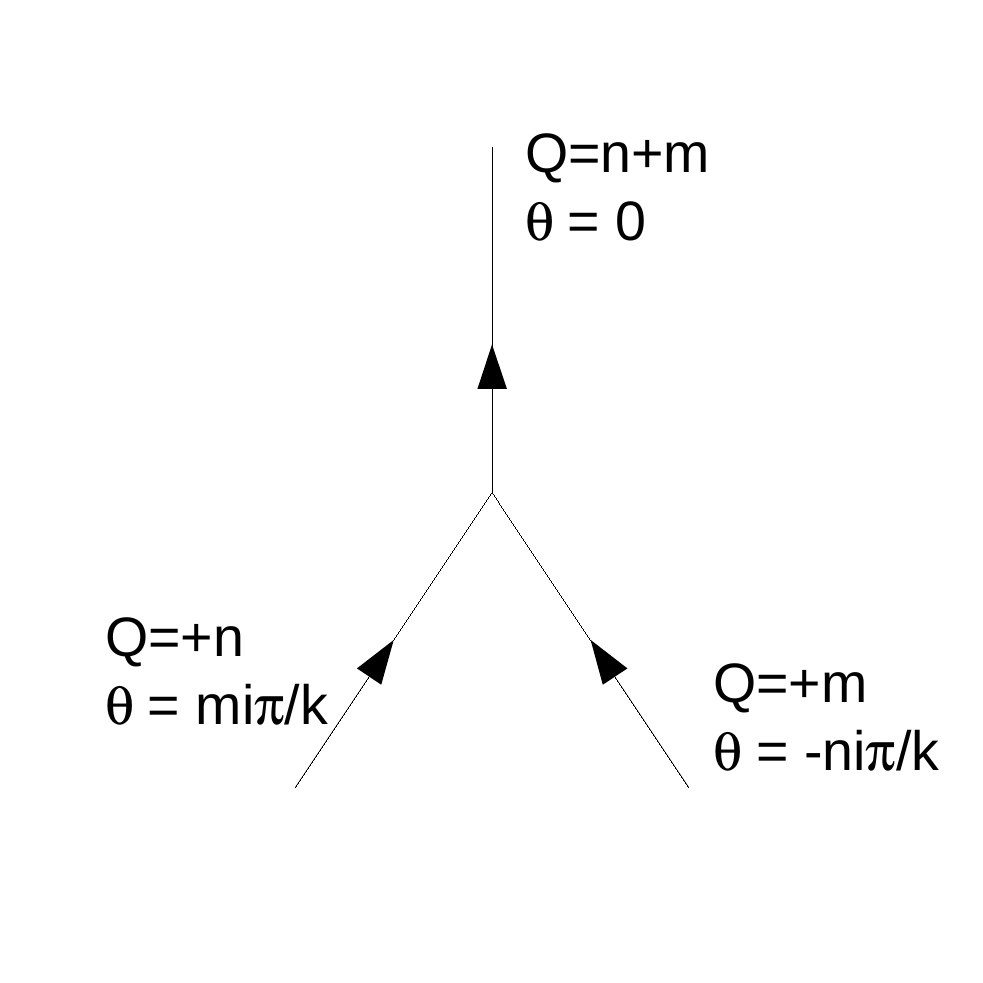}}
\end{center}
\caption[Fusion of CSG solitons.]{The fusing of (a) two $Q=+1$ solitons, (b) a $Q=+n$ and $Q=+m$ soliton.}
\label{fig:SolBound}
\end{figure}

CSG solitons only form bound states when they have very specific relative rapidity. For example, two charge $Q=+1$ solitons bind together to form a charge $Q=+2$ soliton when they have the relative rapidity $\frac{2i\pi}{k}$. Figure \ref{fig:SolBound11} shows two such $Q=+1$ solitons with rapidities $\pm \frac{i\pi}{k}$ joining to become a stationary $Q=+2$ soliton. In this figure and all the ones to follow time flows up the diagram.
We substitute the required charge parameters into the energy formula for the soliton, namely $a=\frac{\pi}{2} - \frac{\pi}{k}$ for the charge $Q=+1$ solitons and $a=\frac{\pi}{2} - \frac{2\pi}{k}$ for the charge $Q=+2$ soliton and use the double angle formula to show that energy is conserved for these rapidities
\begin{equation}
8\sqrt{\beta}\cos\left(\frac{\pi}{2} - \frac{\pi}{k}\right)\left(\cosh\left(\frac{i\pi}{k}\right) + \cosh\left(\frac{-i\pi}{k}\right) \right) = 8\sqrt{\beta}\cos\left(\frac{\pi}{2} - \frac{2\pi}{k}\right)\cosh(0)\, .
\end{equation}
Similarly as shown in figure \ref{fig:SolBoundnm}, two solitons of charge $Q=+n$ and $Q=+m$ fuse to form a bound state of charge $Q=n+m$ at the relative rapidity $\frac{i(n+m)\pi}{k}$. Note that the relative rapidity is always imaginary and in the physical strip $\ 0 < \mathcal{I}m(\theta) < \pi$. The scattering solitons can be given real rapidity, but they must be equal. For example two $Q=+1$ solitons with rapidities $\psi \pm \frac{i\pi}{k}$ fuse to form charge $Q=+2$ soliton travelling with real rapidity $\psi$.

\begin{figure}[!h]
\begin{center}
\subfigure[]{\label{fig:SolScatPole11}\includegraphics[width=0.31\textwidth]{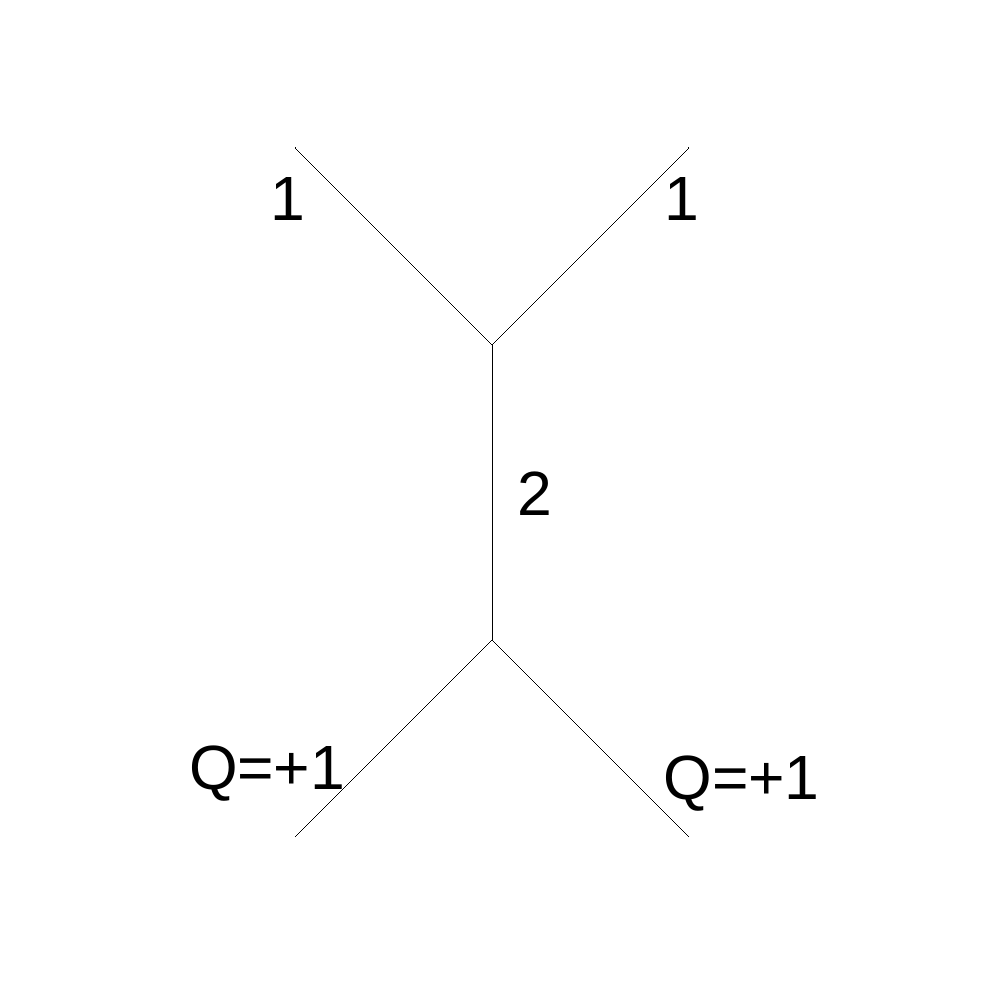}}
\subfigure[]{\label{fig:SolScatPole1n}\includegraphics[width=0.31\textwidth]{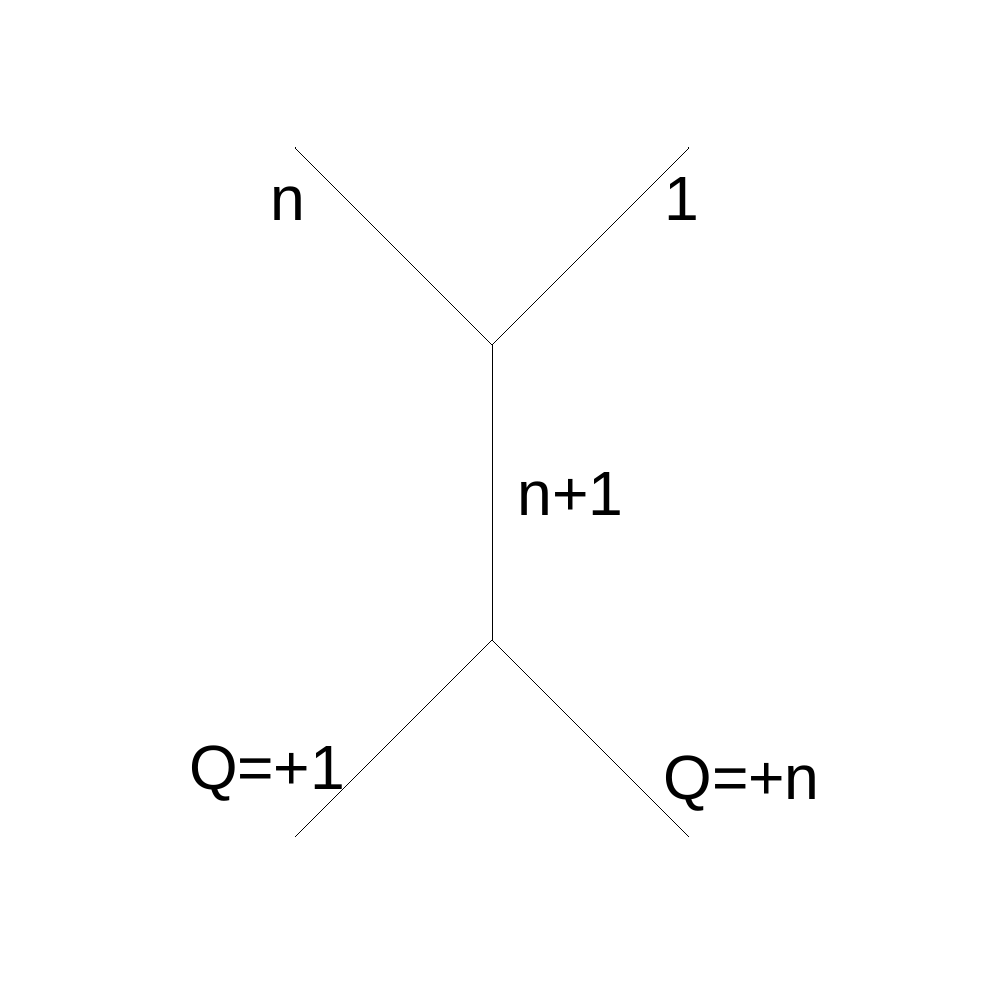}}
\subfigure[]{\label{fig:SolScatPole1nB}\includegraphics[width=0.31\textwidth]{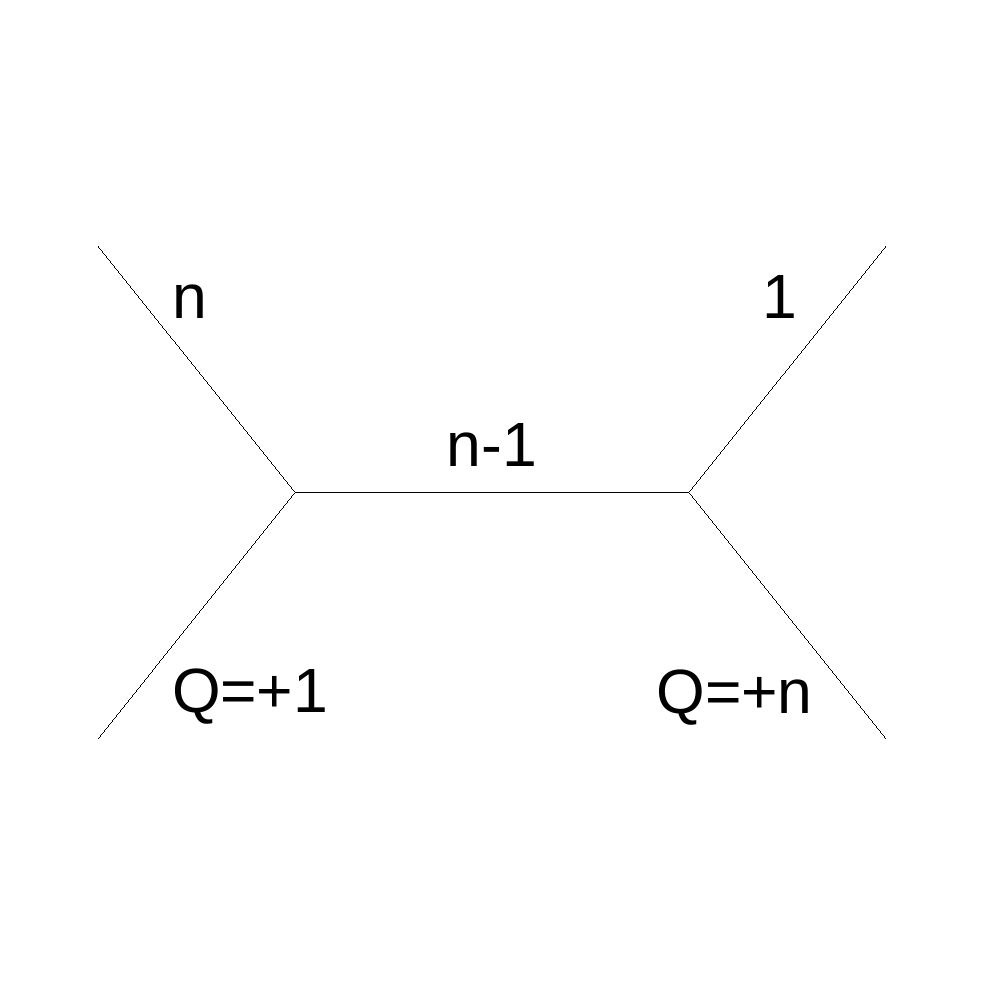}}
\end{center}
\caption[Formation of bound states in forward and cross channels.]{Formation of (a) $Q=+2$ soliton in forward channel, (b) $Q=n+1$ soliton in forward channel, (c) $Q=n-1$ soliton in cross channel.}
\label{fig:SolScatPole11A}
\end{figure}

The $S$-matrix (\ref{eq:SMatrix}) describing the scattering of two charge $Q=+1$ solitons is
\begin{equation}
S_{1,1}(\theta) = F_{0}(\theta)\ F_{2}(\theta) = F_{2}(\theta)\, .
\end{equation}
The $F_{2}(\theta)$ factor has a pole at $\theta = \frac{2i\pi }{k}$ which corresponds to the formation of a charge $Q=+2$ soliton in the forward channel, illustrated in Figure \ref{fig:SolScatPole11}.
Similarly the scattering of charge $Q=+1$ and $Q=+n$ soliton is governed by the $S$-matrix
\begin{equation}
S_{n,1}(\theta) = F_{n-1}(\theta)\ F_{n+1}(\theta)\, ,
\end{equation}
where both of the $F$ factors have poles which correspond to the formation of bound states. As before in the forward channel process, illustrated in figure \ref{fig:SolScatPole1n} and also in the cross channel shown in figure \ref{fig:SolScatPole1nB}.

\begin{figure}[!h]
\begin{center}
\subfigure[]{\label{fig:SolScatPoleQQ}\includegraphics[width=0.31\textwidth]{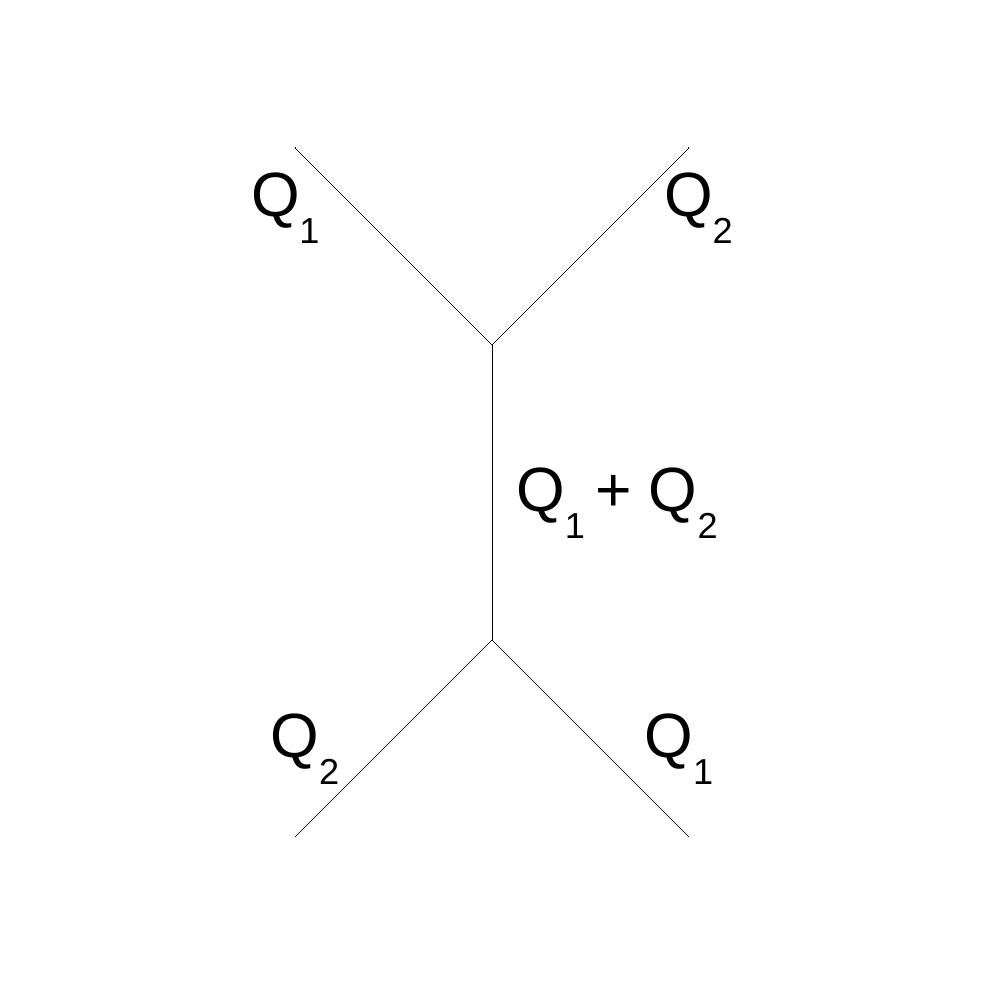}}
\subfigure[]{\label{fig:SolScatPoleQQB}\includegraphics[width=0.31\textwidth]{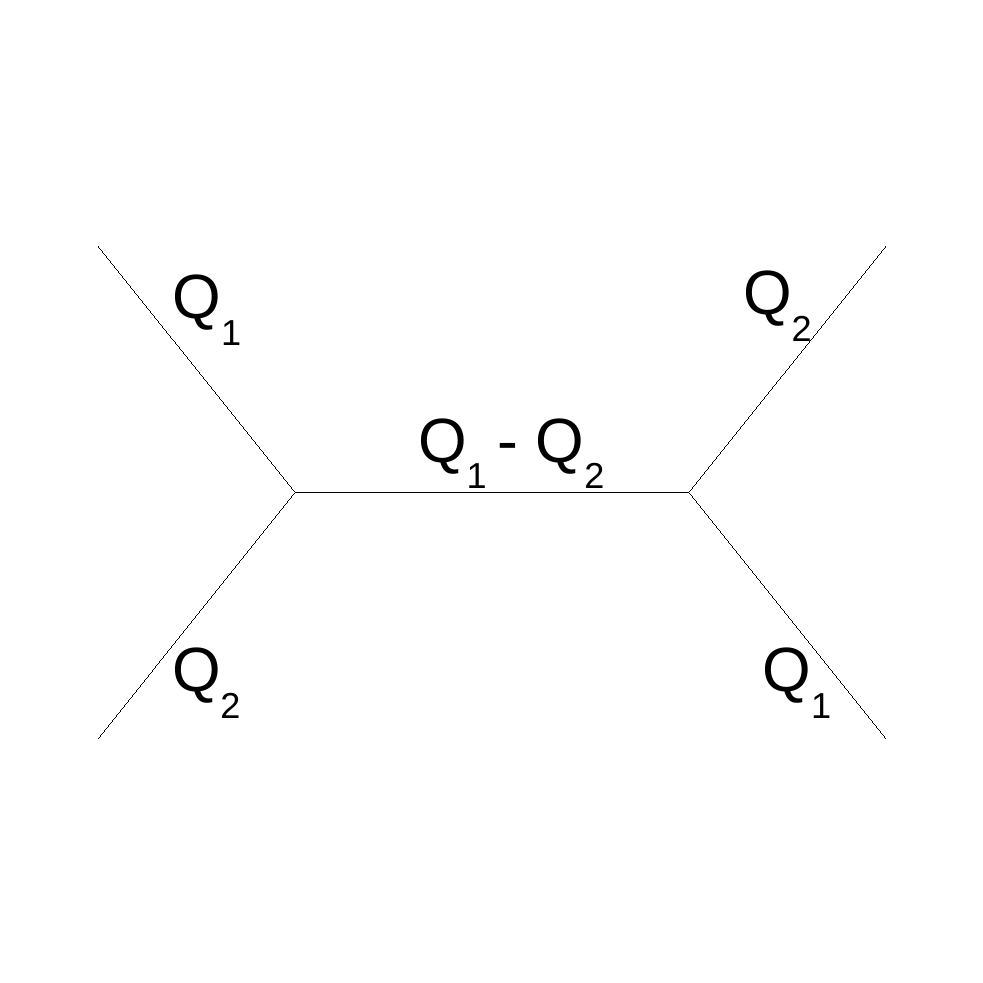}}
\subfigure[]{\label{fig:SolScatPoleQQC}\includegraphics[width=0.31\textwidth]{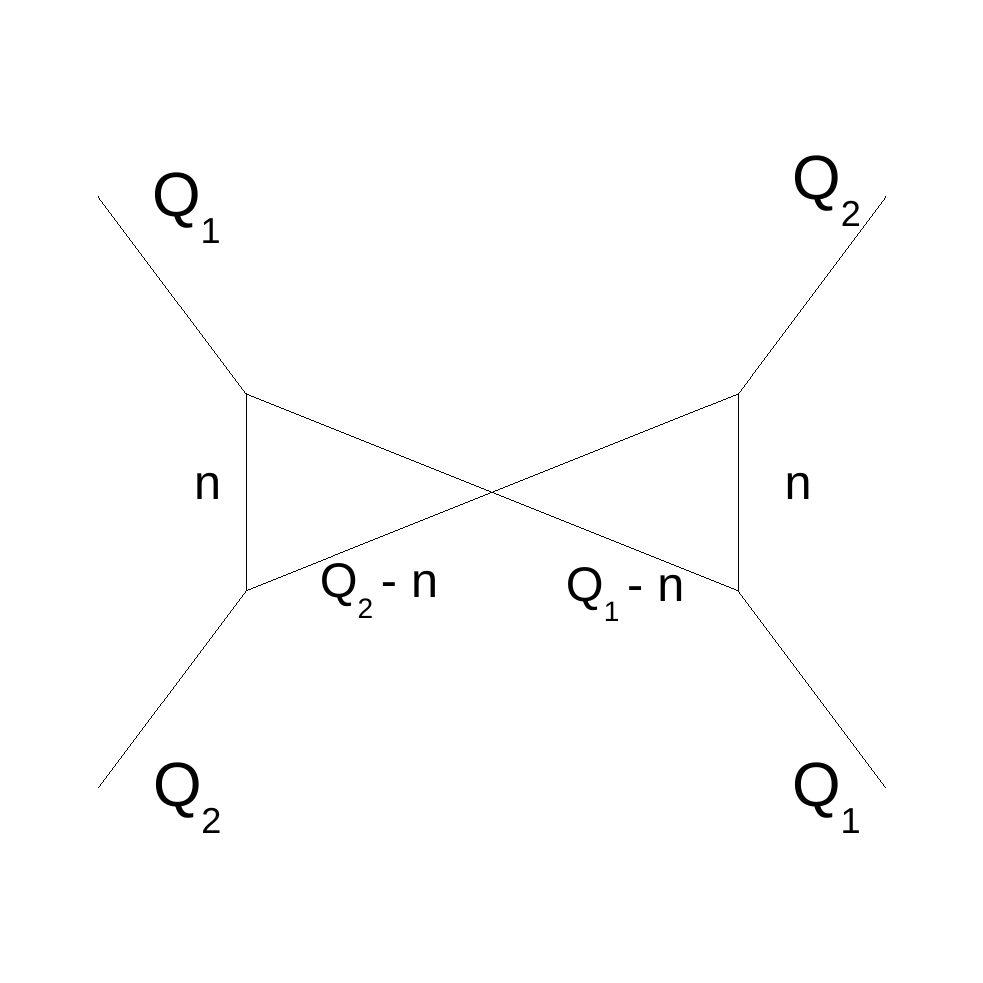}}
\end{center}
\caption[Formation of bound states in $Q_{1}$, $Q_{2}$ soliton-soliton scattering.]{Formation of (a) $Q=Q_{1}+Q_{2}$ soliton in forward channel, (b) $Q=Q_{1}-Q_{2}$ soliton in cross channel, (c) Process with intermediate states of charge $Q_{1}-n$ and $Q_{2}-n$, which results in a double pole due to the two on-shell internal loops.}
\label{fig:SolScatPoleQQX}
\end{figure}

The general $S$-matrix governing the scattering between two solitons of charge $Q=Q_{1}$ and $Q=Q_{2}$ , where $Q_{1} \geq Q_{2}$, has simple poles in the forward and cross channels, shows in figures \ref{fig:SolScatPoleQQ} and \ref{fig:SolScatPoleQQB}
and extra double poles due to Coleman-Thun processes illustrated in figure \ref{fig:SolScatPoleQQC}. There are $Q_{2}-1$ such processes as the stationary intermediate soliton can have charge $Q=n = 1 \rightarrow Q_{2}$. In two-dimensions these processes result in double poles due to the two on-shell internal loops.

This concludes the review of the quantum CSG theory in the bulk. We shall use similar techniques and some of the results in the following sections to investigate quantum aspects of the CSG dressed boundary theory.

\section{Quantum CSG Dressed Boundary}

We start this section by applying a semi-classical method on the bound state to investigate the spectrum of boundaries. The classical bound state solution is periodic therefore, as for the periodic soliton solution, the Bohr-Sommerfeld quantisation condition (\ref{eq:BScond}) can be applied. Using the form of the dressed boundary action (\ref{eq:dressedboundarylagrangianA}) and energy (\ref{eq:boundaryenergy}) the left hand side of the B-S condition becomes
\begin{eqnarray}\label{eq:SeT}
S + E \tau &=& \int^{\tau}_{t=0} dt \ \int^{0}_{-\infty} dx \ \frac{2 \partial_{t}u \partial_{t}u^{*}}{1-uu^{*}} + \left. \bigl[  A_{1} u_{t} +  A_{2} u^{*}_{t} \bigr] \right|_{x=0}\, .
\end{eqnarray}
The computation on the bulk part of this expression works in identical fashion to the calculation for the soliton solution. The boundary term becomes
\begin{equation}
\int^{\tau}_{t=0} dt \ 4\sqrt{\beta}\sin(a)\alpha' = 4\pi \alpha'\, , 
\end{equation}
using that the period is $\tau =\frac{\pi}{\sqrt{\beta}\sin(a)}$ and that $\alpha'$ has no time-dependence when $u$ is the stationary soliton solution. As for the bulk piece we find this boundary term to be equal to $2\pi$ times the boundary term of the charge. Therefore the B-S condition for the dressed boundary bound state becomes
\begin{equation}
 S_{cl}[u_{cl}] + E_{cl}[u_{cl}]\tau = 2\pi Q_{bs} =2\pi n \, ,
\end{equation}
implying that the charge of the bound states is quantised $Q_{bs} = n$. We recall that there are two formulae for the energy and charge of the bound states (\ref{eq:bsenergy}), in this initial analysis we use $E_{bs}^{+}$  and $Q_{bs}^{+}$ which limits to the unexcited boundary when $a=A+b$. We can reinterpret this quantisation of the bound state charge as a quantisation condition on the charge parameter $a$ of the bound soliton, when $\cos(a) >0$
\begin{equation}\label{eq:bsaquant}
a =b-\frac{2\pi n}{k}\, , 
\end{equation}
giving an approximation to the energy spectrum 
\begin{equation}
 E^{+}_{n} = \frac{k \sqrt{\beta}}{\pi} \left(\cos\left(b-\frac{2\pi n}{k}\right) + \sin(A)\sin(b) \right)\, . 
\end{equation}
The energy difference between consecutive states becomes
\begin{equation}\label{eq:energydiff}
 E^{+}_{n+1} - E^{+}_{n} = \frac{k \sqrt{\beta}}{\pi} \left(\cos\left(b-\frac{2\pi (n+1)}{k}\right)- \cos\left(b-\frac{2\pi n}{k}\right) \right)\, ,  
\end{equation}
which we rewrite as
\begin{equation}
E^{+}_{n+1} - E^{+}_{n} = \frac{2 k \sqrt{\beta}}{\pi} \cos\left(\frac{\pi}{2}-\frac{\pi} {k}\right) \cos\left(\frac{\pi}{k}(2n+1) + \frac{\pi}{2} -b\right)\, .
\end{equation}
We note that this the same as the energy formula for a charge $Q=+1$ soliton
\begin{equation}
 E_{sol} (Q=+1) = \frac{2 k \sqrt{\beta}}{\pi} \cos\left(\frac{\pi}{2}-\frac{\pi} {k}\right)\cosh(\theta)\, ,  
\end{equation}
with the imaginary rapidity
\begin{equation}\label{eq:solrapid}
 \theta = i \left(b -\frac{\pi}{k}(2n+1) - \frac{\pi}{2} \right)\, .
\end{equation}
This suggests that the charge $Q_{bs} =n+1$ bound state can be generated by a $Q=+1$ soliton fusing with the charge $Q_{bs} =n$ bound state at this specific rapidity. This semi-classical energy difference agrees with the classical energy curves in figures \ref{fig:bsEQ1}, \ref{fig:bsEQ2}, \ref{fig:bsEQ3a}, where the energy increases as the charge increases with $a$ decreasing from $A+b$. 

In section \ref{ref:chargebound} we found that the unexcited boundary of $Q=+N$ appears as the limit of the bound state where the bound soliton is pushed away to right infinity and has the charge parameter $a=A+b$. The unexcited boundary is described by the charge parameter $A = -\frac{2N\pi}{k}$ and since the bound state charge is quantised and the unexcited boundary can be thought of as a bound state, a quantisation condition on $A$ is implied. Therefore the unexcited boundaries can have charge $Q=N \in \mathbb{Z}$ where $\frac{-k}{2} \leq N \leq \frac{k}{2}$. In the following analysis we denote an unexcited boundary with charge $Q=N$ as $N(0)$. 

\begin{figure}[!h]
\begin{center}
\subfigure[]{\label{fig:CSGBoundaryFusion1}\includegraphics[width=0.31\textwidth]{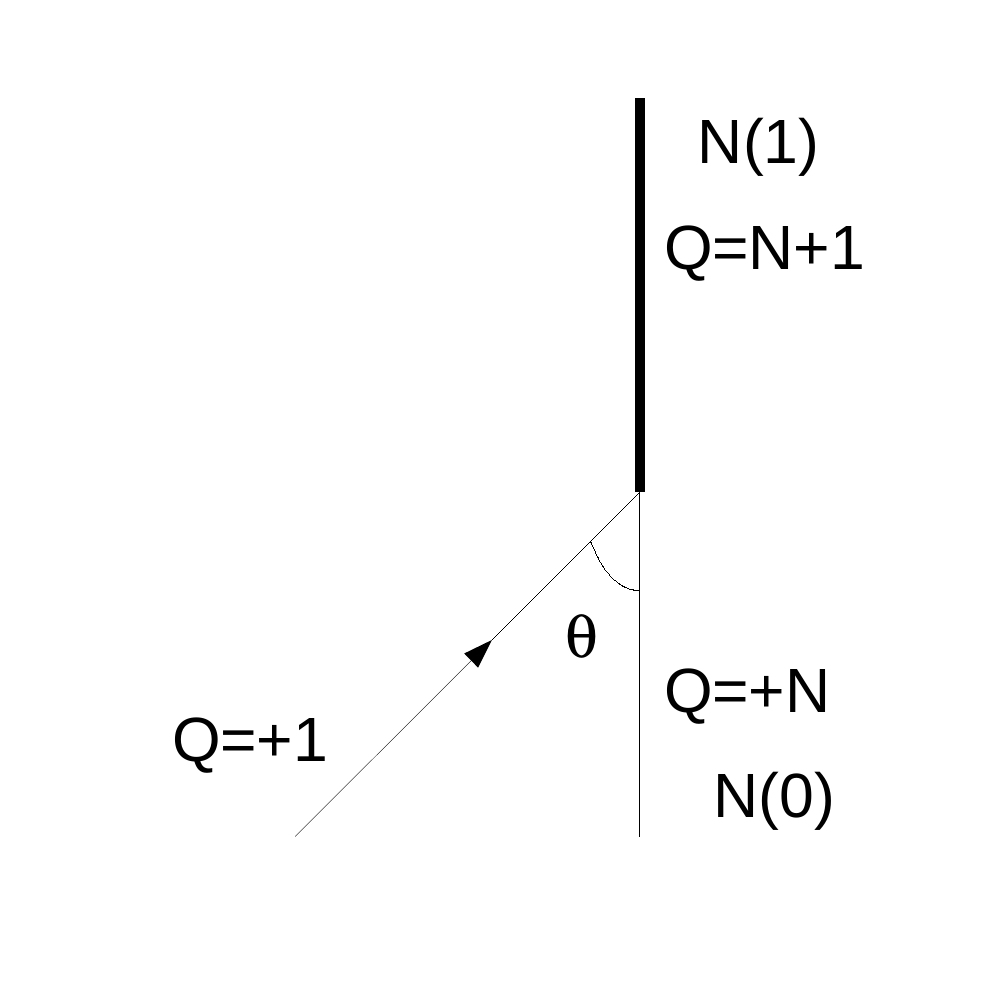}}
\hspace{0.5in}
\subfigure[]{\label{fig:CSGBoundaryFusion}\includegraphics[width=0.31\textwidth]{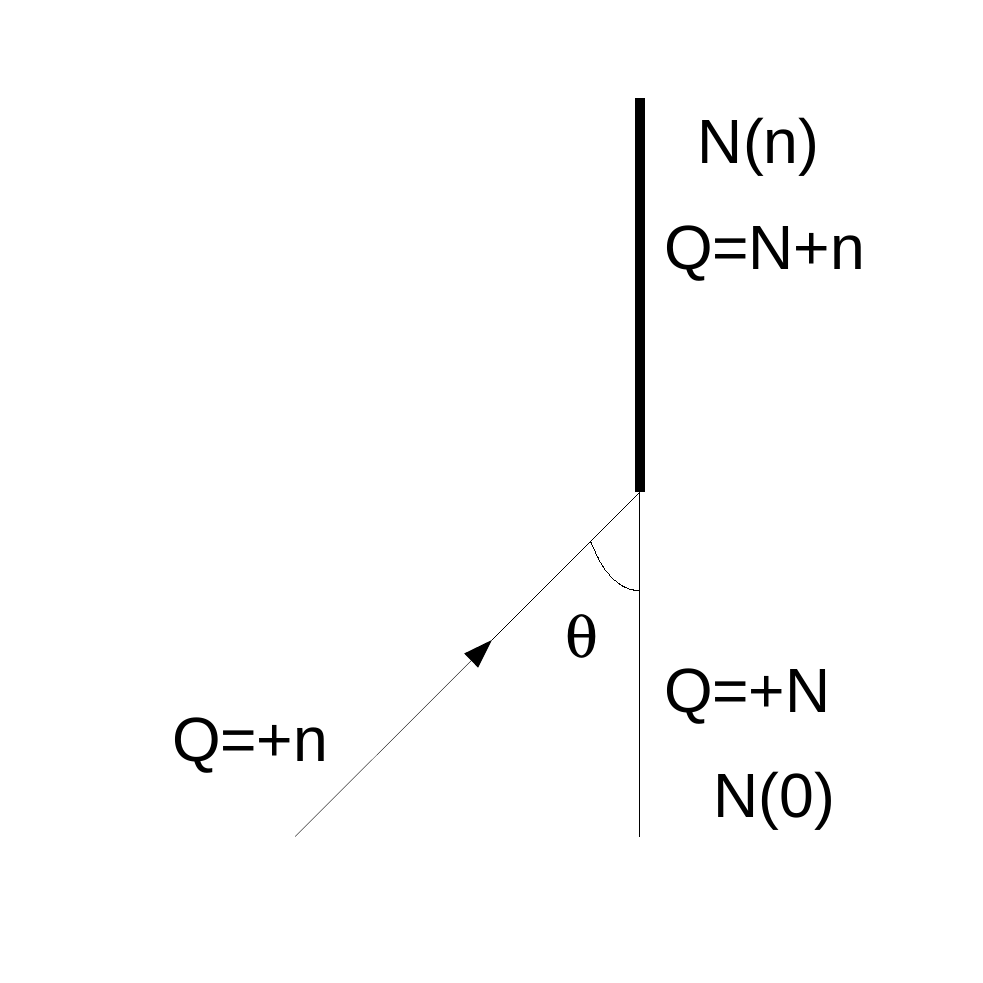}}
\end{center}
\label{fig:BoundaryFusion1}
\caption[CSG solitons fusing with unexcited boundary.]{(a) Charge $Q=+1$ soliton fusing to charge $Q=+N$ boundary, (b) Charge $Q=+n$ soliton fusing to charge $Q=+N$ boundary.}
\end{figure}

Since the unexcited boundary can be described as a particular limit of the bound state, the energy difference formula (\ref{eq:energydiff}) should hold between the unexcited boundary and the first excited bound state. We denote the first excited state above a $Q=N$ unexcited boundary as $N(1)$ and more generally denote the $m^{th}$ excited state by $N(m)$. We consider the process where a charge $Q=+1$ soliton fuses with an unexcited boundary of charge $Q=+N$ to form the first excited bound state with charge $Q=N+1$. The process is shown in figure \ref{fig:CSGBoundaryFusion1}.
A soliton with charge $Q=+1$ is described by $a_{sol} = \frac{\pi}{2} -\frac{\pi}{k}$ and therefore has energy 
\begin{equation}
E_{sol} = \frac{2k \sqrt{\beta}}{\pi} \cosh(\theta)\sin\left(\frac{\pi}{k}\right)\, . 
\end{equation}
The charge $Q=+N$ unexcited boundary has energy
\begin{equation}
E = \frac{k \sqrt{\beta}}{\pi}\cos(b)\cos\left(\frac{2\pi N}{k}\right)\, .
\end{equation}
while the bound state with charge $Q = N+1$ implies that the bound soliton is described by
\begin{equation}
 a= b +A -\frac{2\pi}{k}
\end{equation}
and the bound state has the energy
\begin{equation}
E^{+}_{bs} =  \frac{k \sqrt{\beta}}{\pi} \left(\cos\left(b -\frac{2\pi}{k}(1+N)\right) +  \sin(A)\sin(b)\right)\, . 
\end{equation}
This fusion process is set up so that charge conservation is automatically satisfied, while energy conservation requires the fusing soliton to have the rapidity 
\begin{equation}\label{eq:fusionangle}
\theta = i\left(b -\frac{\pi}{k}(1+2N) - \frac{\pi}{2} \right)\, .
\end{equation}
This fusing rapidity agrees with the rapidity from the semi-classical energy difference (\ref{eq:solrapid}) with $n=N$. The fusion of a charge $Q=+1$ soliton then has the affect of shifting the charge parameter of the stationary soliton in the bound state from
\begin{equation}
a_{0} = b +A \rightarrow a_{1} = a_{0} - \frac{2\pi}{k} \, .
\end{equation}
The quantisation of the bound soliton's charge parameter (\ref{eq:bsaquant}) shows that this sequence continues with $a_{n+1} = a_{n} - \frac{2\pi}{k}$. The semi-classical energy spectrum suggest that the fusion process can be repeated. Namely a $Q=+1$ soliton can fuse with the first excited boundary with charge $Q=N+1$ at the rapidity
\begin{equation}\label{eq:fusionangle2}
\theta = i\left(b -\frac{\pi}{k}(3+2N) - \frac{\pi}{2} \right)\, .
\end{equation}
Continuing the process a $Q=+1$ soliton can fuse with the $m^{th}$ excited boundary with charge $Q=N+m$ at the rapidity
\begin{equation}\label{eq:fusionangle2m}
\theta = i\left(b -\frac{\pi}{k}(1+2m+2N) - \frac{\pi}{2} \right)\, ,
\end{equation}
to form a higher bound state with charge $Q=N+m+1$.

As a generalisation to the process in figure \ref{fig:CSGBoundaryFusion1}, the fusion of a charge $Q=n$ can be considered shown in figure \ref{fig:CSGBoundaryFusion}. We find that the rapidity at which this process occurs is
\begin{equation}
\theta = i\left(b -\frac{\pi}{k}(n+2N) - \frac{\pi}{2} \right)\, ,
\end{equation}
resulting in the same excited boundary than if $n$ $Q=+1$ solitons had been consecutively fused, or in fact any combination of solitons whose charge sum to $n$. The analysis of these fusion processes show that when using $E_{bs}^{+}$, $Q_{bs}^{+}$ the fusion of a soliton steps the bound soliton charge parameter $a$ down from $a=A+b$ in quantum steps and the energy and charge of the bound states increase up the curves illustrated in figures \ref{fig:bsEQ1}, \ref{fig:bsEQ2}, \ref{fig:bsEQ3a}. Closer inspection shows that the energy only increases up to $a=0$. We come back to this point later in the paper.

Similarly we can repeat the analysis using $E_{bs}^{-}$, $Q_{bs}^{-}$ (\ref{eq:bsenergy}), starting by applying the B-S condition to the charge formula $Q_{bs}^{-}$ which implies the quantisation condition on $a$, for $\cos(a)>0$
\begin{equation}
 a = - b - \frac{2\pi n}{k}\, ,
\end{equation}
giving the energy of the charge $n$ state to be
\begin{equation}
 E^{-}_{n} = \frac{k\sqrt{\beta}}{\pi}\left(\cos\left(b + \frac{2\pi n}{k} \right) - \sin(A)\, \sin(b)\right)
\end{equation}
and the semi-classical energy difference 
\begin{eqnarray}
 E^{-}_{n-1} - E^{-}_{n} &=& \frac{k\sqrt{\beta}}{\pi}\left(\cos\left(b + \frac{2\pi (n-1)}{k} \right) - \cos\left(b + \frac{2\pi n}{k} \right)\right)\nonumber \\
&=& -\frac{2k\sqrt{\beta}}{\pi}\cos\left(\frac{\pi}{2}-\frac{\pi}{k}\right)\cos\left(\frac{\pi}{k}(2n-1)+\frac{\pi}{2}+b\right)\, ,
\end{eqnarray}
which corresponds to the negative of the energy of a $Q=+1$ soliton with rapidity
\begin{equation}\label{eq:rapidB}
\theta = i\left(\frac{\pi}{k}(2n-1)+\frac{\pi}{2}+b\right)\, .
\end{equation}
For this choice of energy and charge formulae the bound state reduces to the unexcited boundary when $a=A-b$, which is at the left hand edge of the allowed classical region. Therefore to step into the allowed region the charge parameter has to increase and this coincides with a decrease in the charge, shown in figures \ref{fig:bsEQ1A}, \ref{fig:bsEQ2A}, \ref{fig:bsEQ3A}, the energy can increase or decrease.

\begin{figure}[!h]
\begin{center}
\subfigure[]{\label{fig:Boundaryemission1}\includegraphics[width=0.31\textwidth]{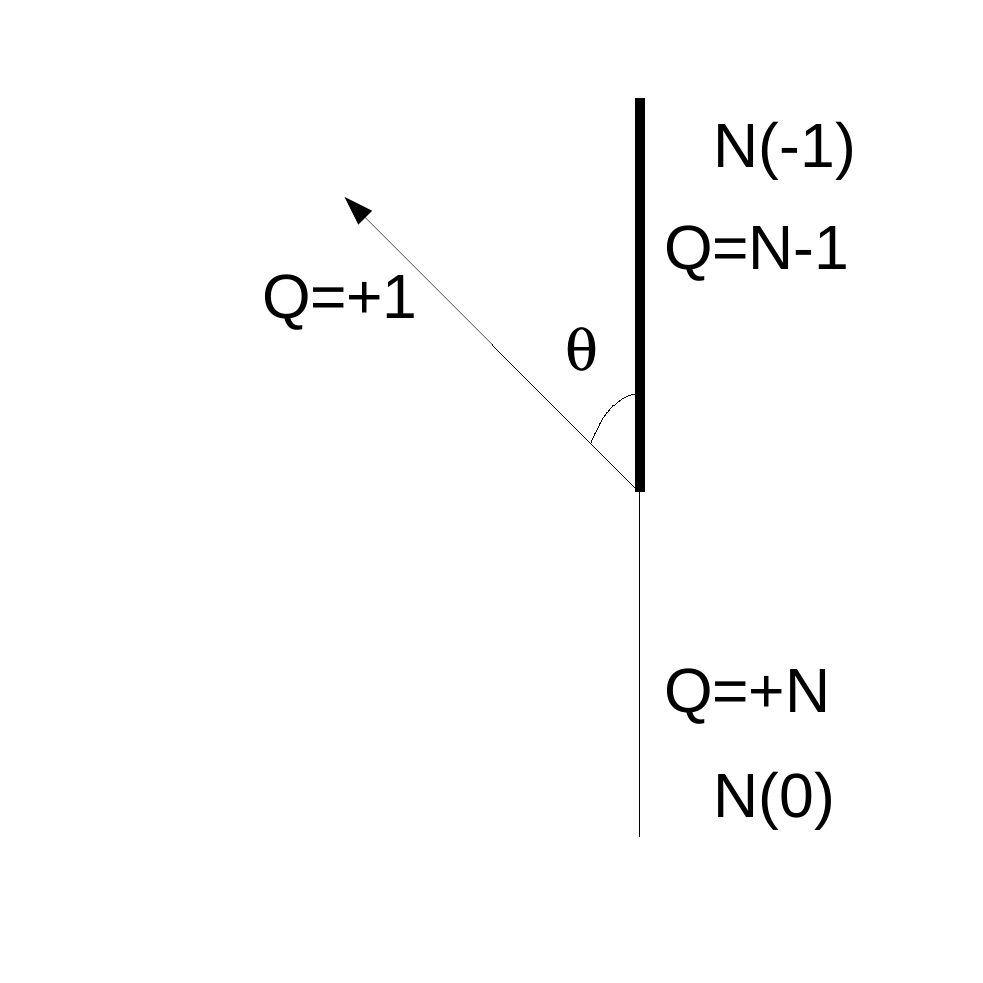}}
\hspace{0.5in}
\subfigure[]{\includegraphics[width=0.31\textwidth]{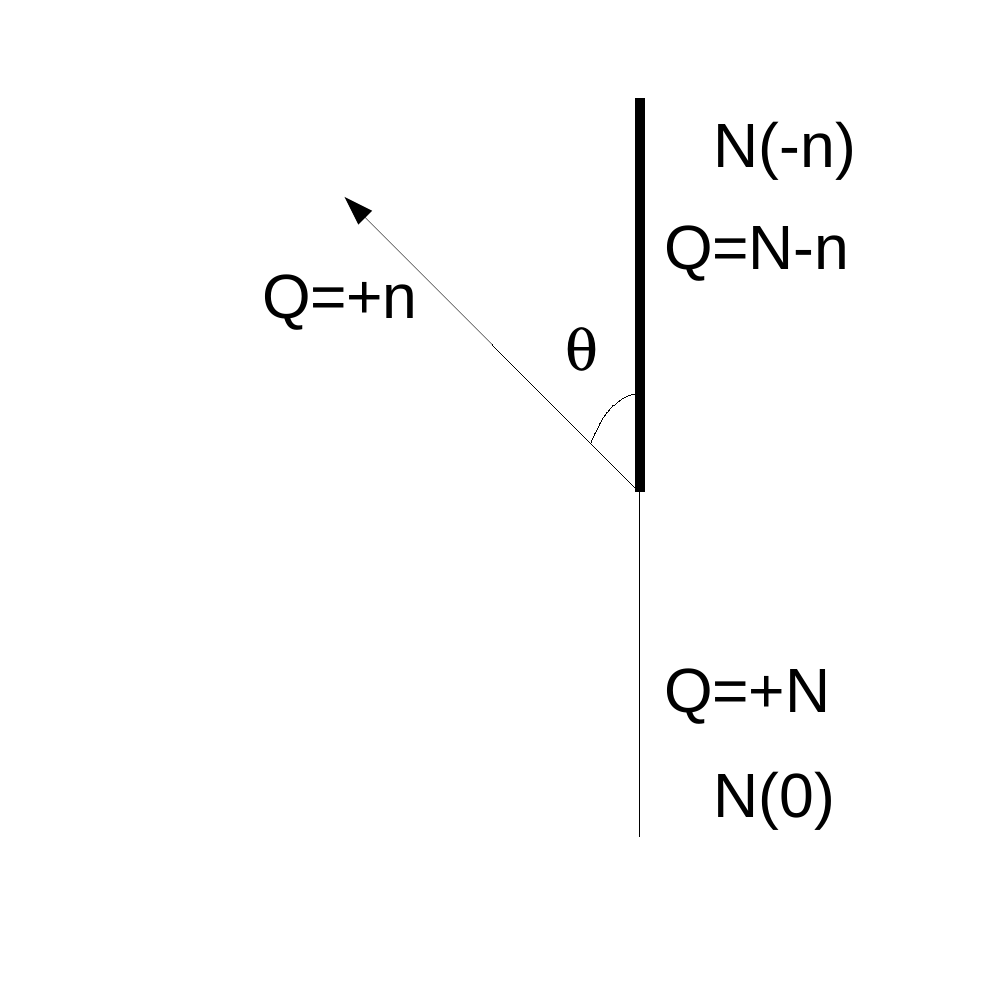}}
\end{center}
\label{fig:Boundaryemission}
\caption[CSG solitons emission from unexcited boundary.]{(a) Charge $Q=+1$ soliton emitting from charge $Q=+N$ boundary, (b) Charge $Q=+n$ soliton emitting from charge $Q=+N$ boundary.}
\end{figure}

For a choice of parameters $(A,b)$ such as in figure \ref{fig:bsEQ3A}, increasing the value of  $a$ from $A-b$  to $A-b + \frac{2\pi}{k}$ decreases the charge and energy of the boundary state whilst moving the bound soliton away from right infinity. This can be interpreted as the emission of a charge $1$ soliton (or
particle) from the unexcited boundary at rapidity given by (\ref{eq:rapidB}). This is illustrated in figure \ref{fig:Boundaryemission}.
This process is also possible for higher charged solitons at the rapidity
\begin{equation}
\theta = i\left(\frac{\pi}{k}(2N-n)+\frac{\pi}{2}+b\right)\, .
\end{equation}
This behaviour is the opposite of the fusion processes described earlier. The difference is due to the bound state charge increasing when $a$ decreases from $A+b$ and decreasing when $a$ increases from $A-b$.
In the next section we use the semi-classical energy spectrum (\ref{eq:energydiff}) and fusing angles (\ref{eq:fusionangle2m}) to help determine the fully quantum reflection matrices.

\subsection{Dressed boundary bootstrap}

The procedure to generate quantum reflection matrices for charged CSG solitons from the dressed boundary is to first conjecture the reflection matrix for the charge $Q=+1$ soliton or particle from a charge $Q=+N$ unexcited boundary. From this reflection matrix we use the reflection bootstrap and boundary bootstrap procedures to generate the general quantum reflection matrix for a charge $Q=+n$ soliton from an excited charge $Q=N+m$ boundary \cite{Cherednik:1985vs,Ghoshal:1993tm,Fring:1993mp}. We make various checks to ensure that the original conjecture makes sense.

As in the previous work on the quantum CSG boundary theory \cite{Bowcock:2006hj}, which covers the subsection of dressed boundaries with $A=0$, it is assumed that the reflection factors are constructed out of $F$ factors (\ref{eqFFactor}). The CSG $S$-matrix (\ref{eq:SMatrix}) is defined as a product of these $F$ factors and since the boundary Yang-Baxter equation relates the reflection matrices and the $S$-matrix this assumption has foundation.

The CSG $S$-matrix is identically the minimal $a_{k-1}$ $S$-matrix which is recovered from the $a_{k-1}^{(1)}$ Affine Toda field theory (ATFT) when the parts with the coupling constant are omitted. Therefore as in the previous work on the CSG boundary theory we use the terms of the reflection matrix of the $a_{k-1}^{(1)}$ ATFT \cite{Delius:1999cs}, which do not include the coupling constant, as a starting point for the CSG dressed boundary reflection matrix. Using the block notation $(x)=F_{x}(\theta)$ the terms in the charge $Q=+n$ soliton reflection matrix are
\begin{equation}
 K_{n}^{\mathrm{base}} = \prod_{c=1}^{n} (c-1)(c-k) .
\end{equation}
The matrix for the reflection of a charge $Q=+1$ soliton from a charge $Q=+N$ unexcited dressed boundary, which we denote by $K_{1}^{N(0)}$, should therefore include the factor $(1-k)$. This cannot be the whole expression as it does not contain a factor that corresponds to the known formation of a bound state discussed in the previous section, where a charge $Q=+1$ soliton fuses with an unexcited boundary of charge $Q=+N$. This process occurs when the incoming soliton has the rapidity $\theta = \frac{i \pi}{k}(1+2N-B)$  where $B = \frac{kb}{\pi}-\frac{k}{2}$. This fusion process indicates the need for block factor $(1+2N-B)$ in the minimal choice for $K_{1}^{N(0)}$.

Delius and Gandenberger \cite{Delius:1999cs} showed that when block factors appear in the pairs $(x)(k-x)$ then the bootstrap is guaranteed to close. In the previous work \cite{Bowcock:2006hj} charge conjugation invariance, i.e. $K^{0}_{1} = K^{0}_{-1}$ was needed. However, in the case of the charged dressed boundary we do not expect invariance under charge conjugation. It is therefore not required for the reflection matrix to have its $F$ factors appear in these pairs. In fact they cannot appear in this way for the charge conjugation symmetry to be broken.

The way forward in this case is to assume that a similar factor to $(k-1-2N+B)$ does accompany $(1+2N-B)$ in $K_{1}^{N(0)}$ and to find the correct factor we check that the classical limit $k\rightarrow \infty$ is correct. Examining the classical reflection factors for a particle and anti-particle reflecting from the dressed boundary
\begin{eqnarray}\label{eq:CSGreflection}
R_{particle} &=& - \frac{(\delta e^{\theta} +ie^{iA})(\delta e^{iA} -i e^{\theta})}{(\delta +i e^{iA}e^{\theta})(\delta e^{iA} e^{\theta}-i)} \, ,\nonumber \\
R_{anti-particle} &=& - \frac{(\delta  -ie^{\theta}e^{iA})(\delta e^{iA}e^{\theta} +i )}{(\delta e^{iA} + i e^{\theta} )(\delta e^{\theta} - i e^{iA})} \, .
\end{eqnarray}
These formulae differ from the ones presented in section \ref{sec:CSGdb} due to a difference in the prescription in the signs of $k$ and $\omega$. We note that in the $A=0$ limit $R_{particle} = R_{anti-particle}$, which confirms the charge conjugation symmetry in this case. Similarly the particle reflection factor from the bound state with these prescriptions is
\begin{equation}
 R_{particle}^{bs} = \frac{(1+ie^{ib}e^{iA}e^{\theta})(e^{iA}-ie^{ib}e^{\theta})(e^{ia}-ie^{\theta})^{2}}{(e^{ib}e^{iA}-ie^{\theta})(ie^{iA}e^{\theta}+e^{ib})(ie^{ia}e^{\theta}+1)^{2}}\, .
\end{equation}
We find that a conjecture for $K_{1}^{N(0)}$ which has the correct classical limit and includes the pole that corresponds to the known bound state is
\begin{eqnarray}\label{eq:K1N0}
 K_{1}^{N(0)}(\theta) &=& K_{1}^{\mathrm{base}} (1+2N-B)(k+B-1+2N)\nonumber\\
  &=& (1-k)(1+2N-B)(k+B-1+2N)\, .
\end{eqnarray}

\subsection{Reflection bootstrap}\label{sec:reflecboot}

\begin{figure}[!h]
\begin{center}
\includegraphics[width=0.65\textwidth]{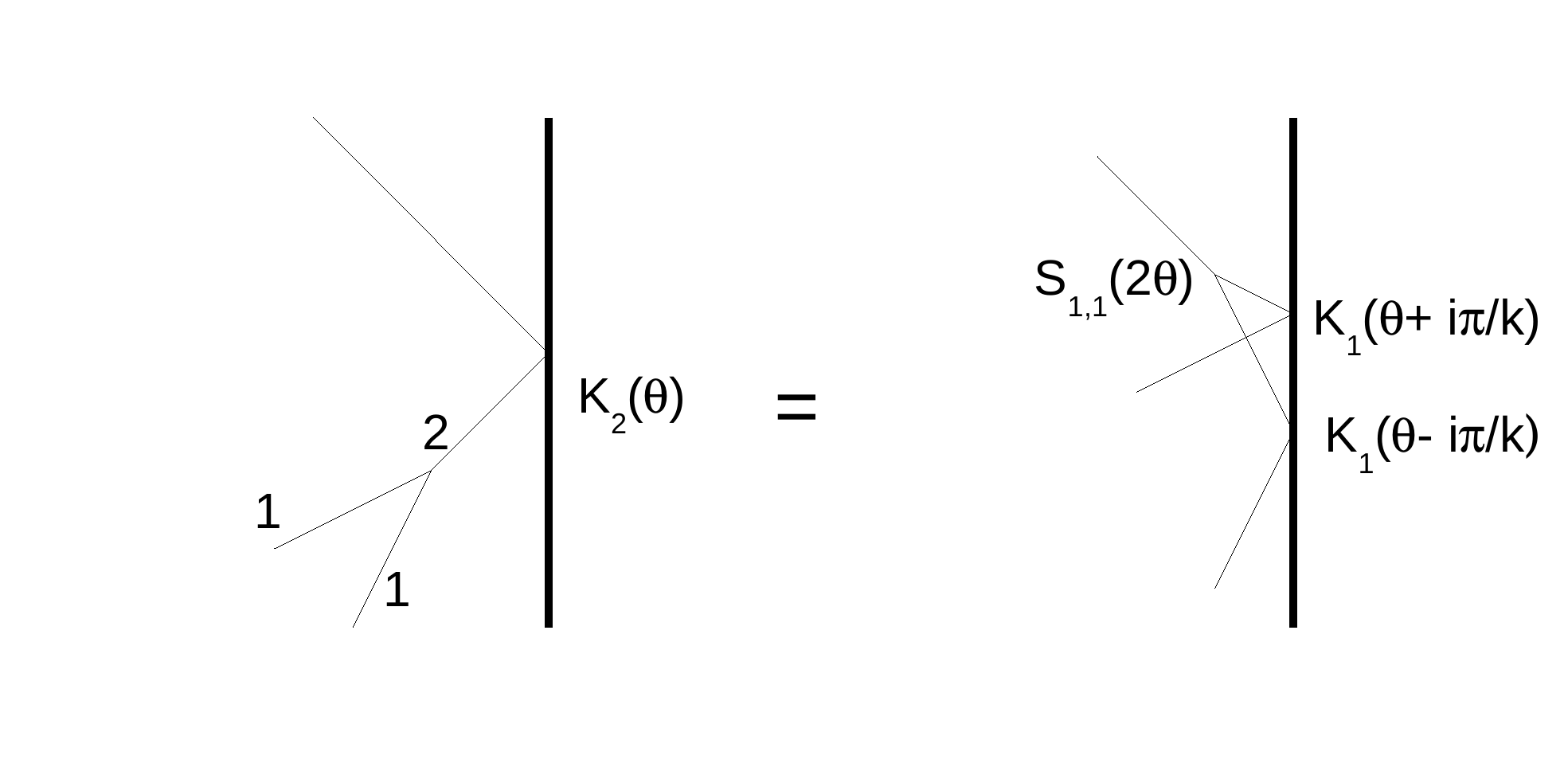}
\end{center}
\caption[Reflection bootstrap relation I.]{Reflection bootstrap for two $Q=+1$ solitons fusing into a $Q=+2$ soliton.}
\label{fig:BoundBootstrap1}
\end{figure}

\begin{figure}[!h]
\begin{center}
\includegraphics[width=0.65\textwidth]{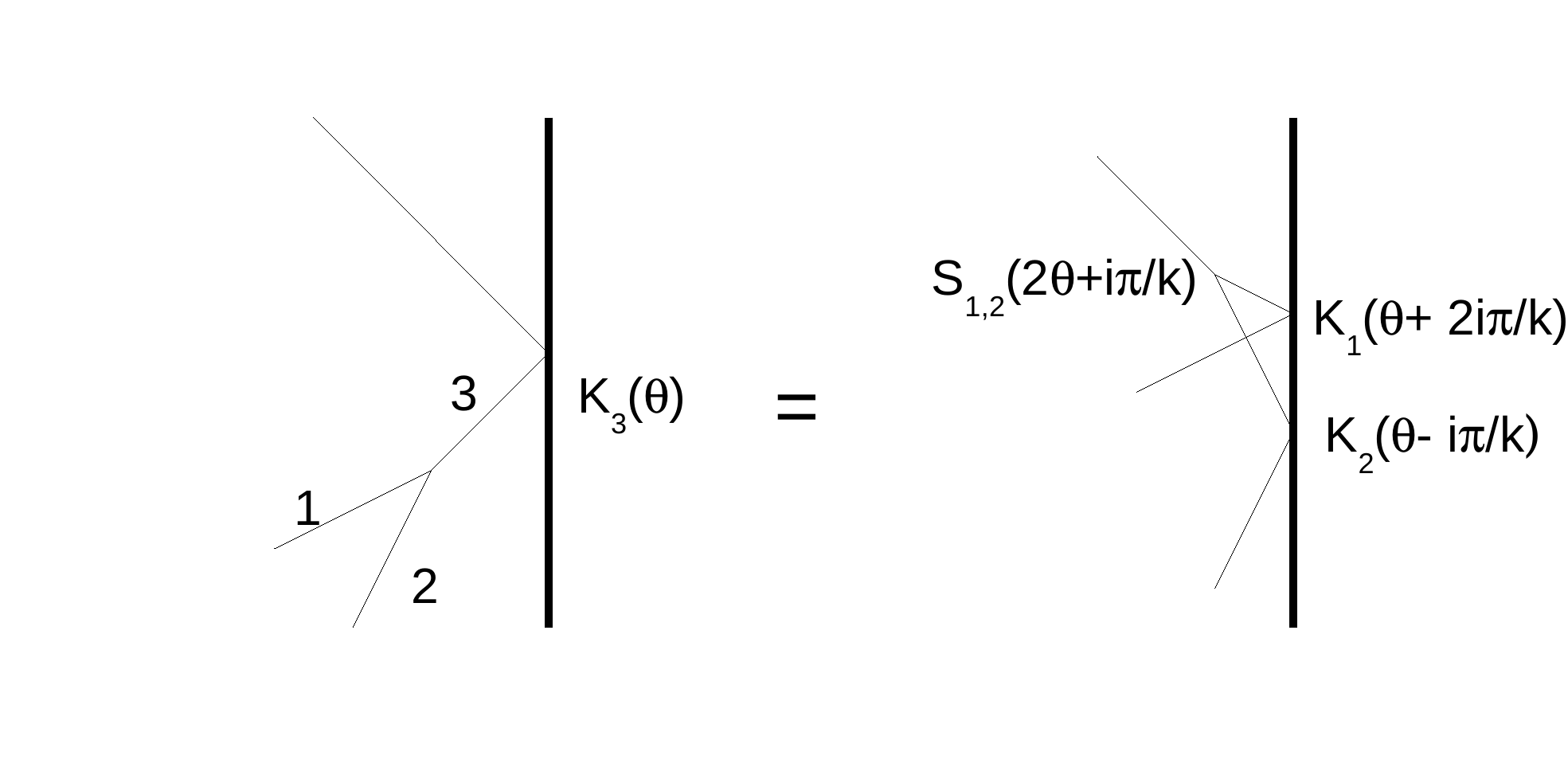}
\end{center}
\caption[Reflection bootstrap relation II.]{Reflection bootstrap for a $Q=+1$ and $Q=+2$ soliton fusing into a $Q=+3$ soliton.}
\label{fig:BoundBootstrap2}
\end{figure}

\begin{figure}[!h]
\begin{center}
\includegraphics[width=0.65\textwidth]{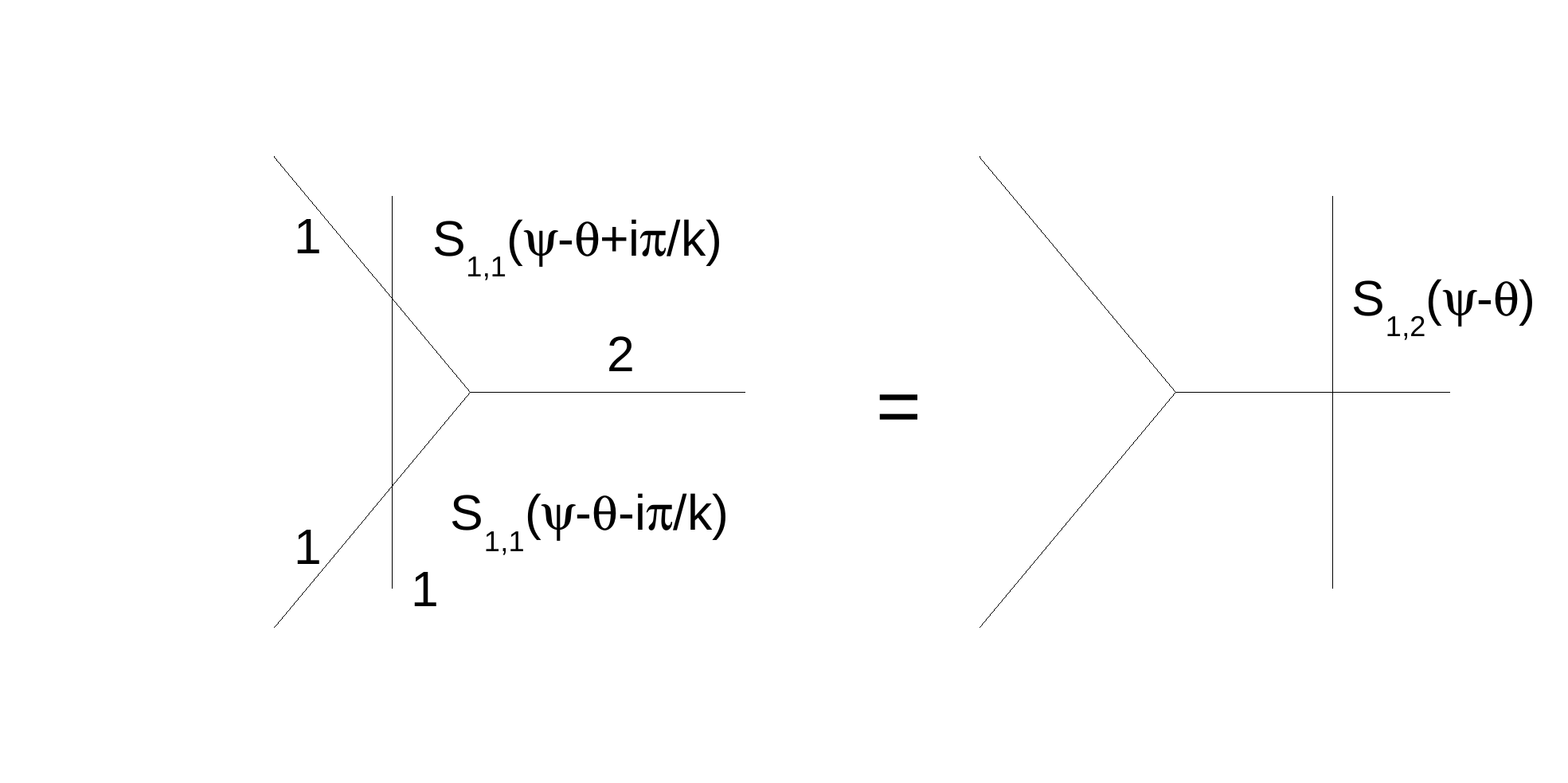}
\end{center}
\caption[Bulk bootstrap relation.]{Bulk bootstrap relation allowing the $S$-matrix between charge $Q=+1$ and $Q=+2$ solitons to be related to the $S$-matrix between two $Q=+1$ solitons.}
\label{fig:Bootstrap1}
\end{figure}

We use the reflection bootstrap mechanism from $K_{1}^{N(0)}$ to generate the reflection matrices for higher charged solitons reflecting from the unexcited boundary, denoted by $K_{n}^{N(0)}$. The reflection bootstrap uses the integrability of the model to equate the fusion of two solitons before and after reflection from the boundary. It allows the reflection matrix for the higher charged soliton to be calculated from known lower charge soliton reflection matrices and $S$-matrices. We illustrate the first step in the reflection bootstrap procedure in figure \ref{fig:BoundBootstrap1},
which gives the relation between a charge $Q=+2$ and charge $Q=+1$ CSG soliton reflecting from the unexcited boundary
\begin{equation}
 K_{2}^{N(0)}(\theta) =  K_{1}^{N(0)}\left(\theta-\frac{i\pi}{k}\right)\ K_{1}^{N(0)}\left(\theta+\frac{i\pi}{k}\right)\ S_{1,1}(2\theta)\, .
\end{equation}
It uses the property illustrated in figure \ref{fig:SolBound11} that the two $Q=+1$ solitons fuse at the relative imaginary rapidity $\frac{2i\pi}{k}$. Explicitly writing the $F$ factors that appear in the two $K_{1}^{N(0)}$
\begin{equation}
F_{x}\left(\theta-i\frac{\pi}{k} \right) = \frac{\sinh\left(\frac{\theta}{2} + \frac{i\pi}{2k}(x-1)\right)}{\sinh\left(\frac{\theta}{2} - \frac{i\pi}{2k}(x+1)\right)}, \ \ \ \ F_{x}\left(\theta+i\frac{\pi}{k} \right) = \frac{\sinh\left(\frac{\theta}{2} + \frac{i\pi}{2k}(x+1)\right)}{\sinh\left(\frac{\theta}{2} - \frac{i\pi}{2k}(x-1)\right)},
\end{equation}
shows that we can combine them using
\begin{equation}
 F_{x}\left(\theta-i\frac{\pi}{k} \right)\ F_{x}\left(\theta+i\frac{\pi}{k} \right) = F_{x+1}(\theta)\ F_{x-1}(\theta)\, .
\end{equation}
Along with the form of the $S$-matrix
\begin{equation}
S_{1,1}(2\theta) = F_{0}(2\theta)F_{2}(2\theta) = -(1)(1-k)\, ,
\end{equation}
this gives
\begin{equation}\label{eq:K2N0}
K_{2}^{N(0)}(\theta) = (1-k)(1)(2-k)(2+2N-B)(2N-B)(k+B+2N)(k+B-2+2N)\, . 
\end{equation}
It is noticed the base factor $K_{2}^{\mathrm{base}}$ appears, allowing it to be rewritten
\begin{equation}
K_{2}^{N(0)}(\theta) = K_{2}^{\mathrm{base}}\prod_{j=0}^{1}(2N-B+2j)(k+B+2N-2j)\, . 
\end{equation}
To find the reflection factors for higher charged solitons we use the fusion process between higher charged solitons and a charge $Q=+1$ soliton. For example for the next step to generate $K_{3}^{N(0)}$ we use the fusion process between a charge $Q=+1$ and $Q=+2$ soliton, illustrated in figure \ref{fig:BoundBootstrap2}.
This gives the relation
\begin{equation}
 K_{3}^{N(0)}(\theta) =  K_{1}^{N(0)}\left(\theta+\frac{2i\pi}{k}\right)\ K_{2}^{N(0)}\left(\theta-\frac{i\pi}{k}\right)\ S_{1,2}\left(2\theta+\frac{i\pi}{k}\right)\, .
\end{equation}
To find the explicit form of $K_{3}^{N(0)}(\theta)$ we use the reflection bootstrap iteratively, namely we use the equation for $K_{2}^{N(0)}$ 
\begin{equation}
 K_{2}^{N(0)}\left(\theta-\frac{i\pi}{k}\right) =  K_{1}^{N(0)}\left(\theta-\frac{2i\pi}{k}\right)\ K_{1}^{N(0)}\left(\theta\right)\ S_{1,1}\left(2\theta-\frac{2i\pi}{k}\right)
\end{equation}
and the bulk bootstrap relation, shown in figure \ref{fig:Bootstrap1},
which gives the relation 
\begin{equation}
S_{1,2}\left(2\theta+\frac{i\pi}{k}\right) = S_{1,1}\left(2\theta+\frac{2i\pi}{k}\right)\ S_{1,1}\left(2\theta\right)\, .
\end{equation}
 We simplify the block factors that appear in $K_{1}^{N(0)}\left(\theta-\frac{2i\pi}{k}\right)\ K_{1}^{N(0)}\left(\theta+\frac{2i\pi}{k}\right)$ using
\begin{equation}
 F_{x}\left(\theta-i\frac{2\pi}{k} \right)\ F_{x}\left(\theta+i\frac{2\pi}{k} \right) = F_{x+2}(\theta)\ F_{x-2}(\theta)\, ,
\end{equation}
to give the form of $K_{3}^{N(0)}(\theta)$
\begin{equation}
K_{3}^{N(0)}(\theta) = K_{3}^{\mathrm{base}}\prod_{j=0}^{2}(2N-B+2j-1)(k+B+2N-2j+1)\, .
\end{equation}
We continue this procedure to generate the reflection matrix for a charge $Q=+n$ soliton reflecting from an unexcited boundary with charge $Q=+N$
\begin{equation}\label{eq:KnNO}
K_{n}^{N(0)}(\theta) = K_{n}^{\mathrm{base}}\prod_{j=0}^{n-1}(2N-B+2j+2-n)(k+B+2N-2j-2+n)\, . 
\end{equation}
We check that the bootstrap closes, namely that
\begin{equation}
K_{1}^{N(0)}(\theta) =  K_{k+1}^{N(0)}(\theta).
\end{equation}
This equation is true only for $k$ even, so we shall restrict ourselves to these values of $k$.
Also as expected, the charge conjugation symmetry is broken
\begin{equation}
K_{k-1}^{N(0)}(\theta) = (1-k)(-2N+B-k-1)(1-2N-B)  = K_{-1}^{N(0)}(\theta)\, .
\end{equation} 
In the classical limit this agrees with the reflection factor for the anti-particle (\ref{eq:CSGreflection}). We leave the details of these checks to appendix \ref{appen:closure}.

\subsection{Boundary bootstrap}
\begin{figure}[!h]
\begin{center}
\includegraphics[width=0.65\textwidth]{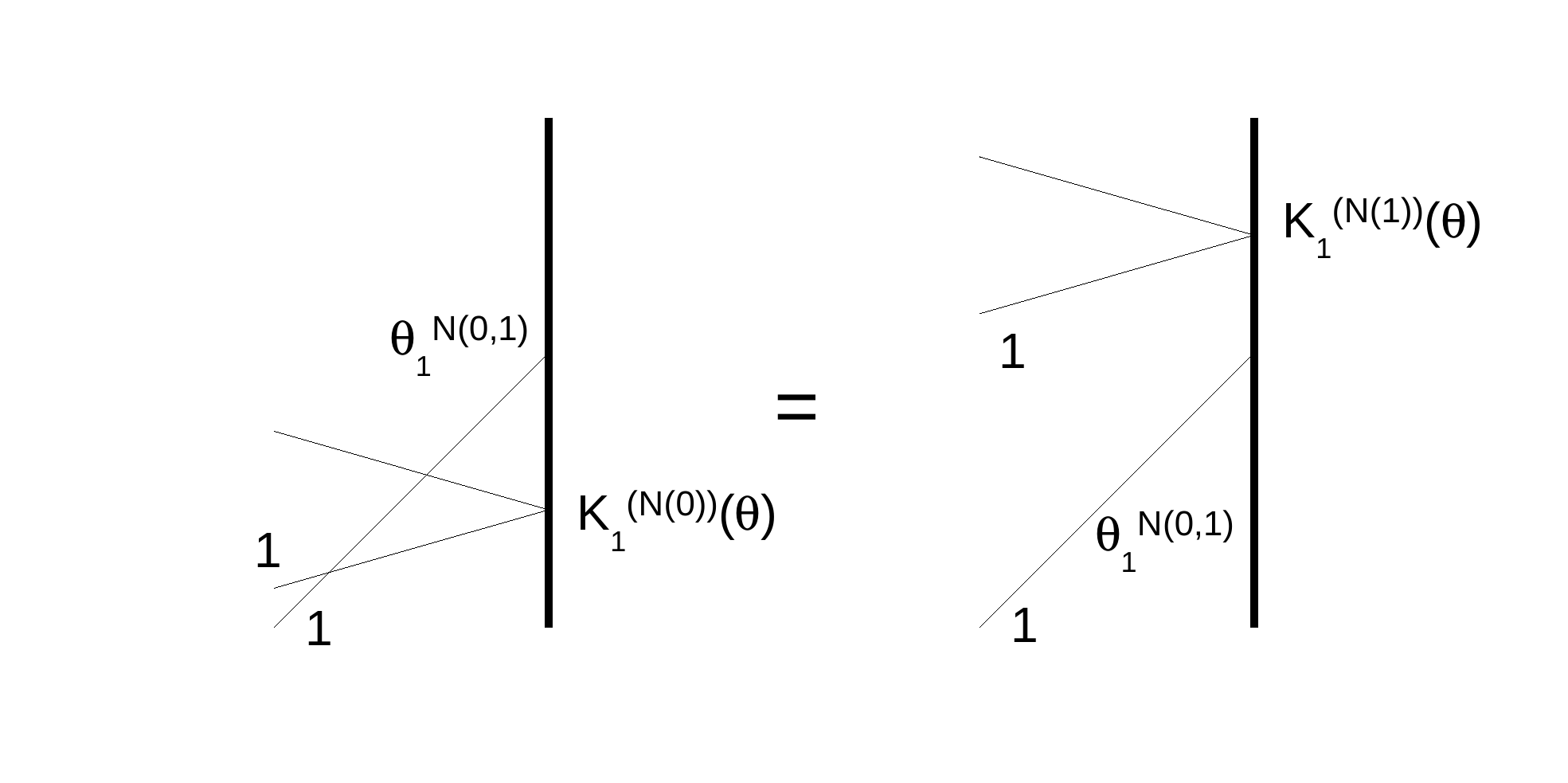}
\end{center}
\caption[Boundary bootstrap relation I.]{Boundary bootstrap giving a relation for the reflection factor from the first excited boundary.}
\label{fig:BoundBootstrap3}
\end{figure}

\begin{figure}[!h]
\begin{center}
\includegraphics[width=0.65\textwidth]{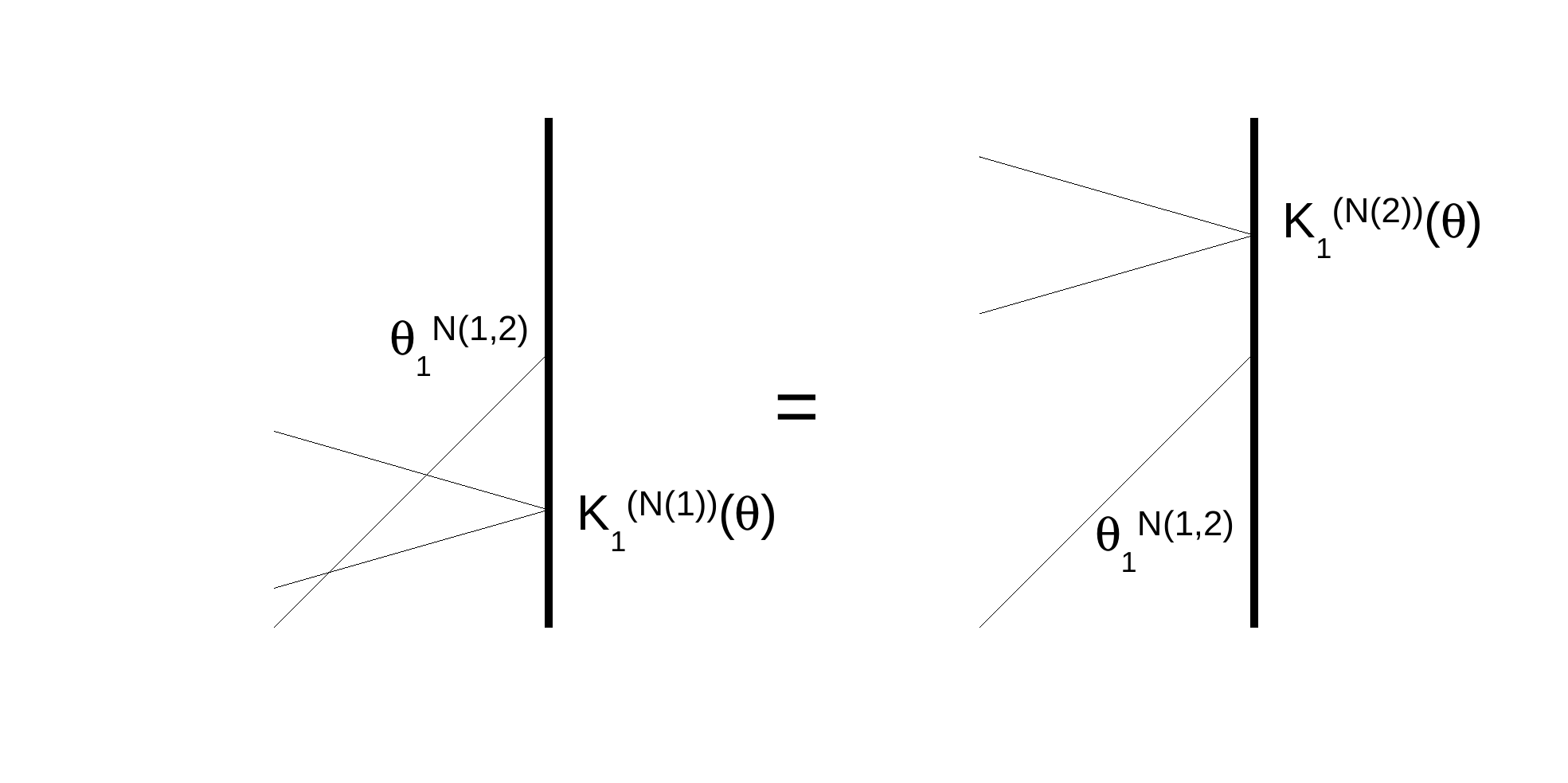}
\end{center}
\caption[Boundary bootstrap relation II.]{Boundary bootstrap giving a relation for the reflection factor from the second excited boundary.}
\label{fig:BoundBootstrap4}
\end{figure}

Using the reflection bootstrap we have constructed the quantum reflection matrices for any charged soliton from the unexcited boundary. Now using the boundary bootstrap mechanism we generate the reflection matrices which describe the reflection from excited boundaries. The first step of this process is illustrated in figure \ref{fig:BoundBootstrap3},
which gives the relation between a charge $Q=+1$ CSG soliton reflecting from the unexcited boundary and the charge $Q=+1$ soliton reflecting from the first excited bound state. Namely
\begin{equation}
S_{1,1}(\theta-\theta_{1}^{N(0,1)})\ K_{1}^{N(0)}(\theta)\ S_{1,1}(\theta+\theta_{1}^{N(0,1)}) = K_{1}^{N(1)}(\theta)\, ,
\end{equation}
where 
\begin{equation}
\theta_{1}^{N(0,1)} = \frac{i\pi}{k}(1+2N-B)\, ,
\end{equation}
is the imaginary rapidity at which a charge $Q=+1$ soliton fuses with an unexcited boundary of charge $Q=+N$ (\ref{eq:fusionangle}). Noticing that the product of $S$-matrices can be simplified
\begin{equation}
S_{1,1}\left(\theta+\frac{i\pi}{k}\psi\right)\ S_{1,1}\left(\theta-\frac{i\pi}{k}\psi\right) = (2+\psi)(2-\psi)\, ,
\end{equation}
then the reflection matrix from the excited boundary is
\begin{equation}\label{eq:K1N1}
K_{1}^{N(1)}(\theta) = (1-k)(1-2N+B)(k+B-1+2N)(1+2N-B)(3+2N-B)\, .
\end{equation}
In the classical limit this agrees with the particle reflection factor from the bound state (\ref{eq:ParticleBoundStateRef}). In $K_{1}^{N(1)}$ there is a new pole which appears in the similar factor $(3+2N-B)$ at
\begin{equation}
\theta_{1}^{N(1,2)} = \frac{i\pi}{k}(3+2N-B)\, ,
\end{equation}
this agrees with the rapidity required for the next bound state to be formed (\ref{eq:fusionangle2}) and therefore we can use this pole to repeat the boundary bootstrap process, illustrated in figure \ref{fig:BoundBootstrap4}.
This gives the relation
\begin{equation}
S_{1,1}(\theta-\theta_{1}^{N(1,2)})\ K_{1}^{N(1)}(\theta)\ S_{1,1}(\theta+\theta_{1}^{N(1,2)}) = K_{1}^{N(2)}(\theta)\, ,
\end{equation}
which can be solved for the quantum reflection matrix for the $Q=+1$ soliton from the second excited bound state
\begin{equation}
K_{1}^{N(2)}(\theta) = (1-k)(1-2N+B)(k+B-1+2N)(3+2N-B)(5+2N-B)\, .
\end{equation}
This again has a new pole, this time in the factor $(5+2N-B)$, which again agrees with the fusion factor (\ref{eq:fusionangle2}). At every step a similar new pole appears and we use it to iteratively apply the boundary bootstrap process. Resulting in the general reflection factor
\begin{equation}\label{eq:K1Nm}
K_{1}^{N(m)}(\theta) = (1-k)(1-2N+B)(k+B-1+2N)(2m-1+2N-B)(2m+1+2N-B)\, ,
\end{equation}
which described the scattering of a charge $Q=+1$ soliton from the charge $Q=N+m$ $m^{th}$ excited bound state. The charge of the boundary has $Z_{k}$ symmetry which is exhibited by this formula since
\begin{equation}
K_{1}^{(N(k))}(\theta) = K_{1}^{N(0)}(\theta) \, .
\end{equation}

The final stage of the bootstrap process to complete the reflection matrices for all possible soliton-boundary reflections is to repeat the reflection bootstrap process starting with $K_{1}^{N(m)}$ (\ref{eq:K1Nm}) for any $m$. The first step is the relation
\begin{equation}
 K_{2}^{N(m)}(\theta) =  K_{1}^{N(m)}\left(\theta-\frac{i\pi}{k}\right)\ K_{1}^{N(m)}\left(\theta+\frac{i\pi}{k}\right)\ S_{1,1}(2\theta)\, ,
\end{equation}
which gives
\begin{equation}
 K_{2}^{N(m)}(\theta) = K_{2}^{\mathrm{base}}\ A_{2}^{N(m)}\ B_{2}^{N(m)} \ C_{2}^{N(m)}\, ,
\end{equation}
where
\begin{eqnarray}
A_{2}^{N(m)} &=& (k+B+2N)(k+B+2N-2) \, ,\nonumber\\
B_{2}^{N(m)} &=& (2-2N+B)(-2N+B) \, ,\nonumber\\
C_{2}^{N(m)} &=& (2m+2N-B-2)(2m+2N-B)^{2}(2m+2N-B+2)\, .\nonumber \\
\end{eqnarray}
Repeating gives
\begin{equation}
 K_{3}^{N(m)}(\theta) = K_{3}^{\mathrm{base}}\ A_{3}^{N(m)}\ B_{3}^{N(m)} \ C_{3}^{N(m)}\, ,
\end{equation}
where
\begin{eqnarray}
A_{3}^{N(m)} &=& (k+B+2N+1)(k+B+2N-1)(k+B+2N-3)\, ,\nonumber\\
B_{3}^{N(m)} &=& (3-2N+B)(1-2N+B)(-1-2N+B)\, , \nonumber\\
C_{3}^{N(m)} &=& (2m+2N-B-3)(2m+2N-B-1)^{2} \nonumber \\ 
&& \ \ \ \ \ \ \mathrm{x} \ \ (2m+2N-B+1)^{2}(2m+2N-B+3)\, .
\end{eqnarray}
We continue this process to give the final general formula for the the quantum reflection matrix for a charge $Q=+n$ soliton from the $m^{th}$ excited boundary with charge $Q=N+m$ 
\begin{eqnarray}\label{eq:KnNm}
K_{n}^{N(m)}(\theta) &=& K_{n}^{\mathrm{base}} \ \ \ \prod_{j=0}^{n-1} (n-2j-2N+B)(k+B+2N-n+2j)\nonumber\\
&& \ \ \ \ \mathrm{x} \ \ (2m-n+2N-B) (2m+n+2N-B)\nonumber \\
&& \ \ \ \ \mathrm{x} \ \ \prod_{j=1}^{n-1}(2m-n+2j+2N-B)^{2}\, .
\end{eqnarray}

From a conjectured form of $K_{1}^{N(0)}$, the reflection factor for the CSG particle from a charge $Q=+N$ unexcited boundary, we have used the bootstrap program to generate the general $K_{n}^{N(m)}$, the reflection factor for a charge $Q=+n$ soliton from the $m^{th}$ excited boundary with charge $Q=N+m$. We have checked that our results agree with known classical formulae and that the bootstrap closes both on the charge of the reflecting soliton and the charge of the boundary.

\section{Physical strip pole analysis}

In this section we perform a preliminary analysis of the poles in the physical strip that appear in the dressed boundary reflection matrices. We first study some specific examples to find out which poles lie in the physical strip $0 < \mathcal{I}m(\theta) < \frac{\pi}{2}$. The examples we use are for values of $A$ and $b$ that we analysed the classical bound state solution in section \ref{sec:CSGdb}, namely $A=0\, , \ b = \frac{\pi}{3}$ and $A=\frac{\pi}{8}\, , \ b = \frac{\pi}{3}$. We use $k=100$ and $k=96$ respectively.

\subsection{Example I: $A=0\, , \ b = \frac{\pi}{3}\, ,\ k=100$} 

The unexcited boundary has charge $Q=0$ and the poles in $K_{1}^{N(0)}$ (\ref{eq:K1N0}) appear at the rapidities 
\begin{eqnarray}
(1-k) && -99\pi i \nonumber  \\
(1+2N-B) && \frac{53}{300} \pi i  \nonumber \\
(k+B-1+2N) && \frac{247}{300}\pi i \, .
\end{eqnarray}
There is one physical pole $(1+2N-B)$, which is the pole we implemented the bootstrap procedure on. The excited boundaries generated by fusing particles to this unexcited boundary have the reflection matrices $K_{1}^{N(m)}$ (\ref{eq:K1Nm}), which has poles at the rapidities
\begin{eqnarray}
(1-k) && -99\pi i \nonumber  \\
(1-2N+B) && -\frac{47}{300} \pi i \nonumber \\ 
(k+B-1+2N) && \frac{247}{300}\pi i \nonumber \\
(2m+1+2N-B) && \frac{53+6m}{300} \pi i \nonumber \\
(2m-1+2N-B) && \frac{47+6m}{300} \pi i \, .
\end{eqnarray}
The pole associated with the factor $(2m+1+2N-B)$ remains in the physical strip until $m=17$, where the pole is at the rapidity $\frac{155}{300}\pi i$. The pole that is used for the bootstrap $(2m-1+2N-B)$ remains in the next higher charge boundary reflection factor, in the next section we show that the existence of this physical pole can be explained by a Coleman-Thun process. Recalling that
\begin{equation}
a_{0} = A + b = \frac{\pi}{3}\, , \ \ \ a_{m} = A + b -\frac{2\pi m}{k} = \frac{\pi}{3} - \frac{m \pi}{50} \, ,
\end{equation}
then explicitly $a_{16} = \frac{2\pi}{150}\, ,\  a_{17} = -\frac{\pi}{150}$. Therefore the charge parameter of the bound soliton for the final bound state actually lies past the maximum of the energy of the bound states at $a=0$ shown in figure \ref{fig:bsEQ1}. However due to the quantisation of $a$ the energy of boundary $N(17)$ is higher than $N(16)$
\begin{equation}
 E_{N(16)} = \frac{k\sqrt{\beta}}{\pi}\cos\left(\frac{2\pi}{150}\right)\, ,\ \ \   E_{N(17)} = \frac{k\sqrt{\beta}}{\pi}\cos\left(\frac{\pi}{150}\right)\, .
\end{equation}
We note that if the pole in $K_{1}^{N(17)}$ was still physical then the next bound state would have reduced energy, so the bootstrap procedure is halted when the highest energy bound state is reached.

\subsection{Example II: $A=\frac{\pi}{8}\, , \ b = \frac{\pi}{3}\, ,\ k=96$} 

The unexcited boundary has charge $Q=-6$ and the poles in $K_{1}^{N(0)}$ (\ref{eq:K1N0}) appear at the rapidities
\begin{eqnarray}
(1-k) && -95\pi i \nonumber  \\
(1+2N-B) && \frac{5}{96} \pi i \nonumber \\
(k+B-1+2N) && \frac{67}{96}\pi i \, ,
\end{eqnarray}
so again there is one physical pole $(1+2N-B)$. The excited boundaries generated by fusing particles to this unexcited boundary have the reflection matrices $K_{1}^{N(m)}$ (\ref{eq:K1Nm}), which has poles at the rapidities
\begin{eqnarray}
(1-k) && -95\pi i \nonumber  \\
(1-2N+B) && -\frac{3}{96} \pi i \nonumber \\ 
(k+B-1+2N) && \frac{67}{96}\pi i \nonumber \\
(2m+1+2N-B) && \frac{5+2m}{96} \pi i \nonumber \\
(2m-1+2N-B) && \frac{3+2m}{96} \pi i \, .
\end{eqnarray}
The pole associated with the factor $(2m+1+2N-B)$ remains in the physical strip until $m=22$, where the pole is at the rapidity $\frac{49}{96}\pi i$. Recalling that
\begin{equation}
a_{0} = A + b = \frac{11 \pi}{24}\, , \ \ \ a_{m} = A + b -\frac{2\pi m}{k} = \frac{11 \pi}{24} - \frac{m \pi}{48} \, ,
\end{equation}
then explicitly $a_{21} = \frac{\pi}{48}\, , a_{22} = 0$. This time the charge parameter of the bound soliton for the final bound state coincides with the maximum of the energy of the bound states at $a=0$ shown in figure \ref{fig:bsEQ2}. Again the bootstrap procedure is halted when the highest energy bound state is reached. These two examples show for parameter choices such as these we correctly implemented the bootstrap methods, albeit for a finite number of steps. In the next section we explain the processes behind the physical poles.

\subsection{Coleman-Thun processes} 
In this section we present the Coleman-Thun type processes that explain the physical poles in the reflection matrices. We limit ourselves to ranges of the parameters where the factors of the form $(x+2N-B)$ and the base factors are the only ones in the physical strip, as in the examples shown above. We note that these are the poles that were used in the  bootstrap, and  are related to the rapidity of the fusing soliton needed to step up the $E_{bs}^{+}$ energy curve, while $a$ decreases from $A+b$.   At the end of the section we comment on the range of parameters for which this is the case and also on whether any of the other poles can be physical. 

\begin{figure}[!h]
\begin{center}
\includegraphics[width=0.31\textwidth]{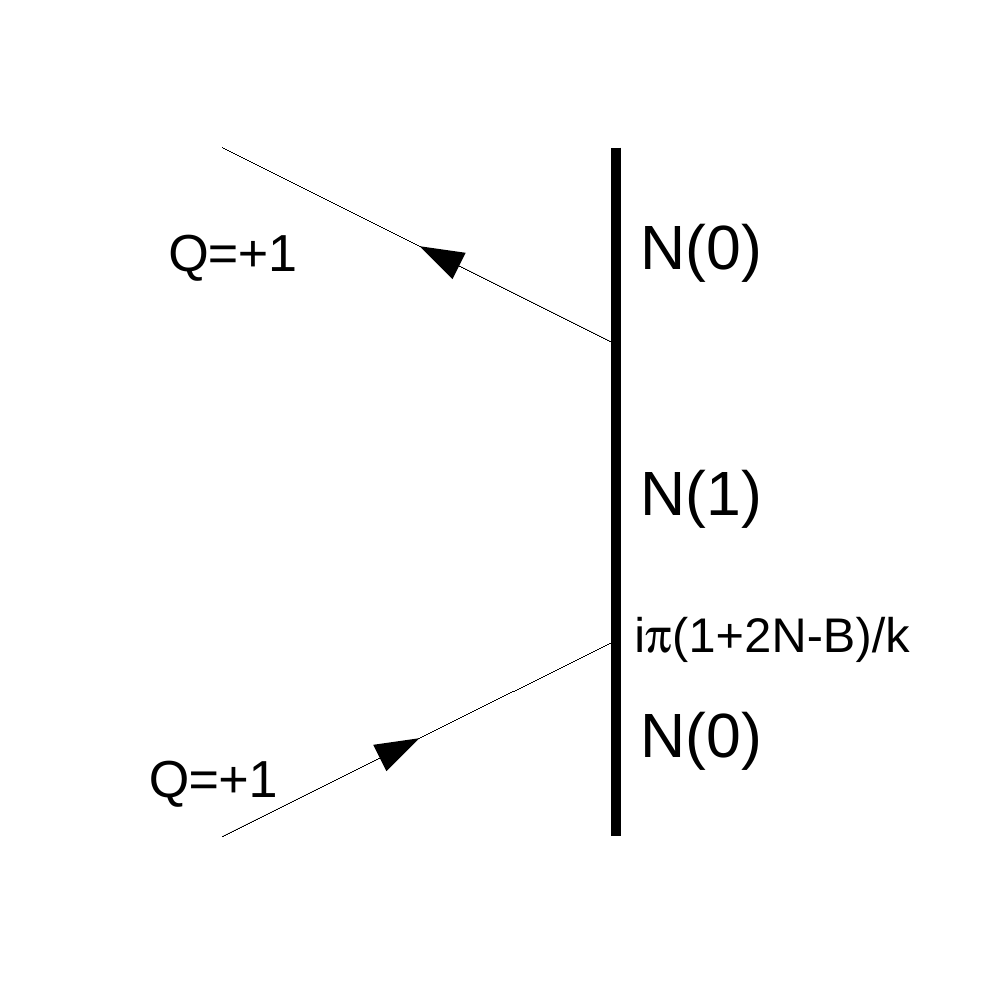}
\end{center}
\caption{Process that explains the physical pole $(1+2N-B)$ in $K_{1}^{N(0)}$.}
\label{fig:BoundaryReflection1}
\end{figure}

Let us begin our analysis with the pole in the physical strip that appears in $K_{1}^{N(0)}$, namely $(1+2N-B)$. As already discussed this pole corresponds to the fusion of an incoming particle with the unexcited boundary.  Figure \ref{fig:BoundaryReflection1} shows the reflection process where the bound state forms and then decays re-emitting the $Q=+1$ soliton.
The label $\frac{i\pi}{k}(1+2N-B)$ indicates the incoming rapidity at which the pole is present. 

\begin{figure}[!h]
\begin{center}
\subfigure[]{\label{fig:BR2}\includegraphics[width=0.31\textwidth]{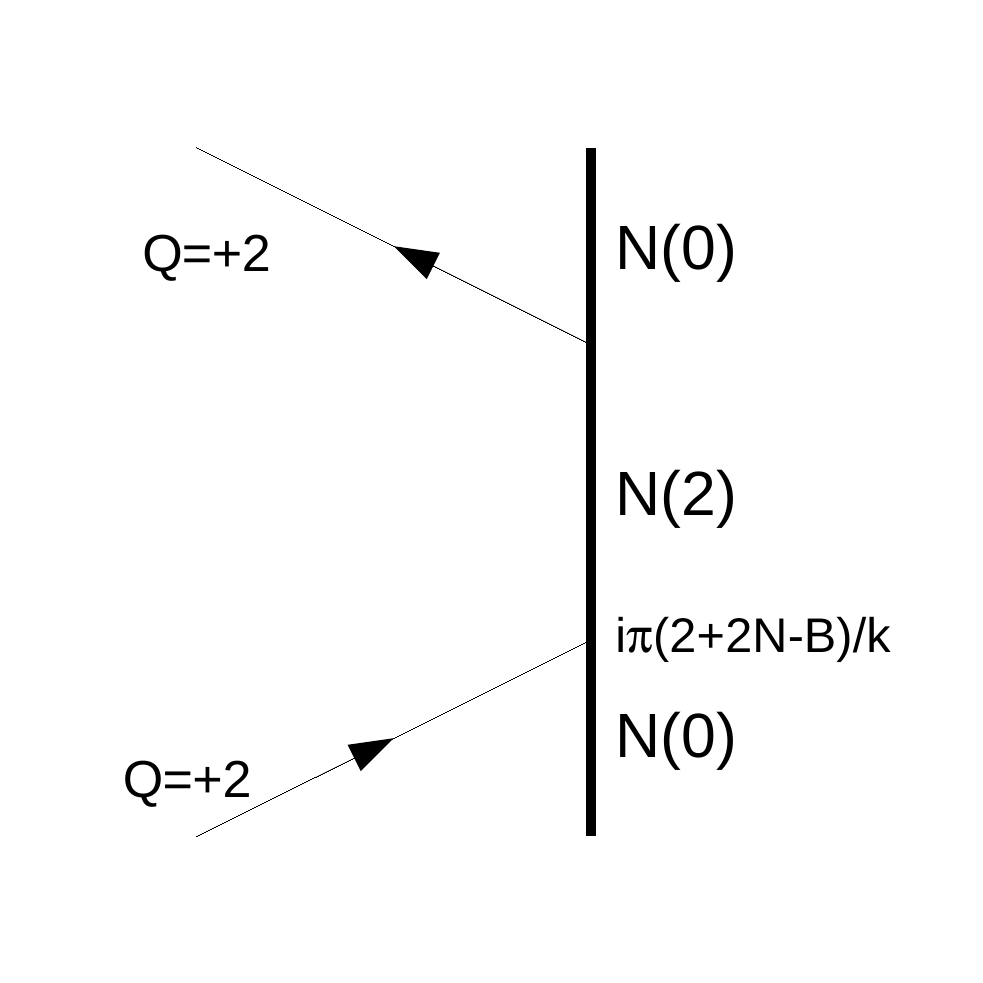}}
\subfigure[]{\label{fig:BR2a}\includegraphics[width=0.31\textwidth]{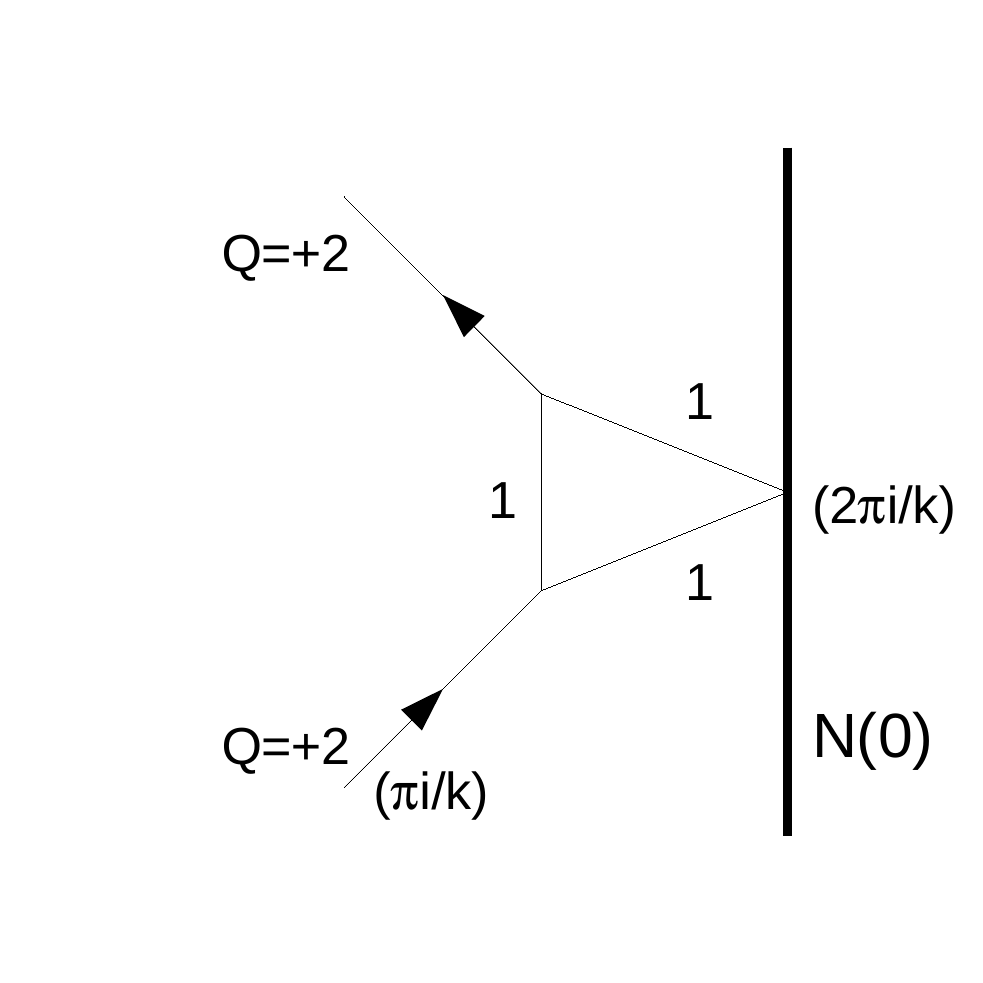}}
\subfigure[]{\label{fig:BoundaryReflection2a}\includegraphics[width=0.31\textwidth]{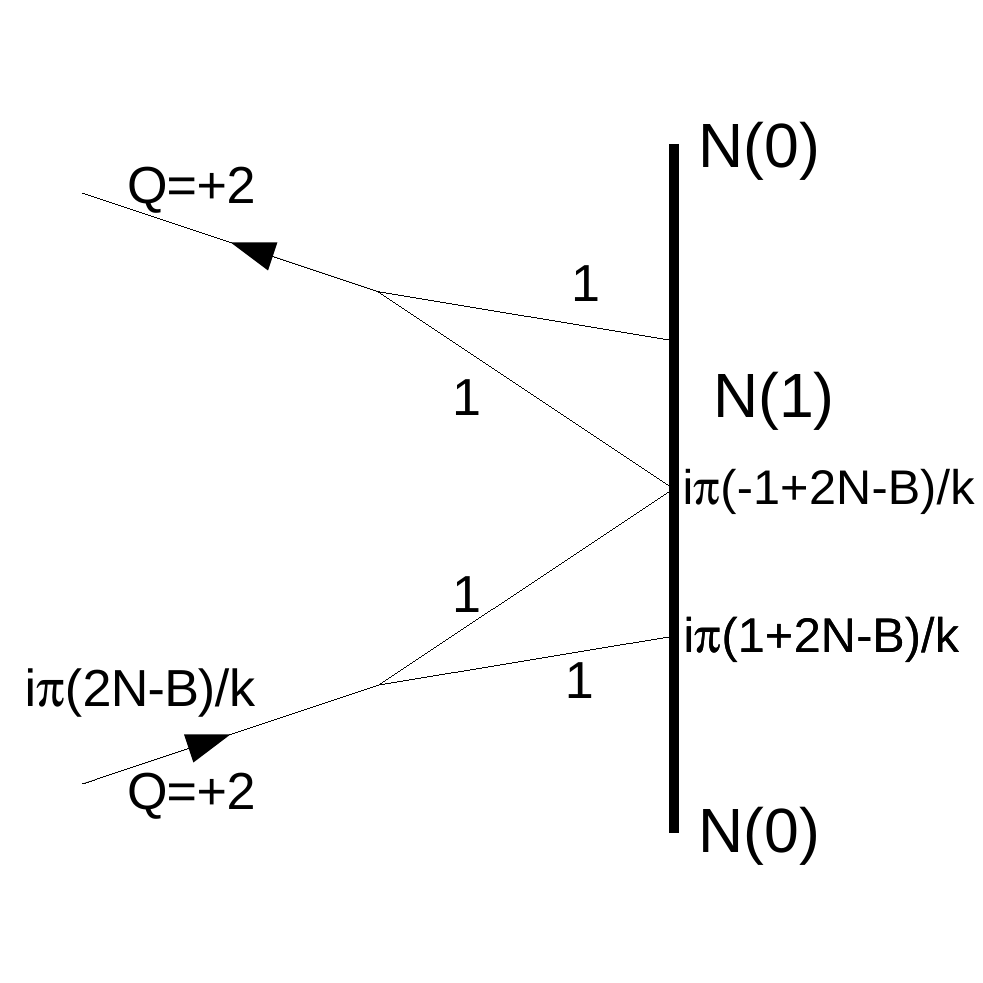}}
\end{center}
\caption{Processes that explain the three physical poles in $K_{2}^{N(0)}$.}
\label{fig:BoundaryReflection1X}
\end{figure}

The reflection factor for a $Q=+2$ soliton reflecting from a unexcited boundary (\ref{eq:K2N0}) has three such poles which for certain parameter choices are in the physical strip
\begin{equation}
(1)(2+2N-B)(2N-B)\, .
\end{equation}
These poles correspond to the processes shown in figure \ref{fig:BoundaryReflection1X}.
 The pole in factor $(2+2N-B)$ is due to diagram \ref{fig:BR2} where the incoming soliton fuses with the boundary, forming a bound state before this excited state decays. The pole in $(1)$ which is in the base factor, in this case $K_{2}^{base}$, is due to a process in which a boundary bound state is not formed. This is a fixed pole as it does not depend on the boundary parameters and arises from the triangular diagram shown in \ref{fig:BR2a} where the internal lines are on shell. It shows the incoming $Q=+2$ soliton decaying and recombining after one of the resultant charge $Q=+1$ solitons has reflected from the boundary. The final pole is from the factor $(2N-B)$ which is due to the process shown in figure \ref{fig:BoundaryReflection2a}, this is a mixture of the two previous processes. The incoming soliton decays before one of the resultant $Q=+1$ solitons fuses with the boundary to form the boundary bound state $N(1)$, the other resultant soliton reflects from the boundary before the bound state decays re-emitting the soliton which fuses with the reflected soliton.

\begin{figure}[!h]
\begin{center}
\subfigure[]{\label{fig:BR3}\includegraphics[width=0.31\textwidth]{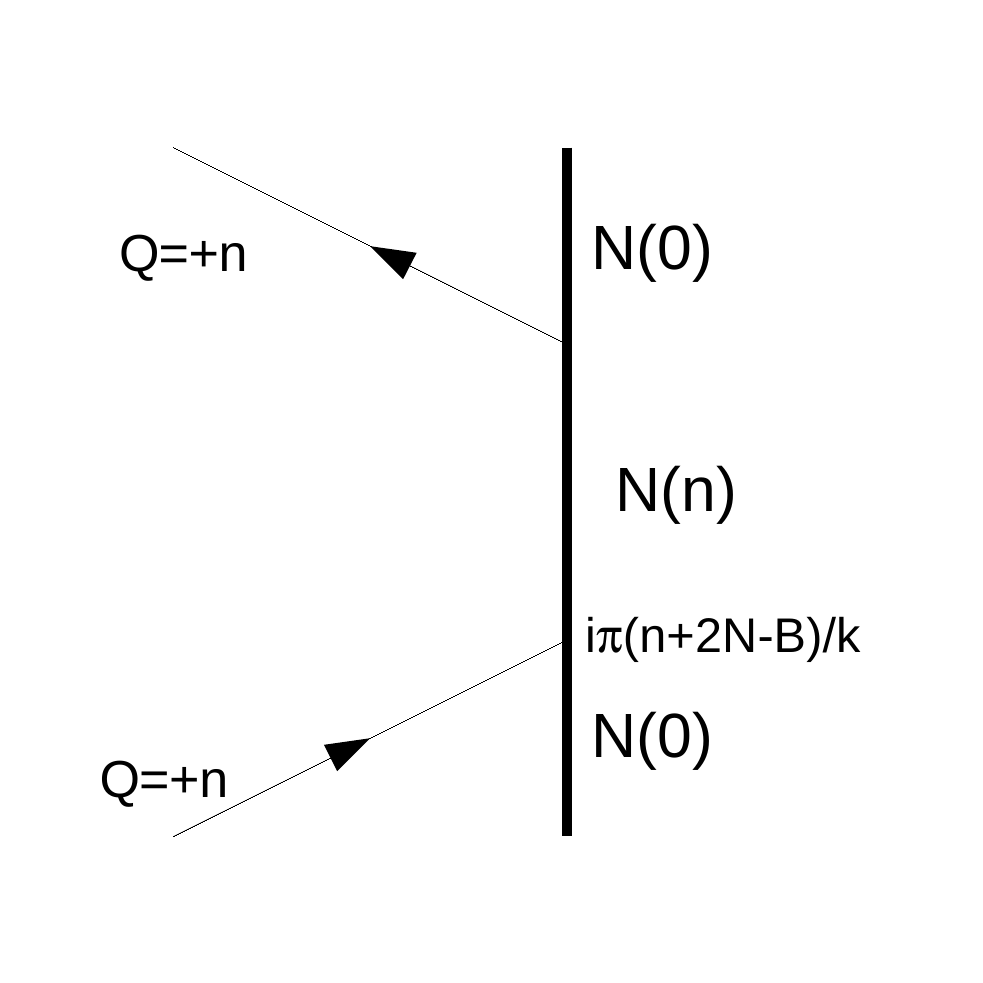}}
\subfigure[]{\label{fig:BR3a}\includegraphics[width=0.31\textwidth]{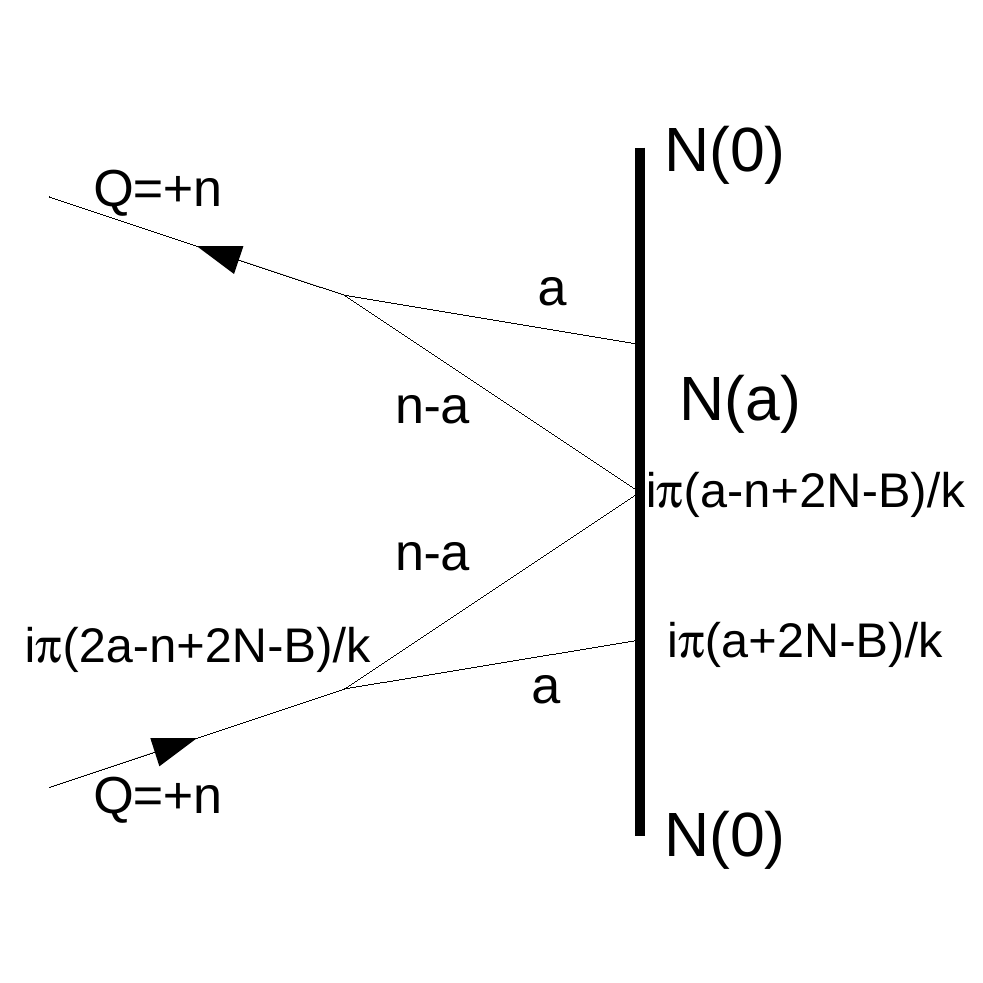}}
\subfigure[]{\label{fig:BoundaryReflection3a}\includegraphics[width=0.31\textwidth]{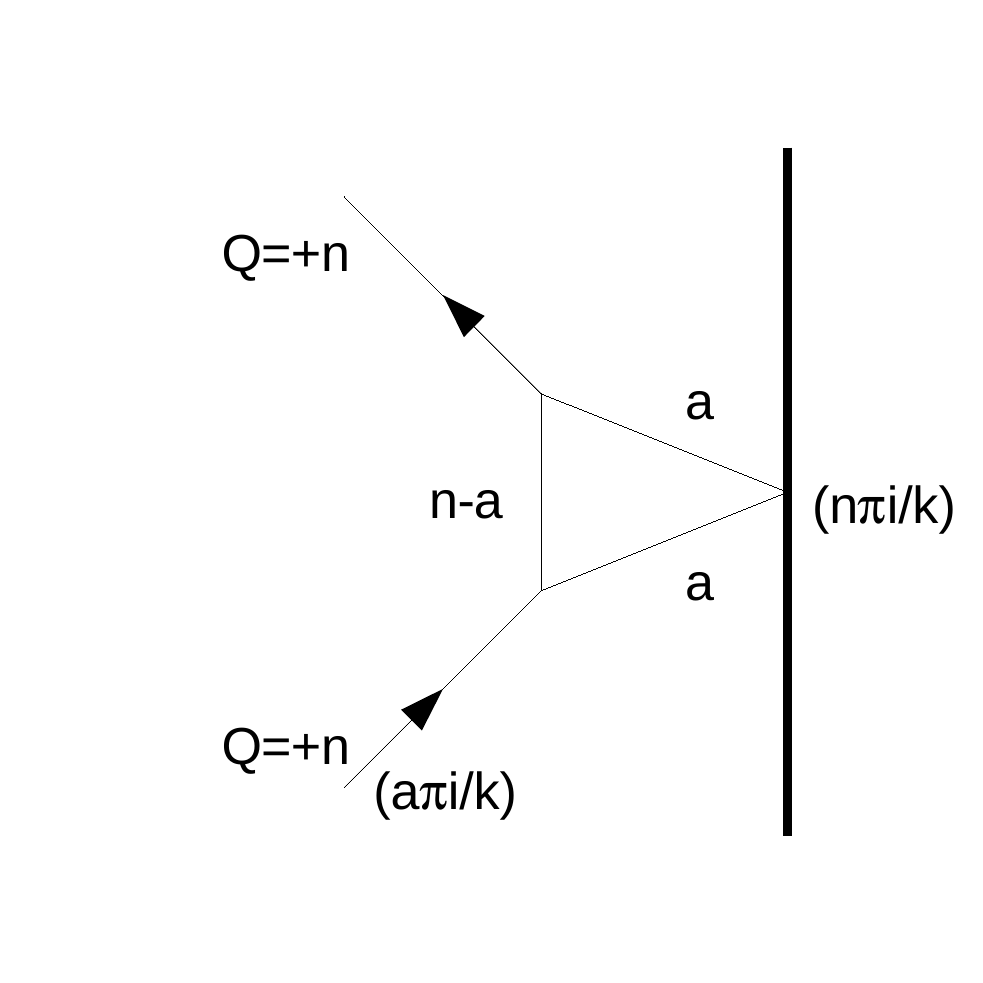}}
\end{center}
\caption{Processes that explain the three physical poles in $K_{n}^{N(0)}$.}
\label{fig:BoundaryReflection2aX}
\end{figure}

Increasing the incoming soliton to any charge $Q=+n$ the poles of the considered form that could be in the physical strip in $K_{n}^{N(0)}$ (\ref{eq:KnNO}) arise from diagrams of the same structure as just described. The pole in the factor $(n+2N-B)$ comes from the process in figure \ref{fig:BR3} and the poles in $(2a-n+2N-B)$ for $a = 1 \rightarrow n-1$ are associated with the process in figure \ref{fig:BR3a}. There are $n-1$ physical poles in $K_{n}^{base}$ which arise due to triangular diagrams shown in figure \ref{fig:BoundaryReflection3a} with $a = 1 \rightarrow n-1$.

The three types of diagrams described above are all that is needed to explain the poles in the reflection matrix for any charged soliton reflecting from a charge $Q=+N$ unexcited dressed boundary. We find that is not the case for the reflection from excited boundaries and more diagrams are needed.

\begin{figure}[!h]
\begin{center}
\subfigure[]{\label{fig:BR4}\includegraphics[width=0.31\textwidth]{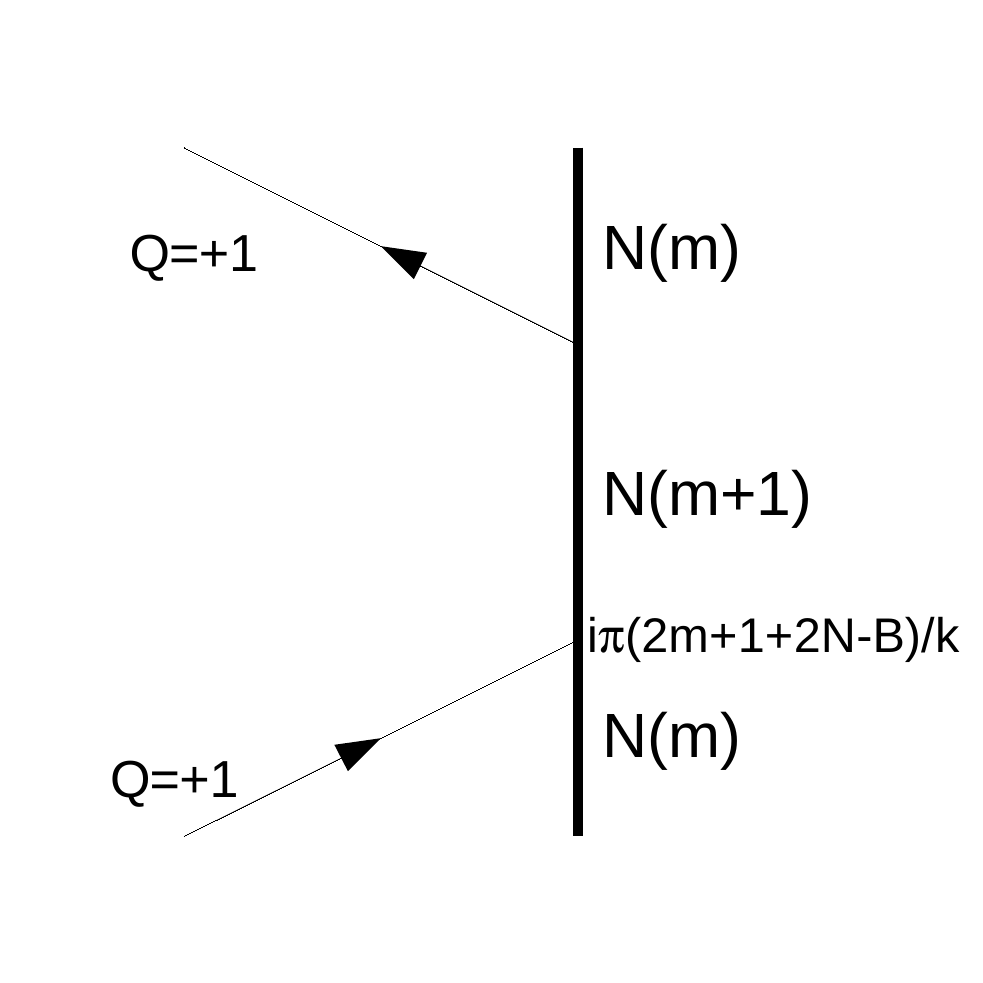}}
\hspace{0.5in}
\subfigure[]{\label{fig:BR4a}\includegraphics[width=0.31\textwidth]{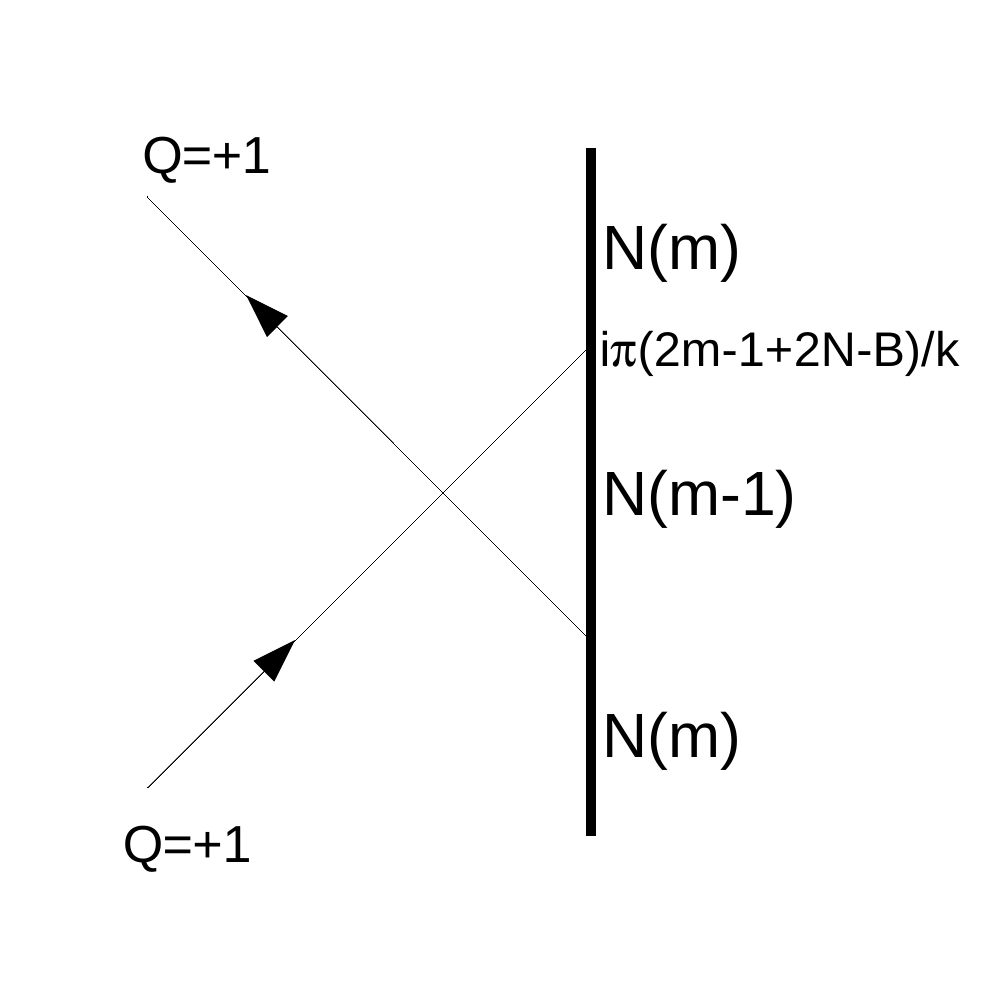}}
\end{center}
\caption{Processes that explain the three physical poles in $K_{1}^{N(m)}$.}
\label{fig:BoundaryReflection4X}
\end{figure}

Considering the reflection matrix $K_{1}^{N(m)}$ for the reflection of a charge $Q=+1$ soliton from an excited boundary with charge $Q=N+m$ (\ref{eq:K1Nm}), we find that it has poles in the factors $(2m-1+2N-B)\, , \ (2m+1+2N-B)$. The pole in $(2m+1+2N-B)$ arises because of the formation of the usual higher bound state, shown in figure \ref{fig:BR4}. The pole in $(2m-1+2N-B)$ is because the excited defect decays before it becomes re-excited, shown in \ref{fig:BR4a}. For this process to occur the excitation state of the original boundary has to be greater or equal to the charge of the scattering soliton, in this example for a $Q=+1$ soliton we need $m>0$.

\begin{figure}[!h]
\begin{center}
\subfigure[]{\label{fig:BR5}\includegraphics[width=0.31\textwidth]{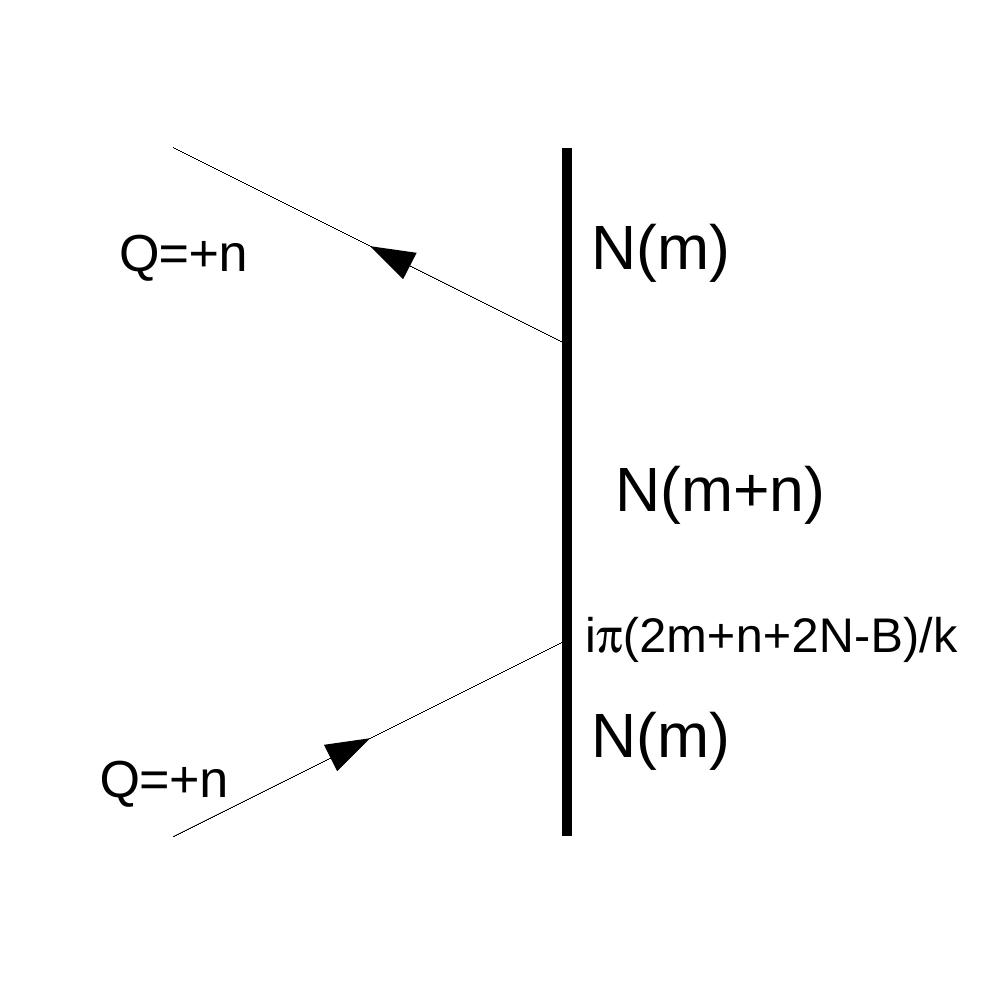}}
\hspace{0.5in}
\subfigure[]{\label{fig:BR5a}\includegraphics[width=0.31\textwidth]{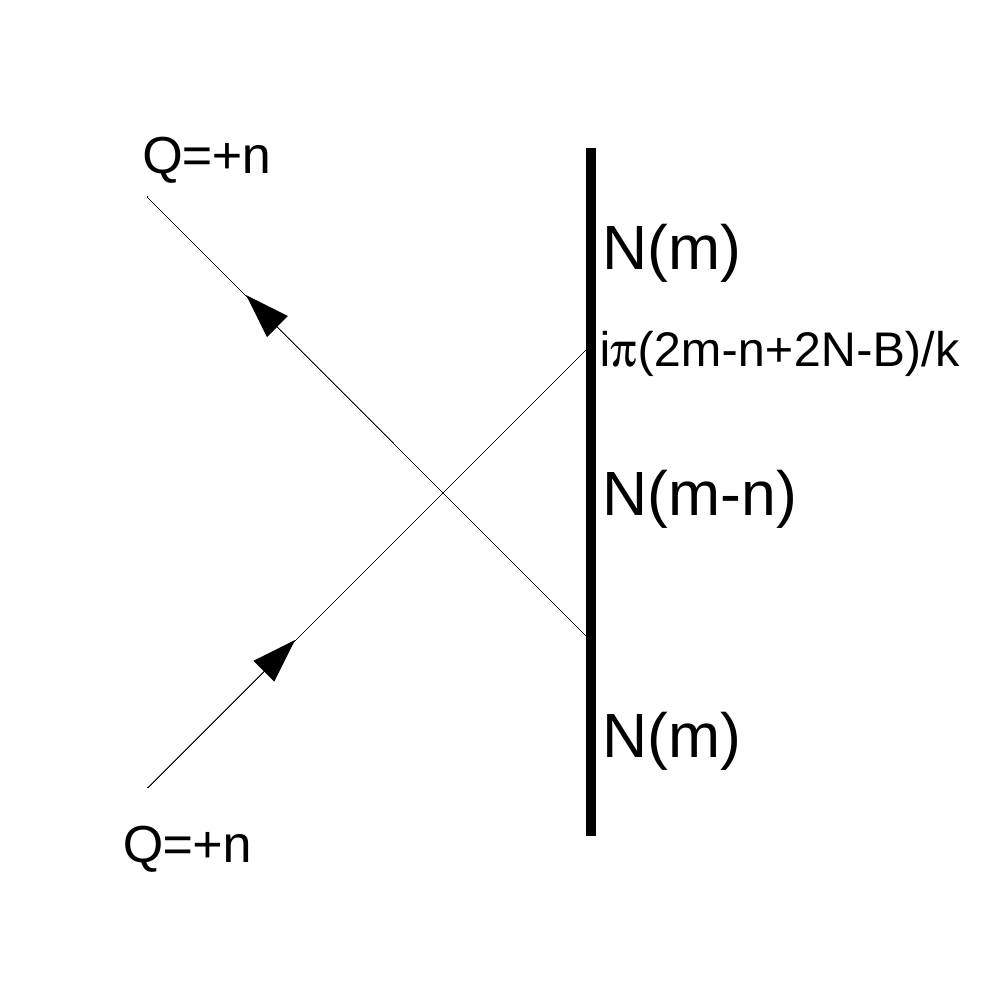}}
\end{center}
\caption[Processes that explain two of the physical poles in $K_{n}^{N(m)}$.]{Processes that explain two of the physical poles in $K_{n}^{N(m)}$, namely the poles in the factors  $(2m+n+2N-B)$ and $(2m-n+2N-B)$.}
\label{fig:BoundaryReflection5X}
\end{figure}

We continue to analyse the reflection factor for any charged soliton $Q=+n$ reflecting from an excited boundary $K_{n}^{N(m)}$ (\ref{eq:KnNm}). It has $n-1$ poles that appear in the $K_{n}^{base}$ factor which are due to the triangular diagrams, shown in figure \ref{fig:BoundaryReflection3a} with $a = 1 \rightarrow n-1$. The pole in the factor $(2m+n+2N-B)$ comes from the standard process of the formation of a higher bound state illustrated in figure \ref{fig:BR5}. Figure \ref{fig:BR5a} shows the process that corresponds to the pole in $(2m-n+2N-B)$, where the boundary decays before fusing with the incoming soliton to reform the original excited boundary. We note that this process only occurs for $n \leq m$. 

\begin{figure}[!h]
\begin{center}
\subfigure[]{\label{fig:BoundaryReflection6}\includegraphics[width=0.45\textwidth]{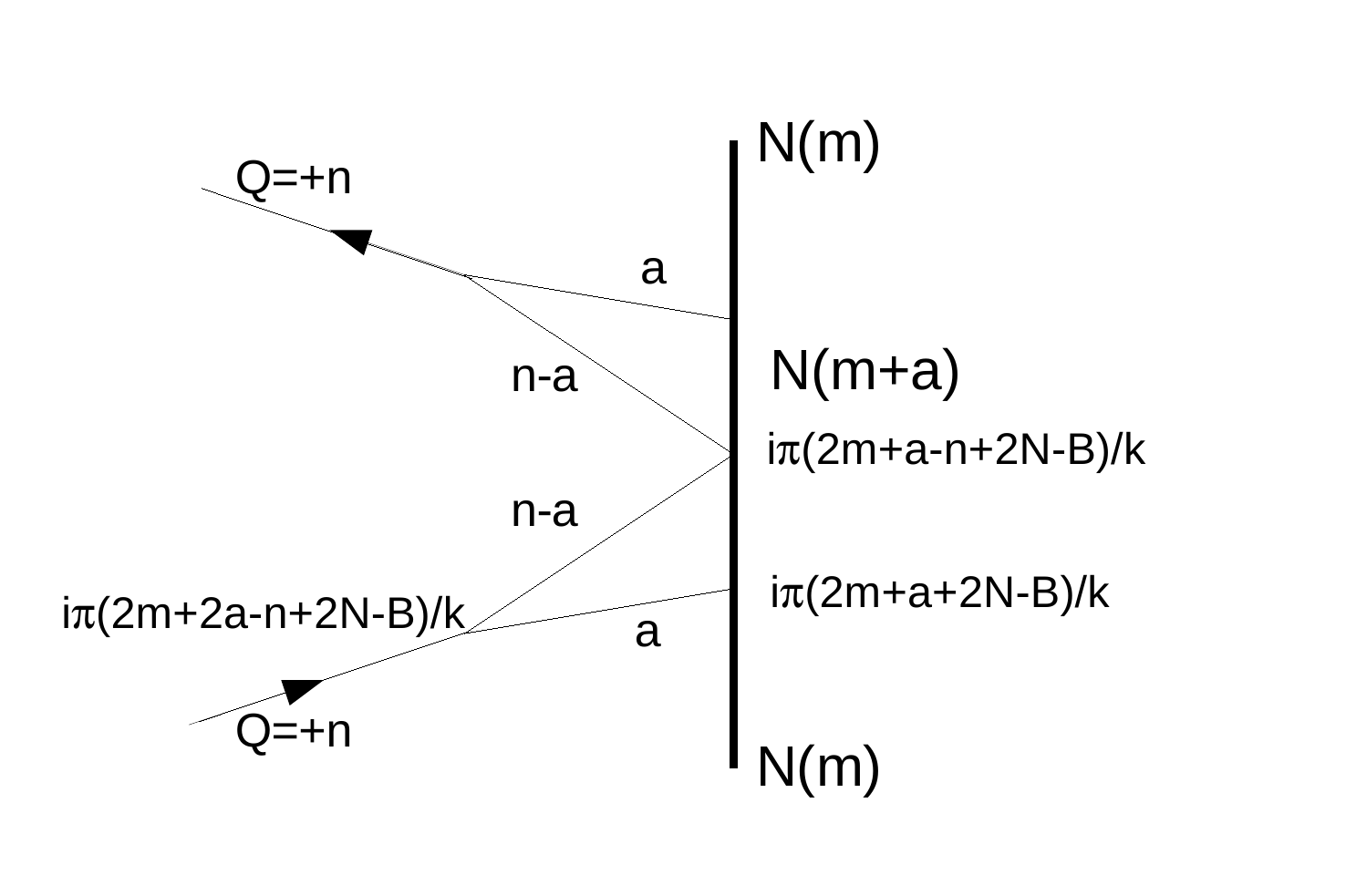}}
\subfigure[]{\label{fig:BoundaryReflection7}\includegraphics[width=0.45\textwidth]{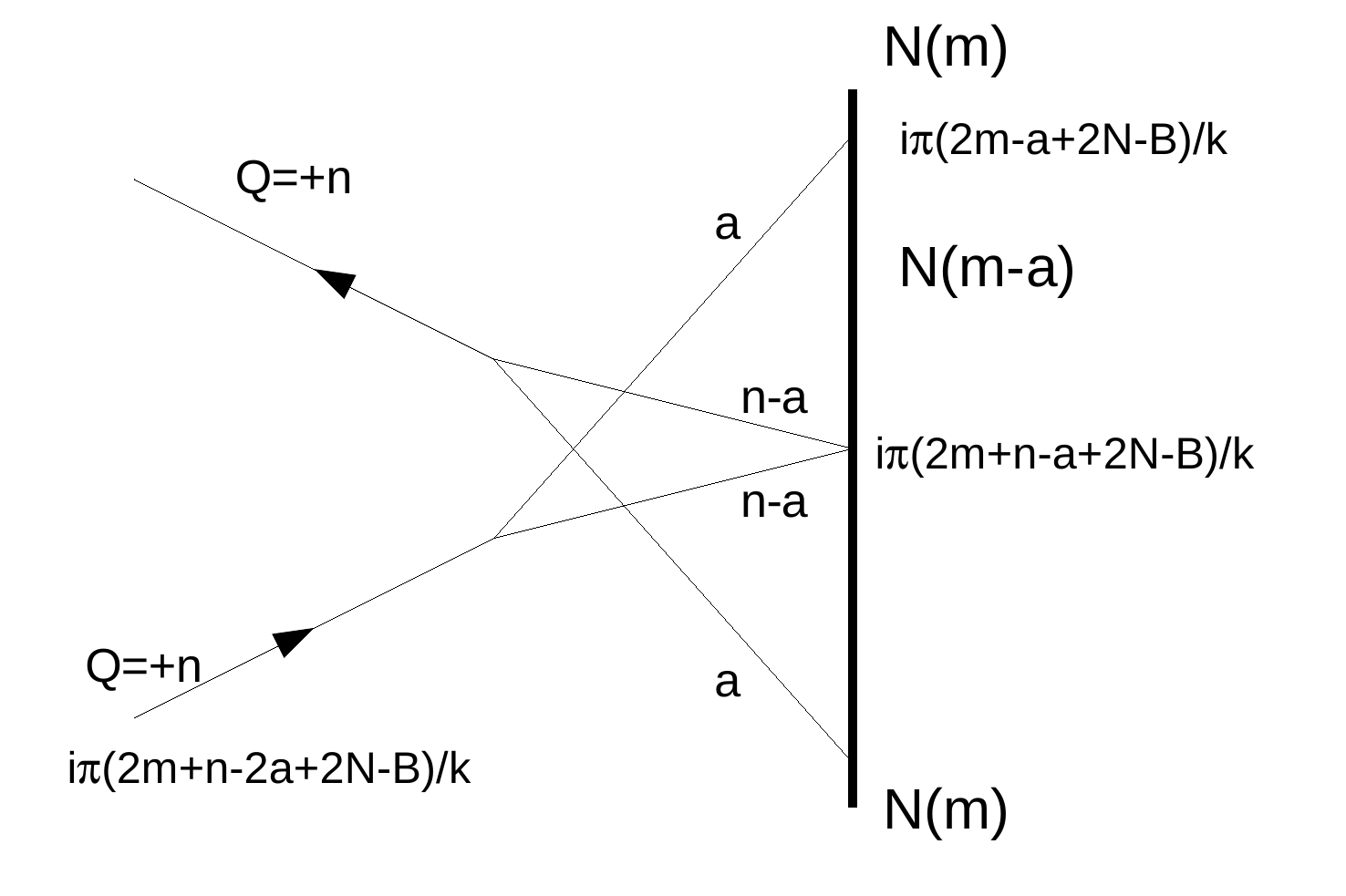}}
\end{center}
\caption[Processes that explain the remaining physical poles in $K_{n}^{N(m)}$.]{Processes that explain the remaining physical poles in $K_{n}^{N(m)}$, namely the double poles in the factor $(2m-n+2j+2N-B)^2$.}
\label{fig:BoundaryReflection6X}
\end{figure}

The double poles from the factors $(2m-n+2j+2N-B)^2$ correspond to two different processes. All $n-1$ poles arise due to the process shown in figure \ref{fig:BoundaryReflection6} with $a=1 \rightarrow n-1$ as long as $n \geq 1$. Whereas $r$ poles arise due to the new process shown in figure \ref{fig:BoundaryReflection7} where $a=1 \rightarrow r$ and $r$ is the lower value of $m$ and $n-1$. In this process the excited boundary decays by emitting a charged soliton which fuses with the reflected soliton, a remnant of the decayed incoming soliton. We note that only when $m > n-1$ is there the maximum number of poles. For $m \leq n-1$ the poles that are not explained by the restrictions on processes in figures \ref{fig:BR5}, \ref{fig:BoundaryReflection7} cancel with zeroes in the product $\prod_{j=0}^{n-1}(n-2j-2N+B)$.

We have described Coleman-Thun processes that explain all the poles of the form $(x+2N-B)$ that appear the various reflection matrices. As we saw with examples I and II the poles that we bootstrapped on do not stay in the physical strip and become unphysical when the bound state with maximum energy has been reached. At this point the bootstrap procedure is halted after a finite number of steps, e.g. $m=17$ in example I. In the examples only the reflection matrix for a $Q=+1$ soliton reflecting from the boundary was considered, we revisit example I to investigate whether a similar pattern exists for higher charged solitons.

\subsection{Example I revisited: $A=0 \, , \ b = \frac{\pi}{3}\, ,\ k=100$}
The pole that we bootstrapped on in the reflection matrix for a $Q=n$ soliton from an unexcited boundary (\ref{eq:KnNO}) appears in the factor $(2N - B +n)$, this corresponds to the formation of a charge $Q=N+n$ bound state and has the value
\begin{equation}
 \theta = \frac{i\pi}{300}(3n + 50)\, ,
\end{equation}
which is physical for $1 \leq n \leq 33$. Therefore a soliton with charge greater than $Q=33$ cannot fuse with this unexcited boundary to form a bound state. Continuing to the reflection matrices (\ref{eq:KnNm}) which describe a charge $Q=n$ soliton scattering with an excited boundary of charge $Q=N+m$, the pole we bootstrapped on appears in the factor $(2N - B + 2m +n)$ which corresponds to the formation of a higher bound state with charge $Q=N+m+n$ and has the value
\begin{equation}
 \theta = \frac{i\pi}{300}(3n + 6m + 50)\, ,
\end{equation}
which is physical for $6m \leq 100 - 3n$. This limits to the original example I when $n=1$ and the pole is physical until $m=17$. For the scattering of the charge $Q=2$ soliton the last boundary it can fuse to is the $15^{th}$ excited state, since from the $16^{th}$ excited state the resultant boundary has lower energy. Similarly a charge $Q=+3$ soliton can fuse to the $15^{th}$ excited state also. These can be deduced by looking at the charge parameter for excited boundaries with different charges and determining whether the fusion of a certain charged soliton will create a higher or lower energy state.
 
We have shown for this example the bootstrap procedure that we implemented is valid for a finite number of steps depending on the charge of the scattering soliton. In fact the pole we started the bootstrap procedure on in (\ref{eq:K1N0}) is the only physical pole in this reflection matrix when
\begin{equation}
\frac{\pi}{k} \leq A + b \leq \frac{\pi}{2} - \frac{\pi}{k} \, , \ \ \ \ \ A < b - \frac{\pi}{k}\, , \ \ \ \ \ \ k \geq 4 \, .
\end{equation}
For these range of parameters the bootstrap procedure we implemented and pole analysis is valid, at least for a finite number of steps. The bound states accessed by the bootstrap procedure are all related to the $E_{bs}^{+}$ energy curve created by fusing solitons to the unexcited boundary with $a=A+b$. We will now briefly examine an example which falls outside of these parameter constraints.

\subsection{Example III: $A=\frac{\pi}{4}\, , \ b = \frac{\pi}{8}\, ,\ k=96$} 

The unexcited boundary has charge $Q=-12$ and the poles in $K_{1}^{N(0)}$ (\ref{eq:K1N0}) appear at the rapidities
\begin{eqnarray}
(1-k) && -95\pi i \nonumber  \\
(1+2N-B) && \frac{13}{96} \pi i \nonumber \\
(k+B-1+2N) && \frac{35}{96}\pi i \, ,
\end{eqnarray}
where two poles lie in the physical strip $(1+2N-B)$ and $(k+B-1+2N)$. This circumstance lies outside all the previous analysis. The residues of these poles have opposite signs, the pole in the factor $(1+2N-B)$ has the residue $1.65\, i$ and $(k-1+2N+B)$ $-3.58\, i$. This second physical pole is at the rapidity for the emission process, shown in figure \ref{fig:Boundaryemission1}, and possibly is explained by a crossed process because it is a remnant pole from the lower energy boundary reflection matrix $K_{1}^{N(-1)}$. 

If we consider our bootstrap procedure then the excited boundaries generated have the reflection matrices $K_{1}^{N(m)}$ (\ref{eq:K1Nm}), which has poles at the rapidities
\begin{eqnarray}
(1-k) && -95\pi i \nonumber  \\
(1-2N+B) && -\frac{11}{96} \pi i \nonumber \\ 
(k+B-1+2N) && \frac{35}{96}\pi i \nonumber \\
(2m+1+2N-B) && \frac{11+2m}{96} \pi i \nonumber \\
(2m-1+2N-B) && \frac{13+2m}{96} \pi i \, .
\end{eqnarray}
where the pole from the factor $(2m-1+2N-B)$ stays in the physical strip until $m=18$. This seems similar to the other examples but the charge parameter of the bound soliton has the value $a_{m} = 0$, which lies outside of the classical region where bound states exist $\left[ \frac{\pi}{8}, \ \frac{3\pi}{8} \right]$. In fact $a_{m}$ moves outside the classical region at $m=13$. This property where the pole stays physical outside the allowed classical region is not explained by our analysis. Examining the residues of the poles we find that there are positive imaginary for $m=0 \rightarrow 11$ but becomes negative imaginary for $m=12$ the exact value which if bootstrapped on would take $a$ outside the classical range. So we have found that the poles we have not been able to explain, both have the feature that their residue is negative imaginary, while all the poles we have explained by the formation of bound states or Coleman-Thun processes have a positive imaginary residue. 

\section{Summary and Discussion}

In this paper we have built on the work in \cite{Bowcock:2008jf} where integrable CSG defect and dressed boundary theories were constructed. We reviewed aspects of the quantum CSG theory, showing that the charge of the CSG soliton is quantised. The quantum $S$-matrix was presented and the existence of poles in the $S$-matrix due to the formation of higher charge bound states and Coleman-Thun processes was explained.

We comprehensively analysed the classical bound states, discovering that there are two limits in which the bound state becomes an unexcited boundary. Both where the bound soliton is positioned at right infinity, but with different charge parameters $a=A+b$ or $A=A-b$. Using the Bohr-Sommerfeld quantisation condition we showed that the charge of the boundary bound states is quantised and through the fact that the unexcited boundary appears as a limit of the bound states, all the boundaries in the quantum theory have integer charge. 

We calculated the semi-classical energy difference between states differing by one unit of charge and showed that the energy difference is exactly that of a soliton with specific rapidity. This discovery prompted the analysis of soliton-boundary fusion processes and we found that a charge $Q=+1$ soliton can fuse with an unexcited boundary to form a excited bound state at the rapidity in agreement with the semi-classical energy difference. This fusion process can be repeated with the effect on the bound soliton of altering its charge parameter $a \rightarrow a - \frac{2\pi}{k}$.

Using the existence of these bound states we conjectured the form of the quantum reflection matrix for a charge $Q=+1$ soliton from the unexcited boundary $K_{1}^{N(0)}$, checking that the conjectured form is in agreement with the classical limit. From $K_{1}^{N(0)}$ we used the reflection bootstrap procedure to generate reflection matrices for any charged soliton from the unexcited boundary $K_{n}^{N(0)}$, checking the classical limit for the anti-particle and that it displayed the charge periodicity property of the soliton. We note that we found that this was only possible for $k$ even. To complete the construction of the remaining reflection matrices which describe the the scattering of a $Q=+1$ soliton from an excited boundary, we used the boundary bootstrap and then to generalise to any charged soliton repeated the reflection bootstrap. We check that the factors are in agreement with the reflection from the classical bound state and that the periodicity of the boundary charge is preserved.

Finally we completed a preliminary analysis of the physical poles that appear in the various matrices. For a range of the parameters we explain the physical poles either by the formation of higher bound states by the fusion process described above or by Coleman-Thun processes. The examples we studied showed that the bootstrap method is only valid for a finite number of steps until the bound state of highest energy is reached, at this point the pole we bootstrapped on becomes unphysical. Further examples showed that there exists physical poles with negative residues that we have not explained. Further analysis is required to complete the picture and to better understand the bound states present on the $E_{bs}^{-}$ energy curve.

\subsubsection{Acknowledgements}
We would like to thank Ed Corrigan, Patrick Dorey and Paul Sutcliffe for discussions. 
JMU thanks the Engineering and Physical Sciences Research Council for a PhD studentship.

\vspace{1in}
\appendix
\begin{Large}{\textbf{Appendix}}\end{Large}

\section{Closure of reflection bootstrap}
\label{appen:closure}
In this appendix we check the closure of the reflection bootstrap procedure implemented in section \ref{sec:reflecboot}. We found the reflection factor for a charge $Q=n$ soliton from a charge $Q=N$ unexcited boundary to be
\begin{equation}
K_{n}^{(N(0))}(\theta) = K_{n}^{\mathrm{base}}\prod_{j=0}^{n-1}(2N-B+2j+2-n)(k+B+2N-2j-2+n)\, . 
\end{equation}
To check that this formula closes means to check that
\begin{equation}
K_{1}^{(N(0))}(\theta) =  K_{k+1}^{(N(0))}(\theta)\, ,
\end{equation}
which is checking that the reflection factors show the property that the soliton's charge is periodic.
First using the crossing symmetry relation
\begin{equation}
 K_{1}(\theta)\ K_{-1}(\theta +i\pi) = S_{1,1}(2\theta)\, ,
\end{equation}
implies
\begin{equation}
K_{-1}(\theta +i\pi) = -(1)(1-k)(k-1)(-1-2N+B)(1-2N-k-B) \, ,
\end{equation}
which becomes
\begin{equation}
K_{-1}(\theta) = (1-k)(k-1-2N+B)(1-2N-B)\, .
\end{equation}
This is promising, as it has the correct pole $(1-2N-B)$ for the known bound state and limits to the classical reflection factor for the anti-particle. To show closure we need $K_{k-1}(\theta)$ to give the same result. First recall $K_{1}^{\mathrm{base}} = (1-k)$ and
\begin{equation}
 K_{k-1}^{\mathrm{base}} = (1-k)(2)(2-k)(3)(3-k)...(k-3)(-2)(k-2)(-1),
\end{equation}
using $(x)(-x) = 1$ shows for $k$ even $K_{k-1}^{\mathrm{base}} = (1-k)$. To check the other factors in
\begin{equation}
K_{k-1}^{(N(0))}(\theta) = K_{k-1}^{\mathrm{base}}\prod_{j=0}^{k-2}(2N-B+2j+3-k)(B+2N-2j-3),
\end{equation} 
we write out terms in the product
\begin{center}
\resizebox{\textwidth}{!}{
\begin{tabular}{ccc||ccc}
$j$&&factor&$j$&&factor\\
\hline
$0$ &&$ (2N-B+3-k)(B+2N-3)$ & $k-2$&&$(2N-B+k-1)(B+2N+1)$\\ 
$1$ &&$ (2N-B+5-k)(B+2N-5)$ &$ k-3$&&$(2N-B+k-3)(B+2N+2)$\\
\vdots&&&\vdots&&\\
$\frac{k}{2}-3$&&$(2N-B-3)(-k+B+2N+3)$&$\frac{k}{2}+1$&&$(2N-B+5)(-k+B+2N-5)$\\
$\frac{k}{2}-2$&&$(2N-B-1)(-k+B+2N+1)$&$\frac{k}{2}$&&$(2N-B+3)(-k+B+N-3)$\\
$\frac{k}{2}-1$&&$(2N-B+1)(-k+B+2N-1)$&&&\\
\end{tabular}\, ,
}
\end{center}
reordering these terms
\begin{center}
\resizebox{\textwidth}{!}{
\begin{tabular}{ccc||ccc}
&&factor&&&factor\\
\hline
&&$ (2N+B-3)(2N-B+3)$ & &&$(2N-B+1)$\\ 
 &&$ (2N+B-5)(2N-B+5)$ &&&$(2N+B+1)(2N-B-1)$\\ 
 &&$ (2N+B-7)(2N-B+7)$ &&&$(2N+B+3)(2N-B-3)$\\
&&\vdots&&&\vdots\\
&&$(2N+B-k+3)(2N-B+k-3)$&&&$(2N+B+k-7)(2N-B-k+7)$\\
&&$(2N+B-k+1)(2N-B+k-1)$&&&$(2N+B+k-5)(2N-B-k+5)$\\
&&$(2N+B-k-1)$&&&$(2N+B+k-3)(2N-B-k+3)$\\
\end{tabular}\, ,
}
\end{center}
shows that the terms appear in pairs expect for two terms, using $(x)(-x)=1$ allows the array of terms to be completed and $K_{k-1}^{(N(0))}$ rewritten
\begin{eqnarray}
K_{k-1}^{(N(0))}(\theta) &=& (1-k)(-2N+B-k-1)(1-2N-B) \nonumber \\
&& \ \ \ \ \ \ \ \ \ \ \ \mathrm{x} \ \ \prod_{j=0}^{k-1} (2N+B+1+2j)(2N-B+1+2j)\nonumber \\
&=&(1-k)(-2N+B-k-1)(1-2N-B) \nonumber \\ 
&=& K_{-1}^{(N(0))}(\theta) \, .
\end{eqnarray} 
The terms in the product cancel as $2N$ is an even integer and therefore the product as the form
\begin{equation}
\prod_{j=0}^{k-1} (B+1+2j)(-B-1-2j) = 1\, .
\end{equation}
We have shown that the reflection bootstrap close for $k$ even.

\bibliographystyle{hunsrt}
\bibliography{QuantumPaper}

\end{document}